\pdfoutput=1

\documentclass[12pt]{article}
\usepackage[T1]{fontenc}
\usepackage{lmodern}
\usepackage{amsmath,array}
\usepackage{graphicx}
\usepackage{enumerate}
\usepackage{natbib}
\usepackage{url} %
\usepackage{setspace}
\usepackage[justification=centering,singlelinecheck=false]{caption}

\usepackage[normalem]{ulem}
\usepackage{lscape}
\useunder{\uline}{\ul}{}
\usepackage{accents}
\usepackage[flushleft]{threeparttable}
\usepackage{bigstrut}
\usepackage{multirow}
\usepackage{adjustbox,appendix}
\usepackage{xcolor}
\usepackage[colorinlistoftodos,textsize=footnotesize]{todonotes}
\usepackage[bookmarks=false,hypertexnames=false]{hyperref}\hypersetup{colorlinks=true, linkcolor=blue, filecolor=gray, urlcolor=blue, citecolor=black}
\newtheorem{thm}{Theorem}

\newtheorem{lemma}{Lemma}
\newtheorem{corollary}{Corollary}
\newtheorem{assum}{Assumption}

\newtheorem{definition}{Definition}

\hypersetup{
  colorlinks=true,linkcolor=blue, citecolor=blue, filecolor=blue, 
  linkbordercolor={0 1 1}, citebordercolor={1 0 0},
  pdfcreator={\LaTeX\ with package \flqq hyperref\frqq},
  pdfsubject={},
}

\begin{document}

\def\spacingset#1{\renewcommand{\baselinestretch}%
{#1}\small\normalsize} \spacingset{1}

\title{A Non-Asymptotic Class of Dense Precision Matrix Estimators: Complexity, Signal, and Double Descent}
\author{\textsc{Mehmet Caner\thanks{
North Carolina State University, Nelson Hall, Department of Economics, NC 27695. Email: mcaner@ncsu.edu. }}
\and \textsc {Agostino Capponi%
\thanks{Department of Industrial Engineering and Operations Research and Columbia Business School, Columbia University. Email: ac3827@columbia.edu.}}
\and \textsc{Maurizio Daniele %
	\thanks{ETH Zurich, KOF Swiss Economic Institute, 8092 Zurich, Switzerland. Email: daniele@kof.ethz.ch}}
\and \textsc{Mihailo Stojnic %
\thanks{Department of Industrial Engineering and Operations Research, Columbia University. Email: flatoyer@gmail.com. 
We thank Hoi Dinh for  research assistance. We thank Jushan Bai, Jianqing Fan, Bernard Salanie, and the participants of the Columbia Econometrics Workshop, Princeton-ORFE Wilks Statistics seminar, Midwest Econometrics Conference, FINML 2025 Erasmus university, University of North Carolina Chapel Hill Econometrics Seminar, Chinese University of Hong Kong Economics Meeting, City University of Hong Kong Econometrics Seminar for constructive feedback and comments.}}}

\maketitle

\begin{abstract}

Precision matrix estimation is a cornerstone problem in statistics, economics, and finance. Estimators that are dense, high-dimensionally consistent, and free of latent factor estimation have so far been available only separately. We introduce a general class achieving all three jointly, extending the projection-based linear predictors developed for a single regression coefficient in a hidden factor model \citep{bbsmw21, bsmw22} to the full precision matrix, while establishing error bounds that hold uniformly across all $p$ rows rather than pointwise for one outcome. We solve the main difficulty of estimating the diagonal term, which does not extend by tools of \cite{bbsmw21} \cite{bsmw22}.
Our analysis isolates three random quantities, complexity, retained signal magnitude, and  projection bias, that jointly govern the estimation error under a signal-to-noise-ratio condition. A tuning-free special case, the ridgeless regression estimator, exhibits potential double descent in estimation error as the number of outcomes increases relative to the sample size, a matrix-valued counterpart to the phenomenon known from linear regression. An S\&P 500 portfolio application illustrates an analogous, doubly ascending pattern in realized Sharpe ratios as portfolio size grows.

\end{abstract}

\noindent {\bf{Keywords}}: Interpolation, Double Descent, Latent Space, Factor Models, Precision Matrix

\newpage
\spacingset{1.8}

\section{Introduction}\label{intro}

Precision matrix estimation is central to portfolio formation, economic network analysis, and inference with many covariates. When the number of outcomes is large relative to the sample size, estimation requires structure. Sparsity and factor representations provide two important approaches, but many economic systems may have numerous small, nonzero links. This motivates studying dense precision matrix estimators and stating explicitly the conditions and loss under which they are consistent.

We study a class of projection-based estimators of the population precision matrix $\Theta=\Sigma^{-1}$. Each outcome is regressed on the remaining outcomes, and the resulting coefficient vector is combined with an estimate of the reciprocal diagonal entry. The estimators need not be sparse and do not require plug-in estimates of latent factors or loadings. A latent factor representation instead supplies the structure used in the analysis. We call this property \emph{factor-model-estimation-free}.

Our construction builds on the latent factor regression predictors of \cite{bbsmw21,bsmw22}. Their framework supplies the projection-based coefficient estimator and the complexity, signal, and projection-bias quantities used in our bounds. Our objective is different: to estimate all rows of a precision matrix simultaneously. This requires uniform coefficient bounds and control of the diagonal normalization with new techniques compared to them. 

The paper makes three contributions. First, it develops uniform error bounds for a class of dense precision matrix estimators under an exact latent factor model, replacing row-sparsity requirements with signal and growth conditions. Second, it studies adaptive principal component regression (PCR) and ridgeless regression (RRE) as specific choices within this construction. The diagonal normalization is essential: in the interpolating regime, the residual-based normalization vanishes, and RRE uses a different estimator under an additional condition discussed in Corollary~\ref{cl6} and Appendix A. Third, the simulations examine how estimation error varies with dimensionality, and an equity portfolio application illustrates the economic performance of the procedures.

The analysis distinguishes complexity, retained signal, and projection  bias. Complexity increases the coefficient-error bound, whereas  retained signal reduces it. Projection bias contributes an additional term. In the overparameterized RRE case, the effective noise dimension can also reduce one component of the bound. These results motivate investigating double descent, but do not prove a non-monotone risk curve at fixed sample size. %

{\bf Previous Literature:} The closest methodological precedents are \cite{bbsmw21,bsmw22}; the distinction from those papers, we investigate the full precision-matrix target with proof of  main diagonal terms, which cannot extend from their papers. Sparse regression-based method, such as \cite{caner2019}, uses lasso based nodewise regression to estimate rows, hence  imposes strong sparsity structure on the precision matrix of outcomes. Factor-based procedures such as \cite{fan2013,caner2022} estimate a factor component with sparsity imposed on covariance matrix of errors, while shrinkage approaches such as \cite{lw2004,lw2017} use regularization and their methods are dense estimation, model-free but analyzes different objective than consistency. The nonlinear observed-factor via deep learning approach of \cite{cd2025} is another relevant comparison.
Their method is only consistent when the number of outcomes are less than the sample size.
Appendix D provides an extended comparison of the articles, including the distinction from the portfolio estimators of \cite{cdsz26}. Our simulations show that our estimators perform well under factor models.
In our empirical application, we report realized risk, returns, and Sharpe ratios of global minimum variance portfolio. We observe a double ascending Sharpe  Ratio for our RRE estimator.

The paper is structured as follows. Section~\ref{model} states the model and assumptions. Section~\ref{lgce} defines the estimators. Section~\ref{unicon} presents the theoretical results, followed by simulations in Section~\ref{sec:mc}. The application is presented in Section~\ref{empirics}. Finally, Section \ref{sec:conclusion} discusses the findings. The Appendix contains proofs, additional simulation results, and an extended discussion.

\section{Model and Assumptions}\label{model}

Let $Y$ be the $n\times p$ matrix of outcomes, with $n$ denoting the sample size and $p$ is the number of outcomes. Under a $K$-factor structure, with $K \ll p$ consider the representation
\begin{equation}
Y=FA'+U,\label{0}
\end{equation}
where $F$ is $n\times K$, $A$ is $p\times K$, and $U$ is $n\times p$. Let $a_j'$ denote the $j$-th row of $A$, and let $A_{-j}$ denote the matrix obtained by removing the $j$-th row from $A$. At observation $t$,
\begin{equation}
y_{jt}=f_t'a_j+u_{jt},\label{1}
\end{equation}
\begin{equation}
Y_{-jt}=A_{-j}f_t+U_{-jt}.\label{2}
\end{equation}
Stacking the observations gives
\begin{equation}
y_j=Fa_j+u_j,\label{3}
\end{equation}
\begin{equation}
Y_{-j}=FA_{-j}'+U_{-j}.\label{4}
\end{equation}
Set $\sigma_j^2:=\operatorname{var}(u_{jt})$, $\Sigma_{U,-j}:=\operatorname{var}(U_{-jt})$, due to zero-mean factor assumption, the covariance matrix of factors $\Sigma_f:=E(f_tf_t')$, $\Sigma:=E(y_ty_t')$, 
with  $y_t:=(y_{1t}, \cdots, y_{pt})': p \times 1$ vector,
and the precision matrix is $\Theta:=\Sigma^{-1}$. Factors and errors have zero means in the theory. 

Below we present our assumptions. Due to space constraints, Appendix E gives their extended interpretation and examples. In particular, Assumption \ref{as1} imposes an exact factor structure, independence of the idiosyncratic errors, and the observations are iid over time.

\begin{assum}\label{as1}
	(i). $y_{jt}, Y_{-jt}$ are zero mean random variable and vector respectively, which are iid across $t=1,\cdots, n$ for each $j=1,\cdots,p$. (ii). $u_{jt}, U_{-jt}$ are zero mean random variable and vector respectively, which are iid across $t=1,\cdots,n$,  for each $j=1,\cdots,p$. (iii). $f_t$ are zero mean random vector,  which are iid across $t=1,\cdots, n$. (iv). $f_t, u_{jt}, U_{-jt}$ are mutually independent across $t=1,\cdots,n $ and across $j=1,\cdots,p$.
	
	\end{assum}

    Assumption \ref{as2} establishes bounds for the noise, which are standard in high-dimensional econometrics.

\begin{assum}\label{as2}
	
	(i). $\min_{1 \le j \le p} \sigma_{j}^2 \ge c > 0$, where $c$ is a positive constant. Also 
	$\max_{ 1 \le j \le p} \sigma_{j}^2 = \sigma^2 < \Gamma < \infty$, with $\Gamma$ as a large positive constant,   (ii). Let $\beta:= \min_{1 \le j \le p} \| \Sigma_{U,-j} \|_2$, and $\beta>0$ is a large positive constant , and also 
	$ \max_{1 \le j \le p } \| \Sigma_{U,-j} \|_2  < \Gamma$, where $\Gamma>0$ is described in (i).  Clearly $\beta < \Gamma$.
	
	\end{assum}

    Assumption \ref{as3} relates $p$ to the sample size $n$. Moreover, it shows that the spectral norm of factor loadings can grow and covariance matrix of the factors has full rank.

\begin{assum}\label{as3}
    
    (i). Define a positive constant $C_p>0$. Then assume the following relation between the number of outcomes, $p$ and the sample size $n$:
    $  p \le n^{C _p/2}.$ (ii). For each $j=1,\cdots,p$, $A_{-j}: (p-1) \times K$ has full rank $K$,  and $p > K+1$. We impose $\max_{1 \le j \le p} \| A_{-j} \|_2 = O (\sqrt{p})$.
	Also, $\Sigma_f$ has full rank $K$, and  $Eigmax (\Sigma_f ) \le C < \infty$, for C a positive constant.
	
    \end{assum}

    Assumption \ref{as4} specifies the distributional conditions required for the quadratic concentration inequalities used in the theory.

\begin{assum}\label{as4}

	(i). For each $j=1,\cdots,p$ let
	 $u_{jt}$ be zero mean  $\sigma_{j} \gamma_e$ sub-Gaussian random variable, which means  for each $j=1,\cdots,p$ $E \exp (t_1 u_{jt}) \le \exp(t_1^2 \sigma_{j}^2 \gamma_e^2/2)$ for all $ t_1 \in R$. 	Define the zero mean vector $U_{-j,t}:= \Sigma_{U,-j}^{1/2} \tilde{U}_{-j,t}$, where $\tilde{U}_{-j,t}$ ($(p-1) \times 1$ vector) is $\gamma_{wj}$ sub-Gaussian vector, which means
	for any unit vector $v \in R^{p-1}$, i.e. any $v$ with $v'v=1$, $v' \tilde{U}_{-j,t} $ is $\gamma_{wj}$ sub-Gaussian.
Also we assume  $E \tilde{U}_{-j,t} \tilde{U}_{-j,t}' = I_{p-1}$. In the proofs we  impose same  $\gamma_w$ constant  for each $j$, instead of  $\gamma_{wj}$ just to save from notation. (ii). Define $ f_t:=\Sigma_f^{1/2} \tilde{f}_t$, where $\tilde{f}_t: K \times 1$,
	is a sub-Gaussian vector with $ E [\tilde{f}_t \tilde{f}_t'] = I_K$, and $\max_{1 \le k \le K } | E f_{tk} | \le C < \infty$, where $C>0$ is a positive constant.
	(iii). We assume 
	 $\| A \|_{\infty} \le C < \infty$.
	 
	 	\end{assum}

        Next we introduce a term that will  be used throughout the paper. Define the uniform signal-to-noise measure
\begin{equation}
\bar\xi=\min_{1\le j\le p}\xi_j,\label{sn1}
\end{equation}
\begin{equation}
\xi_j=\frac{\lambda_K(A_{-j}\Sigma_f A_{-j}')}{\|\Sigma_{U,-j}\|_2}.\label{sn2}
\end{equation}
We provide an example about the signal-to-noise ratio, which we use in the paper. For the  weak-factor loading example in Appendix E.4, the lower bound is
\begin{equation}
\bar\xi\ge \frac{c d_{1n}^2}{\Gamma}.\label{ex1}
\end{equation}
The estimator-dependent quantities $\bar r,\bar\eta,\bar\Psi$ are defined before Theorem~\ref{l1-gl}. Set the rate of convergence for our estimator introduced in Definition 1
\begin{equation}
r_{w1}:=\max\left(\sqrt{\frac{\log n+\bar r}{n\bar\eta}},
\sqrt{\frac{K}{\bar\xi}},\sqrt{\frac{K\bar\Psi}{\bar\xi\bar\eta}},
\sqrt{\frac{K}{\bar\xi\bar\eta}}\right),\label{eqrw1}
\end{equation}
and define the following term related to covariance of factor loadings, $d_{2n}:=\max_j\|a_j'\Sigma_f A_{-j}'\|_2$, which is used below.
The next assumption introduces technical quantities, and can also allow a small number of factors, when $n$ and $p$ are proportional to each other.

\begin{assum}\label{as5}
	 
	 (i).  If $n> p$, then we need $K > \frac{\log n}{\log p}$. \par\noindent (ii). $\sqrt{\frac{K^3}{n}} \to 0, \quad  K \frac{\sqrt{  \max (\log p, \log n)}}{n^{1/2}} \to 0.$ (iii). $K \sqrt{\frac{p(\log p)}{n}} r_{w1} \stackrel{p}{\to} 0$, and $ d_{2n} r_{w1} \stackrel{p}\to 0$, and $d_{2n}$ is nondecreasing in $n$.

	 \end{assum}

Notation:  $Z^+$ is the Moore--Penrose inverse for a generic matrix $Z$.  $\|Z\|_2$ and $\|Z\|_F$ are the spectral and Frobenius norms of $Z$, $\|Z\|_\infty=\max_{i,j}|Z_{ij}|$, and $\|Z\|_{l_\infty}=\max_j\sum_i|Z_{ji}|$ is the maximum-row sum norm.  $\|\cdot\|_2$ is the Euclidean norm. The singular values $\sigma_k$ and the eigenvalues $\lambda_k$ are ordered decreasingly. $Eigmax(.)$ and $Eigmin(.)$ denote the largest and smallest eigenvalues of the matrix. The constants $C,c>0$ are generic.

\section{General Linear  Estimators}\label{lgce}

    To show consistency of our estimators in the high-dimensional context we need to start with a precision matrix formula that can be tied to hidden factors. One of the main obstacles is  establishing a relationship between matrix inversion and linear regression in the hidden factor context. The difficulty is that the formulas for the main diagonal and off-diagonal terms of the precision matrix differ, and this type of approach, tying hidden factors to the general linear class of estimators via precision matrix estimation, has not been done before. 	 
     
     Define  the following parameter vector, which will form a key term in the precision matrix formula:
 $ \alpha_j^*:= \underset{\alpha_{j} \in R^{p-1}}{\arg\min} E [ y_{jt}   - Y_{-jt} ' \alpha_{j}]^2.$
 This definition shows that we can relate to the linear regression context. Note that, the definition of $\alpha_j^*$ is not coming from the hidden factor model in (\ref{1}) and (\ref{2}).
 We also need the following definitions before the next Lemma. Let $\Sigma_f = \Sigma_f^{1/2} \Sigma_f^{1/2}$, since $\Sigma_f$ is a positive definite matrix.
 Set $\bar{A}_{-j}:= A_{-j} \Sigma_f^{1/2}$, $\bar{a}_j:= \Sigma_f^{1/2} a_j$, and 
 $ \bar{G}_j:= I_K + \bar{A}_{-j}' \Sigma_{U,-j}^{-1} \bar{A}_{-j}.$
 Next, we show how they relate to the hidden factor structure. Let $\Theta_{j,j}$ represent the $j$-th main diagonal term, and $\Theta_{j,-j}$ represent the $j$-th row without the $j$-th element.

	\begin{lemma}\label{pm-form}
	Under Assumptions \ref{as1}-\ref{as3}, for $j=1,\cdots,p$
	
	(i). \[ \Theta_{j,j}= \frac{1}{\tau_j^2},\quad
	{\mbox with } \quad
 \tau_j^2=	(\bar{a}_j' \bar{a}_j + \sigma_j^2) - \bar{a}_j' \bar{A}_{-j}' \Sigma_{U,-j}^{-1}  \bar{A}_{-j} \bar{G}_j^{-1} \bar{a}_j	.\]
	
	(ii). \[ \Theta_{j,-j}= \frac{-\alpha_j^{*'}}{\tau_j^2},\quad
	\mbox{with} \quad  \alpha_j^*=	\Sigma_{U,-j}^{-1} \bar{A}_{-j} \bar{G}_j^{-1} \bar{a}_j.\]
    \end{lemma}

Remarks.
1. \textbf{Row assembly.} Combining Lemma 1(i) and (ii), we form the $j$-th row of the precision matrix formula in terms of a hidden factor structure. Stacking each row will produce the entire precision matrix $\Theta$.

2. \textbf{Dense coefficients.} Also, note that in our vector $\alpha_j^*$ some of the vector elements are not zero, nor will be assumed to be zero unlike in \cite{caner2019}, and \cite{mein2006}, \cite{van2014}.
Further, in line with the existing literature, one should note that \cite{bbsmw21} propose linear regression coefficient estimation and analyze the risk of an asset in a hidden factor model. However, neither the precision matrix estimation nor the above formulae are discussed therein.

In this respect, we start with the estimation of $\alpha_j^*$ and then expand it to the estimation of the precision matrix. In the proof of Lemma \ref{pm-form}, equation (B.7) in Appendix B, we  show that $\alpha_j^*$ is a  vector of coefficients that can be derived from a linear regression. A general linear regression type of estimators has been proposed in (2)-(3), and (43) of \cite{bbsmw21} recently. 
It can be seen as a generalization of principal components based regression.

\begin{definition}(\cite{bbsmw21})
 For each $j=1,\cdots,p$,
	\begin{equation}
	 \tilde{\alpha}_j:= \hat{B} (Y_{-j} \hat{B})^+ y_j = \hat{B} ( \hat{B}' Y_{-j}' Y_{-j} \hat{B})^+ ( \hat{B}' Y_{-j}' y_j)	 ,\label{3.1}
	 \end{equation}
	where $(.)^+$ is the Moore-Penrose inverse, $\hat{B}: (p-1) \times q$, with $q \le p-1$, and the second equality is by (134) of \cite{bsmw22}. $\hat{B}$ should have orthonormal columns up to a scalar.
\label{def1}
    \end{definition}
    The proofs use two possible estimators, hence $\hat{B}$ takes two distinct forms. For the first, we set $\hat{B}=I_{p-1}$, corresponding to the ridgeless regression estimator (RRE), which is analyzed separately after this section, so $q=p-1$. The second estimator is adaptive PCR, with $\hat{B}= V_{1,k_j}$, where $V_{1,k_j}$ is the $(p-1)\times k_j$ matrix whose columns are the eigenvectors associated with the $k_j$ largest eigenvalues of $Y_{-j}'Y_{-j}/n$, so $q=k_j$, $j=1,\cdots,p$.

Our estimator is based on the analysis of the precision matrix in Lemma \ref{pm-form}.
    Note that one of the main contributions of this paper is to link the estimator (\ref{3.1}) with $\alpha_j^*$, which is related to the matrix inversion. This part of the paper extends  \cite{bbsmw21}-\cite{bsmw22} from a regression coefficient estimator, to a precision matrix estimator. Hence, we have a different structure, and the estimation-proof  of the main diagonal term in (\ref{14a}) and (\ref{18}) is not in \cite{bbsmw21}-\cite{bsmw22}. Part of the off-diagonal term proof in (\ref{3.1}), is only pointwise in \cite{bbsmw21}-\cite{bsmw22}, and we extend it to a uniform result. We suppress the subscript $j$ on $\hat{B}$ for notational simplicity.\footnote{An optimal $\hat{B}$ may be analyzed, but without knowing its structure, it may be difficult, and it is beyond the current scope of the paper.} We define the reciprocal of the main diagonal terms in the precision matrix based on (B.11)
    \footnote{Note that only in the case of the RRE estimator, as the subcase when $p-1>n$, the estimator simplifies and we benefit from that expression.}
    \begin{equation}
	\tilde{\tau}_j^2:= \frac{y_j' (y_j - Y_{-j} \tilde{\alpha}_j)}{n}.\label{14a}
	\end{equation}
	Then the $j$-th main diagonal term estimator in the precision matrix is:
	\begin{equation}
	\tilde{\Theta}_{j,j} = 1/\tilde{\tau}_j^2,\label{18}
	\end{equation}
	 and 
	 the $j$-th row of the precision matrix estimate without $j$-th element is, via (B.13)
	 \begin{equation}
	 \tilde{\Theta}_{j,-j}= -\tilde{\alpha}_j'/\tilde{\tau}_j^2.\label{19}
	 \end{equation}

\section{Uniform Consistency of General Estimator}\label{unicon}
		 
In this section, we provide upper bounds on the $ l_2$-norm errors for each regression coefficient estimate in (\ref{3.1}). We then consider (\ref{14a}) and obtain upper bounds for the estimation errors.

	 \subsection{Bounding  Estimation Errors: Regression Coefficients}

 We explain three random terms that are essential to understanding our estimation error bounds.
      Let  $P_{\hat{B}}$ be the projection matrix onto the range $\hat{B}$, i.e. $P_{\hat{B}}= \hat{B} [ \hat{B}' \hat{B}]^+ \hat{B}' = \hat{B} \hat{B}^+$,  and the last equality is by Moore-Penrose matrix inverse rules such as in (134) of \cite{bsmw22}.       Define the rank of $Y_{-j} P_{\hat{B}}$ as 
	 \begin{equation}\hat{r}_j:= rank (Y_{-j} P_{\hat{B}}).\label{eqn1}
	 \end{equation}
	 
	\noindent  This term reflects the {\bf complexity}. %
     Define 
	 \begin{equation}
	 \hat{\eta}_j:= \sigma_{\hat{r}_j}^2 (Y_{-j} P_{\hat{B}})/n,\label{eqn2}
	 \end{equation}
	 which is the $\hat{r}_j$-th largest singular value (squared, scaled by $n$) of $Y_{-j} P_{\hat{B}}$ ($n \times (p-1)$ matrix). 	 This last term is obtained when we project $Y_{-j}$ onto the range of $\hat{B}$. 
     It can be considered as {\bf the retained signal}.	 	 Next, we define 
	 \begin{equation}
	 \hat{\Psi}_j:= \sigma_1^2 (Y_{-j} M_{\hat{B}})/n,\label{eqn3}
	 \end{equation}
	  where $M_{\hat{B}}:= I_{p-1} - P_{\hat{B}}$, and $Y_{-j} M_{ \hat{B}}$ is a $n \times (p-1)$ matrix. This term is the largest singular value of $Y_{-j} M_{\hat{B}}$ matrix scaled by $n$, and represents {\bf the projection bias} stemming from using a hidden factor structure to estimate $Y_{-j}$. %
	Define  $\max_{1 \le j \le p } \hat{r}_j:= \bar{r}$, $\min_{1 \le j \le p} \hat{\eta}_j:= \bar{\eta}$, and $\max_{1 \le j \le p} \hat{\Psi}_j:= \bar{\Psi}$.  We  have the following uniform bounds for the estimators of the off-diagonal terms in each row of the precision matrix.%

	 \begin{thm}\label{l1-gl} 
	 
	 (i). Under Assumptions \ref{as1}-\ref{as4}, $K \log n \le c n $, with $c>0$ a positive constant, with probability at least $1 - 2/n^{C_p/2}$
	 
	 \[ \max_{1 \le j \le p} \| \tilde{\alpha}_j - \alpha_{j}^* \|_2^2 \le 2 \gamma_e^2 \Gamma \left( \frac{2 \log n + \bar{r}}{n \bar{\eta}}
 \right) + \frac{C K }{\bar{\xi}} \left( 1 + \frac{1}{\beta} + \frac{\bar{\Psi}}{\bar{\eta} \beta } + \frac{\Gamma}{\bar{\eta} \beta}
 \right).\]

	 (ii). Under Assumptions \ref{as1}-\ref{as4},   
	 \[  \max_{1 \le j \le p} \| \tilde{\alpha}_j - \alpha_{j}^* \|_2^2 = O_p \left( \max( \frac{K}{\bar{\xi}}, \frac{K \bar{\Psi}}{\bar{\xi} \bar{\eta} }, \frac{K }{\bar{\xi} \bar{\eta}}, 
	 \frac{\log n+\bar{r}}{n \bar{\eta}}) \right) = O_p (r_{w1}^2).\]	 
	 
	 \end{thm}

Remarks.     
 1. \textbf{Error components.} The first term on the right side of Theorem \ref{l1-gl}(i) is the variance term, and the second term is the squared bias term. Note that there is no sparsity parameter on the upper bounds and rates.  
  
 2. \textbf{Signal requirement.} Note that $\bar{\xi}$ plays an essential role in getting consistency. A higher signal-to-noise ratio is needed, especially as the number of factors $K$ grows rapidly.

  3. \textbf{Complexity and signal.} $\bar{r}$ represents the complexity and adversely affects the $l_2$-norm error by increasing the estimator's variance.  On the other hand, $\bar{\eta}$ is a positive influence, and a larger signal improves the performance. It should be particularly noted that this positive influence is reflected in the reduction of both the estimator's bias and variance. $\bar{\Psi}$
 has an adverse effect
on the $l_2$ error. This term increases the estimator's bias.

		\subsubsection{Adaptive PCR}
	
	In this subsection, we consider principal components regression (adaptive PCR from now on) estimators. Let us designate $rank (Y_{-j}) \ge k_j  \ge 1$, where $k_j$ is a positive integer and changes with $j=1,\cdots,p$. The adaptive PCR estimators use $\hat{B}= V_{1,k_j}$, where $V_{1,k_j}$ is a $(p-1) \times k_j$ matrix whose columns are the eigenvectors corresponding to the $k_j$ largest eigenvalues of $Y_{-j}' Y_{-j}/n$ for $j=1,\cdots,p$.
 The adaptive PCR estimator is, 
  $\hat{\alpha}_{PCR,j} = V_{1,k_j} (Y_{-j} V_{1, k_j})^+ y_j,$
 where, as before, $(.)^+$ denotes the Moore-Penrose inverse. Before simplifying the terms in the upper bound in Theorem \ref{l1-gl}, we define the quantity $\hat{\lambda}_{k_j}:= \frac{1}{n} \sigma_{k_j}^2 (Y_{-j})$. For example with $k_j=1$ we have the largest singular value of $Y_{-j}$, which is $\sigma_{1}^2 (Y_{-j})/n$ (squared and scaled by $n$). One big issue is  the choice of $k_j$.
     To choose $k_j$, we adapt the so-called elbow method (for details, see (19) of \cite{bbsmw21})	
     \begin{equation}
      \hat{s}_j:= \max \{ k_j \ge 0: \hat{\lambda}_{k_j} \ge C_0 \Delta_{-j} \},\label{17a}
      \end{equation}
where $C_0 >1$ and, $c_{\Delta} >1$ are  positive constants above one, and 
\begin{equation}
 \Delta_{-j}:= c_{\Delta} [ \| \Sigma_{U,-j} \|_2 + tr (\Sigma_{U,-j})/n].\label{eq-delta}
 \end{equation}
 This type of choice is done to minimize the risk of the adaptive PCR in \cite{bbsmw21}.
Here we  need results uniformly over $j=1,\cdots,p$ 
and the choice given in (\ref{eq-delta}). Set $k_j= \hat{s}_j$.
By definition of $\hat{s}_j$ above
\begin{equation}
\hat{\lambda}_{\hat{s}_j+1} < C_0 \Delta_{-j} \le \hat{\lambda}_{\hat{s}_j}.\label{pcl1-4}
\end{equation}
We  show in the proof of Corollary \ref{cl1} in the Appendix the stronger, simultaneous statement
\begin{equation}
P \left(\max_{1 \le j \le p}  \hat{s}_j \le K \right) \to 1,\label{21}
\end{equation}
which will simplify the upper bound for the $l_2$-norm error of our estimators. 	
We prove the following error bounds for an adaptive PCR-based estimator for  the off-diagonal terms in a given row of the precision matrix. With probability approaching one, the three random quantities satisfy $\bar{r}\le K$, $\bar{\eta}\ge \beta >0$, and $\bar{\Psi}/\bar{\eta}\le 1$, yielding Corollary \ref{cl1} from Theorem \ref{l1-gl}.

	\begin{corollary}\label{cl1}
	
	 Under Assumptions \ref{as1}-\ref{as4}  %
     with $0< c' <1$ a positive constant,
     $K \log n \le c n$, with a positive constant $c$, 
     
     (i). 
	 \[
	 \max_{1 \le j \le p} 
	\| \hat{\alpha}_{PCR,j} - \alpha_{j}^* \|_2^2  \le 2 \gamma_e^2 \Gamma \left[ \frac{2 \log n +K}{n \beta}
	\right] 
	+ \frac{C K}{\bar{\xi}} \left[ 1 + \frac{2}{\beta}+ \frac{\Gamma}{\beta^2} \right],
\]
with probability at least 
	$1 - 2/n^{C_p/2} - \exp ((c'-1) n)$. 
	
	(ii). 
			\[ \max_{1 \le j \le p } \| \hat{\alpha}_{PCR,j} - \alpha_{j}^* \|_2^2 = O_p \left( \max ( \frac{ \log n + K}{n }, \frac{K}{\bar{\xi}})\right).\]
	
	\end{corollary}

High-dimensional consistency is possible when $p>n$, with $\bar{\xi}$ increasing with $p$
as shown in Example (\ref{ex1}).
To allow for feasible estimation when $K$ estimated, we follow (56) of \cite{bbsmw21} or (2.7) of \cite{bw19}. Assuming additionally in  Corollary \ref{cl1} that  $tr (\Sigma_{U,-j} ) \ge c'' \| \Sigma_{U,-j} \|_2 (n \wedge p)$, with $c''>0$ a positive constant, for each $j=1,\cdots,p$,  Proposition 8 of \cite{bbsmw21} can be extended with a union bound and the feasible estimation of the unknown number of factors do not change Corollary \ref{cl1}.

	\subsubsection{Ridgeless Regression Estimation}

	In this part, we consider RRE when $p-1<n$ first and then analyze the $p>n$ case.  	
	In that respect $\hat{B}= I_{p-1}$, which simplifies the upper bound in Theorem \ref{l1-gl}. We have 
	 $\hat{\alpha}_{RRE,j} = (Y_{-j})^+ y_j$
	for $j=1,\cdots,p$. Note that when $p-1<n$, RRE is equivalent to OLS, and by using (132) of \cite{bsmw22}, 
	$\hat{\alpha}_{RRE,j} = (Y_{-j})^+ y_j = (Y_{-j}' Y_{-j})^{-1} Y_{-j}' y_j.$
    The three random terms simplify  as $\bar{r} < p$, $\bar{\Psi}=0$, and $\bar{\eta} \ge C >0$ with probability approaching one, where  we obtain Corollary \ref{cl2} from Theorem \ref{l1-gl}.

	\begin{corollary}\label{cl2} (RRE: $p-1<n$)
	
	(i). Under Assumptions \ref{as1}-\ref{as2},  Assumption \ref{as3}(i) with $ 0 < C_p < 2$, Assumptions \ref{as3}(ii)-\ref{as4}
    and $K \log n < c n$, for sufficiently large $n$,
    with probability at least 
	$1  - 1/n^{C_p/2}$, 
	\[
	 \max_{1 \le j \le p} 
	\| \hat{\alpha}_{RRE,j} - \alpha_{j}^* \|_2^2 \le 2 \gamma_e^2 \Gamma \left[ \frac{2 \log n +p}{C n}
	\right] 
	+ \frac{ C K}{\bar{\xi}} \left[ 1 + \frac{1}{\beta}+   \frac{\Gamma}{C \beta }\right].
\]	
	(ii). Under the same Assumptions as for (i)  
			\[ \max_{1 \le j \le p } \| \hat{\alpha}_{RRE,j} - \alpha_{j}^* \|_2^2 = O_p ( \max ( \frac{ \log n + p}{n }, \frac{K}{\bar{\xi}})).\]
	
	\end{corollary}

	In the following, we provide the Corollary for the RRE case when $p-1>n$. First we  find it useful to introduce  the effective rank as 
	$ re (\Sigma_{U,-j}) := \frac{tr (\Sigma_{U,-j})}{\| \Sigma_{U,-j} \|_2}.$ Next, we relax Assumption \ref{as3}(i) that relates $p$ to $n$. We show that for RRE in the $p-1>n$ case, we can allow $p$ to grow exponentially in  $n$. This implies that RRE permits precision matrix estimation in substantially higher dimensions than adaptive PCR.

    {\bf Assumption $3^*(i)$.}
       {\it  We allow $ 
         n< p-1 \le \frac{\exp(C_p n/2)}{4}.$}
        
Now we provide our coefficient estimation result that will be helpful for the precision matrix estimation based on RRE. Define $\delta_n:= \min_{1 \le j \le p}
tr (\Sigma_{U,-j})/n$. $\delta_n$ is nondecreasing in $n$, since $\Sigma_{U,-j}: (p-1) \times (p-1)$ is a diagonal matrix, with positive elements on the diagonal and $p,n$ grow together under Assumption $3^*(i)$ with $p-1>n$. Note that the three random variables in Theorem \ref{l1-gl} simplify in Corollary \ref{cl3} as  $\bar{r}=n$, the bias term is $\bar{\Psi}=0$, and the size of the signal is $\bar{\eta}\ge  \delta_n$ wpa1.	 
	 
	\begin{corollary}\label{cl3} (RRE: $p-1>n$)
	
	Under Assumptions \ref{as1}-\ref{as2}, Assumptions $3^{*}(i)$-\ref{as3}(ii)-\ref{as4}, and for all $j=1,\cdots,p$,
	$  re(\Sigma_{U,-j}) > C n,$
	 where $C$ is a positive constant,  $K \log n < c n$, for some $c>0$,  and $\tilde{U}_{-j}$ has independent entries, 
	 
	 (i). 
	\[
	 \max_{1 \le j \le p} 
	\| \hat{\alpha}_{RRE,j} - \alpha_{j}^* \|_2^2 \le 2 \gamma_e^2 \Gamma \left[ \frac{2 \log n +n}{C n \delta_n }
	\right] 
	+ \frac{C K}{\bar{\xi}} \left[ 1 + \frac{1}{\beta}+ \frac{\Gamma}{\beta \delta_n}\right].
\]
	with probability at least 
	$1 - \exp(-C_p n/2)$, and 	
	
	(ii). 
			\[ \max_{1 \le j \le p } \| \hat{\alpha}_{RRE,j} - \alpha_{j}^* \|_2^2 = O_p ( \max ( \frac{ \log n + n}{n \delta_n  }, \frac{K}{\bar{\xi}})).\]
	
	\end{corollary}

    When $p$ grows polynomially in $n$ with exponent greater than one, RRE is consistent. Clearly,
 $\frac{\log n + n}{n \delta_n} \to 0,$
 since $\delta_n$ may grow larger in $n$. Under Example~\ref{ex1}, $K/\bar\xi$ can be smaller, while $K/d_{1n}^2\to0$. Thus, consistency requires the signal-to-noise ratio to increase with $n$ and dominate the number of factors. We may also observe double descent in RRE for the off-diagonal estimation of $\alpha_j^*$ in the precision matrix. This phenomenon is discussed in detail in Section 4.4.

	     We compare now the RRE in this case with adaptive PCR. In RRE, there is one less bias term compared to adaptive PCR ($\bar{\Psi}=0$ in RRE) in the finite sample upper bound. Since $K < n$, adaptive PCR has a smaller variance term compared to RRE. %
         Adaptive PCR may also have an additional theoretical advantage over RRE. We need the noise $\delta_n$ to be increasing in $n$ for consistency in RRE, whereas in adaptive PCR this is not required.

\subsection{Estimation of Diagonal Precision Matrix Entries}

We have the following Theorem for the reciprocal of the $j$-th main diagonal term in the precision matrix. 
The technique in \cite{bbsmw21} considers only simple linear regression, while precision matrix estimation, particularly the estimation of the diagonal terms, is not handled. %
Note that the rate $r_{w1}$ is defined in equation \eqref{eqrw1} just before Assumption \ref{as5}.

\begin{thm}\label{thma}
(i). Under Assumptions \ref{as1}-\ref{as4} \ref{as5}(i),  
 \begin{eqnarray*}
\max_{1 \le j \le p} | \tilde{\tau}_j^2 - \tau_j^2| & = & 
   O_p \left(K (\sqrt{\frac{p (\log p)}{n}}) r_{w1}\right) \nonumber \\ &+&  O_p \left( d_{2n} r_{w1} \right)
   + O_p \left( \max (\frac{K^{3/2}}{n^{1/2}}, K  \frac{\max(\sqrt{\log p}, \sqrt{\log n})}{n^{1/2}})
   \right).\label{69}
\end{eqnarray*}

(ii).  Under Assumptions \ref{as1}-\ref{as5},   \[ \max_{1 \le j \le p} | \tilde{\tau}_j^2 - \tau_j^2| = o_p (1).\]

(iii).  Under Assumptions \ref{as1}-\ref{as5},   where $c > 0$ is a positive constant
\[ \min_{1 \le j \le p} \tau_j^2 \ge c - o (1).\]

\end{thm}

  Remarks.  
1. \textbf{Dimension regime.} When $p>n$ and $p/n \to (1,\infty)$ it is possible to get consistency since $p$ only enters the estimation errors directly via $\sqrt{\log p}$. With a larger $n$, and via Assumption \ref{as5}(ii) consistency is achieved. 

2. \textbf{Diagonal normalization.} Theorem \ref{l1-gl}(ii) clearly shows that the rate of estimation error for off-diagonal-row elements is $r_{w1}$. In contrast, Theorem \ref{thma} gives the rate for the reciprocal of the main diagonal term, which is slower. Specifically, consistency additionally requires $d_{2n}r_{w1}=o_p(1)$ by Assumption \ref{as5}(iii), together with the remaining parts in Assumptions \ref{as5}(iii) and Assumption \ref{as5}(ii).

3. \textbf{Interpolation exception.} Note that Theorem \ref{thma} excludes the RRE case with $p-1>n$, since the expression in (\ref{14a}) is equal to zero due to interpolation. We provide an alternative estimator for the RRE case with $p-1>n$ in Corollary A.3 of Appendix A.

\subsection{Consistent Estimation of Rows}

	In the following, we provide our theorem for estimating the rows of the precision matrix. This shows that, in the general case, each row of the precision matrix can be estimated uniformly and consistently by our estimator.  
  Specific results regarding adaptive PCR and RRE will be discussed in the subsections \ref{apcr} and \ref{RRE}.
  Note that  the main diagonal terms determine  the rate of convergence. The off-diagonal term estimation is consistent but converges faster than the main diagonal terms.
 This can be seen by comparing the rates in Theorem \ref{l1-gl} and Theorem \ref{thma}.  Denote each row of $\Theta$ as $\Theta_{j}'$, which is a  $1 \times p$ vector.

\begin{thm}\label{thm1}

Under Assumptions \ref{as1}-\ref{as5},  
\begin{eqnarray*}
 \max_{1 \le j \le p} \| \tilde{\Theta}_j' - \Theta_{j}' \|_2 &=& O_p ( K \frac{\sqrt{p \log p}}{\sqrt{n}} r_{w1}) + O_p ( d_{2n} r_{w1})\\
&+& O_p \left( \max(\frac{K^{3/2}}{n^{1/2}}, K \frac{\sqrt{\max(\log p, \log n)}}{n^{1/2}})\right)= o_p (1),
\end{eqnarray*}
where $r_{w1}$ is  defined in (\ref{eqrw1}).
\end{thm}

 Remarks.
1. \textbf{Dimension dependence.} First, we see that if $p/n \to [0,\infty)$, we obtain consistency and the price that we pay in terms of the number of variables, $p$, is just $\sqrt{\log p}$ at the first glance, as can be seen in Assumption \ref{as5}. However, we have terms such as $\bar{r}, \bar{\Psi}, \bar{\eta}$ that may depend on $p$, and we will analyze them in the adaptive PCR and RRE cases.

2. \textbf{Signal requirement.} We see that through the definition of $r_{w1}$ one of the keys to consistent estimation is the ratio $K/\bar{\xi}$, and this ratio needs to go to zero. So we cannot have a large, hidden factor system and expect consistency.

3. \textbf{Scope of the theorem.} Note that the rate for Theorem \ref{thm1} is obtained for all linear estimators in Section 3, except for RRE when $p-1>n$, the interpolating estimator. This is explained in detail below in the Corollary 
\ref{cl6} and in  Corollary A.3 in Appendix A.

4. \textbf{Loading restrictions.} $d_{2n} r_{w1} =o_p(1)$ condition  may imply weak factor loadings in certain cases.  The details are in  Appendix B.

\subsubsection{Adaptive PCR Estimator}\label{apcr}

In case of the adaptive PCR, the random terms in the upper bound in Theorem \ref{thm1} can be bounded  as:  $\bar{r} \le K$, $\bar{\Psi}/\bar{\eta} \le1$, and $\bar{\eta} \ge \beta$ with wpa1. Hence, the rate in Theorem \ref{thm1} can be simplified in the adaptive PCR, with $\hat{\Theta}_{PCR,j}'$ representing the estimator of the $j$-th row of $\Theta$.  The following Corollary follows from Theorem \ref{thm1} by replacing $r_{w1}$ with 
$r_{w1,PCR}$, the rate that is shown before Corollary A.1 in Appendix A.

\begin{corollary}\label{cl4}

Let $p/n \to [0,\infty)$ and with Assumptions \ref{as1}-\ref{as5}, 
\begin{eqnarray*}
 \max_{1 \le j \le p} \| \hat{\Theta}_{PCR,j}' - \Theta_{j}' \|_2 &=& O_p \left(K \sqrt{\log p} \max (  \frac{\sqrt{\log n + K}}{\sqrt{n }}, \frac{\sqrt{K}}{\sqrt{\bar{\xi} }})\right)\nonumber \\
&+& O_p \left(  d_{2n} \max ( \frac{\sqrt{\log n + K}}{\sqrt{n }}	, \frac{\sqrt{K}}{\sqrt{\bar{\xi}}})\right)\\
& + &O_p \left( \max(\frac{K^{3/2}}{n^{1/2}}, K \frac{\sqrt{\max(\log p, \log n)}}{n^{1/2}})\right) = o_p (1) .
\end{eqnarray*}

\end{corollary}

Remarks. 
1. \textbf{Consistency conditions.} The price of consistency is $\sqrt{\log p}$ by number of variables, even when $p>n$. So with $p/n \to (1,\infty)$, to get consistency, we need the signal-to-noise ratio to dominate the number of factors.

2. \textbf{Factor-based comparisons.} In principal components method of \cite{fan2013}, which assumes a model based on latent factors, and estimates them,  their rate of convergence is $s^* (\frac{1}{\sqrt{p}}+ \frac{\sqrt{\log p}}{\sqrt{n}})$, where $s^*$ is the maximum number of nonzeros across all rows of  the covariance matrix of errors. The rate of convergence of \cite{fan2013} is smaller than the one of our method unless $p$ is small. %
We should note that their method assumes a latent factor structure and estimates that. %
\cite{fan2011} uses a sparse covariance matrix of errors as in \cite{fan2013} observed factor model to estimate precision matrix estimators. Their rate of convergence is $s^* \sqrt{\log p/n}$.

\subsubsection{Ridgeless Estimators}\label{RRE}

 When we consider the sub case of RRE with $p<n-1$, we set wpa1 $\bar{r} <p$, $\bar{\Psi}=0$, and $\bar{\eta} \ge c$, where $ c $ is a positive constant in Theorem \ref{thm1} above. Let the $j$-th row of the estimator of the precision matrix  be $\hat{\Theta}_{RRE,j}'$.
 Corollary \ref{cl5} follows from Corollaries \ref{cl2} and A.2, with $r_{w1}$ in Theorem \ref{thm1} replaced by the $r_{RRE-l}$ rate given after Corollary A.1 in Appendix A.

\begin{corollary}\label{cl5}

Set $p/n = a_{2n} \to 0$, and under Assumptions \ref{as1}-\ref{as2}, and have 
Assumption \ref{as3}(i) with $0 < C_p < 2$,
Assumptions \ref{as3}(ii)-\ref{as5}
\begin{eqnarray*}
(i). \max_{1 \le j \le p} \| \hat{\Theta}_{RRE,j}' - \Theta_{j}' \|_2 &=& O_p \left(\max (K \sqrt{a_{2n} \log p},1) \max (  \frac{\sqrt{\log n + p}}{\sqrt{n }}, \frac{\sqrt{K}}{\sqrt{\bar{\xi}}})\right)\nonumber \\
&+& O_p \left( d_{2n} \max ( \frac{\sqrt{\log n + p}}{\sqrt{n }}	, \frac{\sqrt{K}}{\sqrt{\bar{\xi}}})\right)\\
& + & O_p \left( \max(\frac{K^{3/2}}{n^{1/2}}, K \frac{\sqrt{\log n}}{n^{1/2}})\right) = o_p (1).
\end{eqnarray*}

(ii). With a $p \times 1$ vector a with $\| a\|_2=1$,
\begin{eqnarray*}
 \max_{1 \le j \le p} | (\hat{\Theta}_j' - \Theta_j') a | 
&=& O_p \left(\max (K \sqrt{a_{2n} \log p},1) \max (  \frac{\sqrt{\log n + p}}{\sqrt{n }}, \frac{\sqrt{K}}{\sqrt{\bar{\xi}}})\right)\nonumber \\
&+& O_p \left( d_{2n} \max ( \frac{\sqrt{\log n + p}}{\sqrt{n }}	, \frac{\sqrt{K}}{\sqrt{\bar{\xi}}})\right)\\
& + & O_p \left( \max(\frac{K^{3/2}}{n^{1/2}}, K \frac{\sqrt{\log n}}{n^{1/2}})\right) = o_p (1).
\end{eqnarray*}

\end{corollary}

Remarks. 1.
\textbf{Low-dimensional regime.} Note that unless $p/n \to 0$,  RRE will not be consistent.  

2. \textbf{Linear functionals.} (ii) is obtained from (i) by Cauchy-Schwartz inequality, and shows with a scalar constant target $\Theta_j' a,$ the rates will not change. %

3. \textbf{Double-descent interpretation.} Compared with the adaptive principal components estimator, we include the extension in part (ii) to link Corollaries \ref{cl5} and \ref{cl6} to the double descent phenomenon.

Next, we consider the RRE estimator when $p-1>n$. Note that in this scenario, the main diagonal term is estimated differently because RRE is an interpolating estimator. We prove this after  Corollary A.2. In the proof of Theorem \ref{thm1}, it is clear that the rate of convergence should be the slower of the main diagonal or off-diagonal terms. 
Let the $j$-th row of
the precision matrix be $\hat{\Theta}_{INT,j}' $.  Corollary 6 below proves $\hat{\Theta}_{INT,j}$'s consistency  by combining Corollaries \ref{cl3} and A.3.

\begin{corollary}\label{cl6}

Under Assumptions \ref{as1}-\ref{as2}, Assumption $3^{*}(i)$ Assumptions \ref{as3}(ii)-\ref{as4}-\ref{as5}(ii),  with signal-to-noise condition $ \sqrt{K} d_{2n}/\sqrt{\bar{\xi}} \to 0$,
and $\min_{ 1 \le j \le p} re (\Sigma_{U_{-j}}) > C n$ and $\tilde{U}_{-j}$ matrix with independent entries,  

(i). \begin{eqnarray*}
\max_{1 \le j \le p} \| \hat{\Theta}_{INT,j}' - \Theta_{j}'\|_2 &= &O_p \left( \max (\frac{\sqrt{\log n + n}}{\sqrt{n \delta_n}}, \frac{\sqrt{K} d_{2n}}{\sqrt{\bar{\xi}}})\right)\\
&+&O_p \left( \max(\frac{K^{3/2}}{n^{1/2}}, K \frac{\sqrt{\log p}}{n^{1/2}})\right) = o_p (1).
\end{eqnarray*}

(ii).  With a $p \times 1$ vector, $a$, and the $l_2$ norm of that vector is normalized, $\|a \|_2=1$
\begin{eqnarray*}
 \max_{1 \le j \le p} | (\hat{\Theta}_j' - \Theta_j') a |
&= &O_p \left( \max (\frac{\sqrt{\log n + n}}{\sqrt{n \delta_n}}, \frac{\sqrt{K} d_{2n}}{\sqrt{\bar{\xi}}})\right)\\
&+&O_p \left( \max(\frac{K^{3/2}}{n^{1/2}}, K \frac{\sqrt{\log p}}{n^{1/2}})\right) = o_p (1).
 \end{eqnarray*}

\end{corollary}

Remarks.
1. \textbf{Linear functionals.} Corollary \ref{cl6}(ii) is obtained from (i) by Cauchy-Schwartz inequality, and shows with a scalar constant target $\Theta_j' a,$ the rates will not change. %

2. \textbf{Signal condition.} First of all, we still get consistency with $p-1>n$,
as long as the signal-to-noise ratio dominates the number of factors multiplied by $d_{2n}$. %

3. \textbf{Practical trade-off.} It is not clear from the rates of convergence that RRE or adaptive PCR may do better. However, with a small number of factors, adaptive PCR may yield a better performance. But there is no tuning parameter estimation in RRE.

4. \textbf{Shrinkage comparison.} A comparative method can be the non-linear shrinkage method of \cite{lw2017}, which is dense and model-estimation-free. However, their precision matrix estimator is not consistent, and there is no rate of convergence attached. Their aim is to balance the bias-variance tradeoff in high dimensions.

\subsection{Double Descent}\label{sec:rre_dd}

The double descent phenomenon can be understood in terms of the relative contributions of the bias and variance components to the estimation error, both for the off-diagonal and diagonal elements. We begin with the analysis of the off-diagonal terms in Corollaries \ref{cl2}--\ref{cl3}. In Corollary \ref{cl2}(i), we consider the squared $l_2$-norm of the estimation error. For the OLS estimator, the variance component decreases as $p$ declines, with its magnitude determined by $\gamma_e^2 \Gamma$, where $\gamma_e$ denotes the sub-Gaussian constant of the error term and $\Gamma$ is the maximum eigenvalue of the error covariance matrix. 
In contrast, the squared bias component, $K/\bar{\xi}$, is weighted by $\max_{1 \leq j \leq p}\|a_j\|_2^2$, corresponding to the maximum squared Euclidean norm of the factor loading vectors across outcomes, as shown in (A.27). Hence, the relative magnitude of $\gamma_e^2\Gamma$ and $\max_{1 \le j \le p} \|a_j\|_2^2$ determines the relative importance of the variance and bias components in the overall estimation error. 
For the RRE estimator, analogous arguments apply to the bias component in Corollary \ref{cl3}(i). The variance component, however, differs in an important respect. Although the leading constant multiplying the variance term is the same as in Corollary \ref{cl2}(i), an increase in $p$ may lead to a decline in the variance through the potentially increasing noise term $\delta_n$. Corollaries \ref{cl5}-\ref{cl6} provide the overall picture, in $l_2$-norm, taking into account both the main diagonal and off-diagonal terms. %

In sum, the theoretical results indicate that, to the extent that a double descent pattern emerges, it is likely to be driven primarily by a reduction in the variance component. For the OLS estimator, Corollary \ref{cl5} shows that decreasing $p$ reduces both variance-related terms $K \sqrt{p \,\log p/n} \sqrt{(\log n+ p)/n}$, and $d_{2n} \sqrt{(\log n + p)/n}$. For the interpolating estimator, Corollary \ref{cl6} yields an analogous reduction in the variance component as $p$ increases, through the term $\sqrt{(\log n + n)/n \delta_n}$, where the increase in the noise $\delta_n$ contributes to the reduction in variance. In the Monte Carlo section \ref{sec:mc_rre_dd}, we therefore report the values of $p, d_{2n}, \bar{\xi}, \delta_n$ under different specifications. We additionally report the corresponding weights $\gamma_e^2 \Gamma, \max_{1 \le j \le p} \| a_j \|_2^2$,  in order to illustrate the relative contribution of the different components and provide a more transparent assessment of the mechanisms underlying the observed double descent behavior.

Thus, our theoretical results establish the possibility of a double descent pattern in precision matrix estimation, rather than providing a formal proof. %
Note that even in the case of a simple regression of a single outcome on multiple covariates, the prediction error of that regression displays double descent only in certain cases as shown in \cite{hmrt22}. There it crucially depends on whether the covariance of the errors are identity or not, and the signal-to-noise ratio associated with the individual outcome variable. These conditions differ from those governing the signal-to-noise structure in our precision matrix setting. Corollary 3 of \cite{hmrt22} also has a target with dimension increase (i.e. in $p$), and Corollary 4 of \cite{hmrt22} has an upper risk bound not a closed form solution for the risk of a single outcome like our  upper bounds for estimation error for precision matrix.

\section{Monte Carlo Evidence}\label{sec:mc}

 First, guided by our theoretical results, we evaluate whether estimation error diminishes when $\gamma = p/n$  is fixed and both $p$ and $n$ jointly increase. %
 Second, we investigate whether the RRE estimator exhibits the double descent phenomenon.

\subsection{Monte Carlo Design and Models}\label{sec:mc_design}

To evaluate our theoretical results, we simulate data for $t=1,...,n$ from the following latent factor model:
\begin{align}
	y_{t}= A  f_t + u_t, \label{sim_gdp}
\end{align} 

where $y_t = (y_{jt},Y_{-jt}')'$ is a $p \times 1$ vector of outcomes, $A =(a_j,A'_{-j})$ is a $p\times K$-dimensional matrix of factor loadings, and $u_t=(u_{jt},U_{-jt})'$ is a $p\times 1$ vector of idiosyncratic errors. The latent factors are drawn from the normal distribution $f_t \sim \mathcal{N}(\mu_f, \Sigma_f)$, where $\mu_f$ is the $K\times 1$-dimensional mean vector, with each element equal to 0.5, and $\Sigma_f$ is the covariance matrix of the latent factors. We allow nonzero mean for the factors, and this design also shows robustness of our method to nonzero mean. 
Following \cite{bbsmw21}, we set the diagonal elements of $\Sigma_f$  as the $K$-length sequence from 2.5 to 3 with equal increments. For example, if $K=3$, then $\Sigma_f = diag(2.5,2.75,3)$. The $i,j$-th off-diagonal element of the factor covariance matrix is defined as:
$
	[\Sigma_f]_{i,j} = (-1)^{i+j}  (0.3)^{|i-j|} \min( [\Sigma_{f}]_{i,i}, [\Sigma_{f}]_{j,j} ).
$
Each element of $A$ is drawn independently from the normal distribution $\mathcal{N} (0.5,1/\sqrt{K})$. The idiosyncratic errors, $u_t$, are drawn from the normal distribution, $\mathcal{N}(0, \Sigma_U)$. In our simulation design, we define $\Sigma_U$ as a $p\times p$ diagonal matrix whose diagonal entries are independently drawn from the uniform distribution with $\text{Unif}(1,3) \times \log(p)$. In particular, it facilitates an assessment of how noise affects the performance of the adaptive PCR and RRE estimators, especially in regimes where $p-1>n$.

The time dimension $n$ is set to 100, 200 and 400 for all simulations. Moreover, we consider $p = n/2$ and $p = 3n/2$, corresponding to low dimensional and high-dimensional settings, respectively. The number of true latent factors is set to $K = 3$ and 5.\footnote{We also examined the case $K=20$ to assess the impact of a larger number of latent factors. As the results are qualitatively similar to those for $K=3, 5$, they are omitted for brevity and are available from the authors upon request.} The number of replications is 500. 
The simulation study considers the following set of estimators. Among our proposed methods, we include PCR-3, PCR-5, PCR-20, and PCR-adaptive, corresponding to principal component regression with fixed numbers of latent factors equal to 3, 5, or 20, and to the adaptive procedure for selecting the number of factors. Furthermore, we include the ridgeless regression estimator (RRE). For comparison, we include several benchmark methods from the literature: the POET estimator of \cite{fan2013}, the nonlinear shrinkage estimator of \cite{lw2017} (NL-LW), and the nodewise regression (NDW) estimator of \cite{caner2019}. We do not include the residual nodewise estimator of \cite{caner2022}, as it relies on observed factors and is not applicable in our latent factor setting.

\subsection{Monte Carlo Results}

The simulation results are reported in Table~\ref{tab:sim_results_gaussian}, where each entry corresponds to the estimation error of the precision matrix measured by the maximum row-wise $l_2$-norm. 
Overall, the results provide clear support for the theoretical consistency result established in Corollary \ref{cl4} for the adaptive PCR estimator. Focusing on the case $K=3$, Table~\ref{tab:sim_results_gaussian} shows that when $p=n/2$ and $n=100$, the adaptive PCR estimator attains an estimation error of 0.0745. As the sample size increases to $n=400$ while maintaining $p=n/2$, the error decreases to 0.0306. 
In the high-dimensional setting for the adaptive PCR estimator, where $p=3n/2$, with $n=100$, and $K=5$, the error is 0.0771. As the sample size increases to $n=400$, the error decreases to 0.0309. Similarly, the estimation errors associated with the RRE estimator decline as both $p$ and $n$ increase.
Our PCR based methods have the smallest estimation errors among others.

\begin{table}[!t]
	\centering
	\caption{Simulation Results for the Precision Matrix Estimation}
	\begin{adjustbox}{max width=\textwidth}
	\begin{threeparttable}
    \begin{tabular}{c|cccccccc}
			\hline
			\hline
			\multicolumn{1}{c}{Method} & \multicolumn{2}{c}{$n = 100$} &       & \multicolumn{2}{c}{$n = 200$} &       & \multicolumn{2}{c}{$n = 400$} \bigstrut\\
			\cline{2-3}\cline{5-6}\cline{8-9}
			 \multicolumn{1}{c}{}  & $p = n/2$ & $p = 3n/2$ &       & $p = n/2$ & $p = 3n/2$ &       & $p = n/2$ & $p = 3n/2$ \bigstrut\\
			\hline
			\multicolumn{9}{c}{$K = 3$} \bigstrut\\
			\hline
			PCR-3F & 0.0745 & 0.0718 &       & 0.0472 & 0.0469 &       & 0.0306 & 0.0296 \bigstrut[t]\\
    		PCR-5F & 0.0816 & 0.0785 &       & 0.0501 & 0.0492 &       & 0.0319 & 0.0308 \\
    		PCR-20F & 0.1607 & 0.1416 &       & 0.0787 & 0.0714 &       & 0.0438 & 0.0397 \\
    		PCR-adaptive & 0.0745 & 0.0718 &       & 0.0472 & 0.0469 &       & 0.0306 & 0.0296 \\
    		RRE   & 0.5811 & 0.1947 &       & 0.4896 & 0.1729 &       & 0.4048 & 0.1543 \bigstrut[b]\\ \hline
    		POET  & 0.0979 & 0.0948 &       & 0.0942 & 0.0890 &       & 0.0885 & 0.0859 \bigstrut[t]\\
    		NL-LW & 0.1016 & 0.0951 &       & 0.0923 & 0.0886 &       & 0.0840 & 0.0785 \\
            NDW   & 0.1031 & 0.1041 &       & 0.0749 & 0.0824 &       & 0.0601 & 0.0647
            \bigstrut[b]\\
			\hline
			\multicolumn{9}{c}{$K = 5$} \bigstrut\\
			\hline
			PCR-3F & 0.0987 & 0.0906 &       & 0.0929 & 0.0841 &       & 0.0873 & 0.0791 \bigstrut[t]\\
    		PCR-5F & 0.0785 & 0.0771 &       & 0.0488 & 0.0477 &       & 0.0314 & 0.0309 \\
    		PCR-20F & 0.1567 & 0.1390 &       & 0.0768 & 0.0696 &       & 0.0427 & 0.0398 \\
    		PCR-adaptive & 0.0781 & 0.0771 &       & 0.0488 & 0.0477 &       & 0.0314 & 0.0309 \\
    		RRE   & 0.5540 & 0.1706 &       & 0.4776 & 0.1522 &       & 0.3996 & 0.1372 \bigstrut[b]\\ \hline
    		POET  & 0.0873 & 0.0787 &       & 0.0730 & 0.0679 &       & 0.0701 & 0.0609 \bigstrut[t]\\
            NL-LW & 0.0980 & 0.0950 &       & 0.0902 & 0.0878 &       & 0.0827 & 0.0867  \\
    		NDW   & 0.1104 & 0.1133 &       & 0.0814 & 0.0901 &       & 0.0658 & 0.0701 
            \bigstrut[b]\\
			\hline
			\hline
		\end{tabular}
		\vspace*{-0.5cm}
		\begin{tablenotes}
            \footnotesize
			\singlespacing
			\item \leavevmode\kern-\scriptspace\kern-\labelsep 
			Note: The quantities in the table correspond to estimation error for the precision matrix in terms of the maximum row-wise $l_2$-norm.
		\end{tablenotes}
	\end{threeparttable}
    \end{adjustbox}%
	\label{tab:sim_results_gaussian}%
\end{table}%

In Appendix C, we additionally report Table C.1, which presents results obtained under errors following a $t$-distribution with five degrees of freedom. The results are very similar to those reported in the main analysis.

\subsection{Analysis of the Double Descent Phenomenon for the RRE}\label{sec:mc_rre_dd}

\begin{figure}
    \centering
    \caption{Maximum Absolute Linear-Combination Error of the RRE Precision Matrix Estimator for Different Values of $p$ ($n=200$, $K=3$)}
    \includegraphics[width=0.75\linewidth]
    {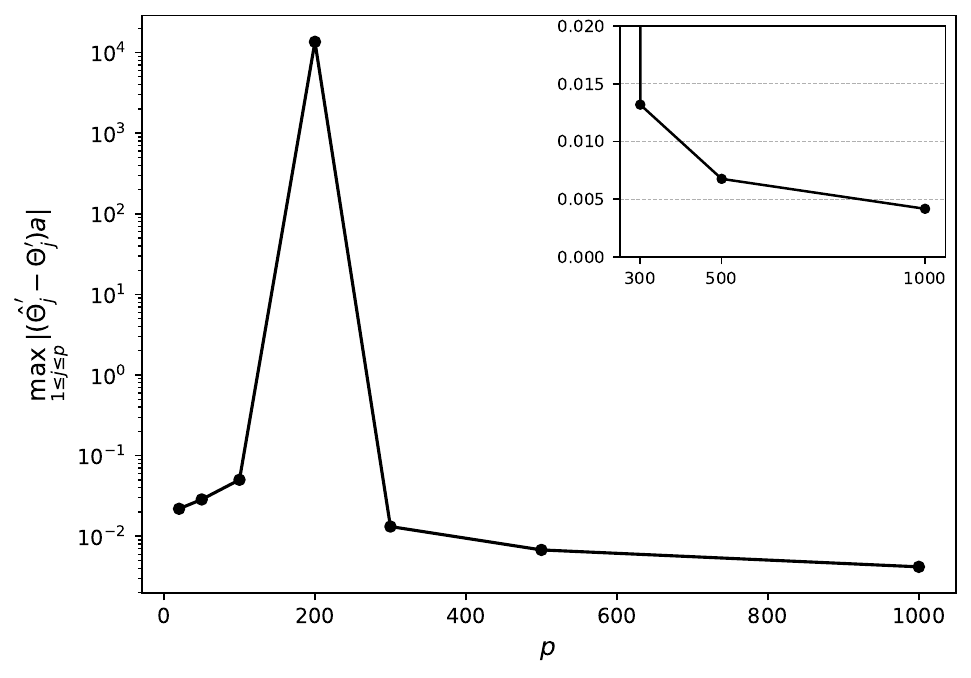}
    \label{fig:rre_dd_mc1}
\end{figure}

This section provides a more thorough analysis of the double descent phenomenon for the RRE and offers a finite-sample assessment of the theoretical discussion of double descent presented in Section \ref{sec:rre_dd}. We first rely on the same data generating process as outlined in Section \ref{sec:mc_design}, set $n=200$ and $K=3$, and vary $p\in\{20,50,100,200,300,500,1000\}$. Table \ref{tab:rre_dd_mc1} documents the behavior of the main quantities entering the theoretical convergence rates for different values of $p$, providing insights into the mechanisms underlying the observed double-descent pattern. This pattern is illustrated in Figure \ref{fig:rre_dd_mc1}, which shows the maximum row-wise-scaled absolute estimation error in Corollary 6(ii), and an in-depth sub-graph focusing on the beyond 300 variables ($p>n$) regime.
\begin{table}[!t]
	\centering
	\caption{Quantities Entering the Convergence Rate of the RRE Precision Matrix Estimator for Different Values of $p$ ($n=200$, $K=3$)}
    \begin{adjustbox}{max width=\textwidth}
	\begin{threeparttable}
        \begin{tabular}{cccccc}
			\hline
			\hline
			$p$ & $\max_{1 \le j \le p} \| a_j\|_2^2$ & $Eigmax(\Sigma_U)$ & $d_{2n}$ & $\bar{\xi}$ & $\delta_n$ \bigstrut\\ 
            20 & 7.062  & 8.706  & 26.523  & 1.733  & 0.558 \\
            50 & 8.655  & 11.585 & 49.354  & 4.801  & 1.899 \\
            100 & 9.840 & 13.722 & 76.831  & 9.055  & 4.540 \\
            200 & 11.262 & 15.849 & 118.222 & 16.675 & 10.545 \\
            300 & 11.354 & 17.070 & 146.097 & 23.920 & 17.039 \\
            500 & 12.357 & 18.617 & 200.374 & 37.018 & 30.943 \\
            1000 & 13.329 & 20.709 & 295.649 & 67.376 & 68.987 \bigstrut[b]\\
			\hline
			\hline
		\end{tabular}
		\vspace*{-0.3cm}
		\begin{tablenotes}
			\footnotesize
			\singlespacing
			\item \leavevmode\kern-\scriptspace\kern-\labelsep 
			Note: The table reports the maximum row norm of the factor loadings, the largest eigenvalue of the idiosyncratic covariance matrix $Eigmax(\Sigma_U)$, and the quantities $d_{2n}$, $\bar{\xi}$, and $\delta_n$ appearing in the theoretical convergence rate.
		\end{tablenotes}
	   \end{threeparttable}
     \end{adjustbox}%
	
	\label{tab:rre_dd_mc1}%
\end{table}%

The results are consistent with the theoretical mechanism described in Corollaries \ref{cl5}--\ref{cl6}. The maximum Euclidean norm (squared) of the factor loading vector increases only moderately with $p$, from $7.062$ at $p=20$ to $13.329$ at $p=1000$. Similarly, $Eigmax(\Sigma_U)$ increases from $8.706$ to $20.709$. Thus, the variance component of the estimation error receives a comparatively large weight from the idiosyncratic error covariance. At the same time, the noise-related quantity $\delta_n$ in the error bound of the interpolating estimator increases sharply with the dimensionality, from $10.545$ at $p=200$ to $68.987$ at $p=1000$. 

This behavior is important for understanding the double descent pattern. In the low-dimensional regime, reducing $p$ lowers the variance-related components of the error bound, consistent with the OLS result in Corollary \ref{cl5}. As $p$ approaches $n$, however, the estimator enters the interpolation region, where estimation becomes increasingly sensitive to the dimensionality. Beyond this region, the RRE is governed by the bound in Corollary \ref{cl6}. Here, the increasing $\delta_n$ can reduce the variance contribution through the term involving
$\sqrt{\frac{\log n+n}{n\delta_n}}$. Hence, sufficiently rapid growth in $\delta_n$ can offset the adverse effect of increasing dimensionality and lead to a reduction in the variance component for $p>n$.
The other quantities reported in Table \ref{tab:rre_dd_mc1} are also consistent with this interpretation. The  bias-related term  $d_{2n}$ increases from $26.523$ to $295.649$, while $\bar{\xi}$ (signal-to-noise) increases from $1.733$ to $67.376$. Consequently, the bias-related terms involving $d_{2n}\sqrt{K/\bar{\xi}}$ do not exhibit an explosive increase over the considered range. %
Thus, the relatively large weight on the variance component, together with the strong increase in $\delta_n$, provides a plausible explanation for why the reduction in variance can become sufficiently pronounced to generate the observed double descent pattern.

Appendix C reports two additional designs: one with idiosyncratic variances drawn from $\mathrm{Unif}(0.1,0.3)$ and one with $tr(\Sigma_U)/n=2$. They show how the reported error pattern changes when the noise scale and $\delta_n$ vary differently with $p$. These experiments are descriptive comparisons of changing data-generating processes, rather than a proof that growth in $\delta_n$ alone determines double descent.

\section{Empirical Application}\label{empirics}

 The analysis focuses on the global minimum variance portfolio (GMVP), whose vector of portfolio weights, $\omega = \frac{\Theta 1_p}{1_p' \Theta 1_p}$, depends only on the precision matrix of asset returns. We cover tables with transaction costs. Appendix C reports the results without transaction costs and for a shorter evaluation period. We focus on the Sharpe Ratio, the portfolio's standard deviation, and its average return.

\subsection{Data and illustration of the forecasting experiment}

For our analysis, we use monthly excess returns for stocks in the S\&P 500 index. Excess returns are calculated by subtracting the corresponding one-month Treasury bill rate from the individual stock returns. The sample period spans January 1995 to December 2025, yielding $n = 372$ return observations.

We conduct the out-of-sample forecasting experiment using a rolling-window approach with an in-sample window of ten years, corresponding to $n_I = 120$ monthly observations. Consequently, the out-of-sample evaluation period spans from January 2005 to December 2025, yielding $n - n_I = 252$ out-of-sample portfolio returns. At each investment date $h$, the model is estimated using the most recent 120 monthly return observations. Based on the resulting portfolio weight estimates $\hat{\omega}_h$, the out-of-sample portfolio return for period $h+1$ is computed as $\hat{y}^{P}_{h+1} = \hat{\omega}_h' y_{h+1}$. Subsequently, we roll the estimation window forward by one observation. The portfolio weights $\hat{\omega}_{h+1}$ are then re-estimated using the updated in-sample window, and the corresponding out-of-sample portfolio return is computed. This procedure is repeated iteratively until the end of the sample is reached. All portfolios are rebalanced monthly.

The out-of-sample average return and variance in the absence of transaction costs are computed as $\hat{\mu}_{os} = \frac{1}{n - n_I} \sum_{h = n_I}^{n-1} \hat{\omega}_h' y_{h+1}$ and $\hat{\sigma}^2_{os} = \frac{1}{n - n_I - 1} \sum_{h = n_I}^{n-1} \left(\hat{\omega}_h' y_{h+1} - \hat{\mu}_{os}\right)^2$. Based on these quantities, the Sharpe ratio without transaction costs is given by $SR = \frac{\hat{\mu}_{os}}{\hat{\sigma}_{os}}$.
In addition, we evaluate portfolio performance in the presence of transaction costs. Following \cite{Li2015}, the out-of-sample excess returns with transaction costs are computed as
$
\hat{y}^{tc}_{P,h+1} = \hat{\omega}_{h}' y_{h+1} - c \left(1 + \hat{\omega}_{h}' y_{h+1}\right) \sum_{j = 1}^{p} |\hat{\omega}_{h+1,j} - \hat{\omega}_{h,j}^+|,
$
where $\hat{\omega}_{h,j}^+ = \hat{\omega}_{h,j}(1+y_{h+1,j})/(1+y_{P,h+1})$ denotes the portfolio weight of asset $j$ prior to rebalancing. The parameter $c$ represents proportional transaction costs, which are set to 10 basis points per transaction. Based on these adjusted returns, the mean, variance, and Sharpe ratio of the out-of-sample portfolio returns with transaction costs are defined as $\hat{\mu}^{tc}_{os} = \frac{1}{n - n_I} \sum_{h = n_I}^{n-1} \hat{y}^{tc}_{P,h+1}, \hat{\sigma}^{2, tc}_{os} = \frac{1}{n - n_I - 1} \sum_{h = n_I}^{n-1} \left(\hat{y}^{tc}_{P,h+1} - \hat{\mu}^{tc}_{os}\right)^2$, and $SR^{tc} = \frac{\hat{\mu}^{tc}_{os}}{\hat{\sigma}^{2, tc}_{os}}$. Finally, the portfolio turnover is measured as
$ PT =\frac{1}{n - n_I} \sum_{h = n_I}^{n-1}\sum_{j = 1}^{p} |\hat{\omega}_{h+1,j} - \hat{\omega}_{h,j}^+|.$

To assess the performance of the different approaches across varying asset dimensions, we consider portfolios of sizes $p \in \{10, 50, 100, 200\}$. At each investment date, the $p$ largest assets, as measured by market capitalization, are selected, provided they have a complete return history over the preceding $n_I$ periods, following the procedure of \cite{lw2017}. The portfolio performance of each method is evaluated using the annualized out-of-sample average return (AV), standard deviation (SD), Sharpe ratio (SR), and portfolio turnover (PT).

Additionally, we employ the circular bootstrap procedure proposed by \cite{ledoit2008robust} to test for statistically significant differences in Sharpe ratios across models. Specifically, we test the hypotheses
$
H_0: \quad \text{SR}_{max} \leq \text{SR}_0 \, \text{vs} \, H_a: \quad \text{SR}_{\text{max}} > \text{SR}_0,
$
where $\text{SR}_{max}$ denotes method with the highest Sharpe ratio for a specific portfolio size $p$ and $\text{SR}_0$ is the Sharpe ratio of an alternative method. The reported pairwise $p$-values are unadjusted. Because the benchmark is selected as the sample winner and several comparisons are reported, they should be interpreted descriptively; they do not provide selection-adjusted or familywise evidence of superiority.

Similar to set of models analyzed in the Monte Carlo study in Section~\ref{sec:mc}, we incorporate the following approaches in the empirical study: EW: The equally weighted portfolio; PCR-3, PCR-5, PCR-adaptive: Our principal component regression with fixed number of latent factors equal to 3, 5  as well as the adaptive procedure for the selecting the number of factors; RRE: the ridgeless regression estimator; POET: The principal orthogonal complement thresholding covariance matrix estimator from \cite{fan2013}, with latent factors and the number of factors is estimated based on the $IC_1$ criterion by \cite{Bai2002}; NL-LW: The non-linear shrinkage estimator of \cite{lw2017};  RN-FM: Residual based nodewise regression factor model of \cite{caner2022} using the Fama-French three factors by \cite{Fama1993} as observed factors.\footnote{We further analyzed the forecasting performance of the Fama-French three factor model, as well as the nodewise regression estimator of \cite{caner2019}, and SF-NL-LW: The single factor non-linear shrinkage estimator of \cite{lw2017}. However, since both approaches do not provide better performance in terms of SR, we omit their results from the result tables.}

\subsection{Out-of-sample forecast for GMV portfolio results}

The out-of-sample results that account for transaction costs are presented in Table~\ref{tab_appl_tc_p2}. For small- to medium-sized portfolios ($p \leq 100$), i.e. in the low-dimensional regime where $p<n$, our PCR estimator attains the highest SR. Specifically, the five-factor principal component estimator (PCR-5) performs best at $p=10$, whereas the three-factor specification (PCR-3) yields the highest SR for $p=50$ and $p=100$. Between $p = 10$ and $p = 50$, a substantial improvement in performance is observed. For $p = 50$, the Sharpe ratio exceeds one, indicating very strong risk-adjusted performance. In particular, the highest SR increases from 0.588 (for $p = 10$ with PCR-5) to 1.144 (for $p = 50$ with PCR-3). At $p=50$, PCR-3 has the highest sample Sharpe ratio, but several pairwise $p$-values exceed 0.05, and the tests are not adjusted for selection or multiplicity. The Sharpe ratio of PCR-3, at 1.144, is approximately 65\% larger than that of the strongest external comparator in this table, NL-LW, which attains an SR of 0.692. %
For $p=100$, PCR-3 continues to report the highest SR among the models examined, but the SR falls to 0.65 relative to its $p=50$ value. These results underscore the importance of judicious portfolio-size selection.  We should also stress that PCR-adaptive performs well in comparison with fixed PCR methods. 
\begin{table}[!t]
	\centering
	\caption{Out-of-sample forecasting results with transaction costs: January 2005--December 2025}
    \begin{adjustbox}{max width=\textwidth}
	   \begin{threeparttable}
        \begin{tabular}{l|cccccccc}
			\hline
			\hline
			\multicolumn{1}{c|}{Model} & EW    & PCR-3 & PCR-5 & PCR-adaptive & RRE   & POET  & NL-LW & RN-FM \bigstrut\\
			\hline
			\multicolumn{9}{c}{$p = 10$} \bigstrut\\
			\hline
			\multicolumn{1}{c|}{AV} & 0.088 & 0.101 & 0.107 & 0.094 & 0.100 & 0.088 & 0.042 & 0.094 \bigstrut[t]\\
			\multicolumn{1}{c|}{SD} & 0.166 & 0.184 & 0.183 & 0.176 & 0.182 & 0.166 & 0.139 & 0.176 \\
			\multicolumn{1}{c|}{SR} & 0.526 & 0.549 & \textbf{0.588} & 0.534 & 0.550 & 0.526 & 0.300 & 0.535 \\
			\multicolumn{1}{c|}{p-value SR} & 0.375 & 0.328 & -     & 0.328 & 0.086 & 0.370 & 0.055 & 0.217 \bigstrut[b]\\
			\hline
			\multicolumn{9}{c}{$p = 50$} \bigstrut\\
			\hline
			\multicolumn{1}{c|}{AV} & 0.077 & 0.230 & 0.212 & 0.217 & 0.151 & 0.078 & 0.084 & 0.091 \bigstrut[t]\\
			\multicolumn{1}{c|}{SD} & 0.142 & 0.201 & 0.200 & 0.198 & 0.174 & 0.141 & 0.122 & 0.139 \\
			\multicolumn{1}{c|}{SR} & 0.547 & \textbf{1.144} & 1.059 & 1.096 & 0.865 & 0.553 & 0.692 & 0.651 \\
			\multicolumn{1}{c|}{p-value SR} & 0.016 & -     & 0.227 & 0.472 & 0.089 & 0.013 & 0.049 & 0.033 \bigstrut[b]\\
			\hline
			\multicolumn{9}{c}{$p = 100$} \bigstrut\\
			\hline
			\multicolumn{1}{c|}{AV} & 0.079 & 0.203 & 0.166 & 0.149 & -0.028 & 0.080 & 0.034 & 0.086 \bigstrut[t]\\
			\multicolumn{1}{c|}{SD} & 0.146 & 0.313 & 0.294 & 0.267 & 0.270 & 0.143 & 0.129 & 0.142 \\
			\multicolumn{1}{c|}{SR} & 0.540 & \textbf{0.650} & 0.564 & 0.557 & -0.103 & 0.557 & 0.266 & 0.609 \\
			\multicolumn{1}{c|}{p-value SR} & 0.724 & -     & 0.328 & 0.345 & 0.002 & 0.749 & 0.109 & 0.888 \bigstrut[b]\\
			\hline
			\multicolumn{9}{c}{$p = 200$} \bigstrut\\
			\hline
			\multicolumn{1}{c|}{AV} & 0.082 & 0.133 & 0.082 & 0.059 & 0.092 & 0.071 & 0.062 & 0.086 \bigstrut[t]\\
			\multicolumn{1}{c|}{SD} & 0.150 & 0.434 & 0.309 & 0.275 & 0.146 & 0.131 & 0.111 & 0.145 \\
			\multicolumn{1}{c|}{SR} & 0.544 & 0.305 & 0.267 & 0.215 & \textbf{0.631} & 0.539 & 0.560 & 0.590 \\
			\multicolumn{1}{c|}{p-value SR} & 0.614 & 0.259 & 0.182 & 0.113 & -     & 0.588 & 0.683 & 0.806 \bigstrut[b]\\
			\hline
			\hline
		\end{tabular}
		\vspace*{-0.3cm}
		\begin{tablenotes}
        \footnotesize
			\singlespacing
			\item \leavevmode\kern-\scriptspace\kern-\labelsep 
			Note: Our principal component regression with a fixed number of latent factors equal to 3 or 5, as well as the adaptive procedure for selecting the number of factors (PCR-3, PCR-5, PCR-adaptive) and the ridgeless regression estimator (RRE) are compared to the equally weighted portfolio (EW), the POET estimator of \cite{fan2013} (POET), the non-linear shrinkage estimator (NL-LW) of \cite{lw2017} and the residual based nodewise regression factor model of \cite{caner2022} (RN-FM). The table shows the portfolio average return (AV), standard deviation (SD), Sharpe ratio (SR) and the p-value of the \cite{ledoit2008robust} test for equal Sharpe ratios, where for each portfolio size $p$ the method with the largest SR is used as benchmark and tested against all remaining methods (p-value SR).
		\end{tablenotes}
	\end{threeparttable}
    \end{adjustbox}%
	\label{tab_appl_tc_p2}%
\end{table}%
For the large-dimensional case ($p=200$), the RRE estimator attains the highest SR. As the portfolio dimension increases, the RRE estimator's SR follows a pattern that is qualitatively consistent with the double descent behavior of its estimation error in Corollary \ref{cl6}: in the low-dimensional case with $p=10$, the SR starts at 0.550, declines to -0.103 as the portfolio size increases to 100 stocks, and rises again to 0.631 in the high-dimensional case $p=200$, where the number of stocks exceeds the in-sample length of 120 periods. We describe this as an illustrative, non-monotonic pattern rather than a direct empirical confirmation of Corollary \ref{cl6}: the portfolio (and hence the population precision matrix and Sharpe ratio being targeted) changes composition as $p$ varies across $\{10,50,100,200\}$, so part of the pattern reflects a changing target rather than estimation error alone, and Corollary \ref{cl6} itself only bounds estimation error from above. From a practical perspective, this behavior indicates that RRE enables the construction of portfolios with $p>n$ without requiring tuning parameters (in contrast to several alternative methods), while still delivering competitive out-of-sample performance.

Among the competing approaches considered (POET, NL-LW, SF-NL-LW, and RN-FM), RN-FM performs competitively across most portfolio sizes (with the exception of $p=50$), frequently yielding the highest SR among the alternatives. With respect to out-of-sample risk, the nonlinear shrinkage estimators of \cite{lw2017} produce the lowest SD across the portfolio sizes examined. Results obtained without accounting for transaction costs are available in Table C.2 in  Appendix C and show qualitatively very similar results.
The Appendix C  further includes the analysis of a shorter out-of-sample period from January 2005 to December 2019. The results are qualitatively very similar to results here  as illustrated in Table C.3.

\section{Conclusion} \label{sec:conclusion}

In this paper, we have introduced a general class of dense precision matrix estimators, extending projection-based linear predictors for a single regression coefficient \citep{bbsmw21,bsmw22} to the estimation of the precision matrix, uniformly over rows and without requiring a plug-in estimate of the underlying latent factors or sparsity assumptions on the matrix itself. Under certain conditions, our RRE estimator exhibits double descent in the estimation error.

\spacingset{1.1}
\bibliographystyle{chicagoa}
\bibliography{interpolation}

\clearpage
\appendix
\setcounter{section}{0}
\setcounter{figure}{0}

\setcounter{equation}{0}
	\setcounter{lemma}{0}
	\setcounter{table}{0}
	\renewcommand{\theequation}{A.\arabic{equation}}
	\renewcommand{\thelemma}{A.\arabic{lemma}}
	\renewcommand{\thecorollary}{A.\arabic{corollary}}
	\renewcommand{\thetable}{A.\arabic{table}}
	\renewcommand{\thethm}{A.\arabic{thm}}

{\Large \bf  Appendix}

\spacingset{1.8}

 The Appendix has five parts.   Appendix A covers the proofs of Theorem 1-3, Corollaries A.1-A.3.  Appendix B covers population precision matrix formula, the proof of Lemma 1, and two tail inequalities with discussion about a remark in Theorem 3.  Appendix C covers additional simulations and empirics, including two additional double-descent designs. Appendix D contains the extended literature comparison, and Appendix E discusses the assumptions.

\begin{center}
{\bf Appendix A}
\end{center}

{\bf  Proof of Theorem 1}:

 	Using (4) and (10) we get  the following inequality
	\begin{equation}
	\| \tilde{\alpha}_j - \alpha_{j}^* \|_2^2   \le  3 \| \hat{B} (Y_{-j} \hat{B})^+ u_j \|_2^2 + 
	3 \| \hat{B} (Y_{-j} \hat{B})^+ F a_j \|_2^2 
	 +  3 \| \alpha_{j}^* \|_2^2.\label{pl1-gl1}	
	\end{equation}
	
	The proof  contains a detailed transformation of results that can be obtained in linear regression framework to precision matrix
estimation.
Since we ultimately aim to use our results for high-dimensional outcomes, we need them to hold uniformly over $j = 1, \dots, p,$. It is important to note that this is in striking contrast with pointwise results (as are those from, e.g., proof of Lemma 17 \cite{bbsmw21} that would be applicable to only one outcome.    Each of the right side terms  in (\ref{pl1-gl1}) will be analyzed in each of Steps 1-3.

	{\bf Step 1.} Start with the first right side term in (\ref{pl1-gl1})
	\begin{equation}
	\| \hat{B} (Y_{-j} \hat{B})^+ u_j \|_2^2 = u_j' M_j u_j,\label{pl1-gl2}
	\end{equation}
	with $M_j: n \times  n$ matrix defined as 
	\begin{equation}
	M_j:= (Y_{-j } \hat{B})^{+'} \hat{B}' \hat{B} (Y_{-j } \hat{B})^+.\label{pl1-gl2a}
	\end{equation}

To simplify the proof,  we consider  the following trace term: 
		\[ tr (M_j)
		= tr [ (Y_{-j} \hat{B})^{+'} \hat{B}' \hat{B} (Y_{-j} \hat{B})^+],\]
		which can be upper bounded as
		\begin{eqnarray}
		tr ( M_j ) & \le & rank (M_j) \| M_j \|_2 = rank (Y_{-j} \hat{B})^+  \| M_j \|_2 \nonumber \\
		& = & rank (Y_{-j } \hat{B}) \| M_j \|_2  =  rank (Y_{-j} P_{\hat{B}}) \| M_j \|_2, = \hat{r}_j \| M_j \|_2 \label{pl1-gl5}
		\end{eqnarray}
		where the first inequality is due to $M_j$ being positive semidefinite, a $n \times n$ matrix and Exercise 7.49 of \cite{abamag2005}. The first equality is by Exercise 8.27a and 4.24a of \cite{abamag2005}, the second equality is by Exercise 10.28 of \cite{abamag2005}, and the third equality is again by Exercises 4.13d, 4.15b, 4.25b, and 10.28 of \cite{abamag2005} with $P_{\hat{B}} = \hat{B} \hat{B}^+$  in p.8 of \cite{bbsmw21}. The last equality is just the definition of the complexity $\hat{r}_j$. Now we consider, as proved in (44) of \cite{bbsmw21} 
		\begin{equation}
		\| M_j \|_2 \le (n \hat{\eta}_j)^{-1},\label{pl1-gl6}
				\end{equation}
		where $\hat{\eta}_j:= \hat{\sigma}_{\hat{r}_j}^2 (Y_{-j} P_{\hat{B}})/n$ which is the $\hat{r}_j$-th largest singular value, squared, of the matrix $Y_{-j} P_{\hat{B}}$, and divided by $n$. Define $C_p>0$ a positive constant.
        
        {
Then  by (\ref{pl1-gl5})(\ref{pl1-gl6}) and $\bar{\eta}$, $\bar{r}$ definitions
\begin{equation}
\gamma_e^2 \sigma_{j}^2 [ 2 C_p\| M_j \|_2 \log n + tr (M_j)]
\le \gamma_e^2 \Gamma [ \frac{2 C_p \log n }{n \bar{\eta}} + \frac{\bar{r}}{n \bar{\eta}}]  .\label{pt1-s1}  
\end{equation}

Now we provide a uniformity proof across $j=1,\cdots, p$ outcomes, and this is theoretical extension of pointwise, one outcome proof in \cite{bbsmw21}. Next, define the following events 
\begin{equation}
    E_{1j}:=\{u_j' M_j u_j \le 2 \gamma_e^2 \sigma_{j}^2 [ 2 C_p \| M_j \|_2 \log n + tr (M_j) ]\},\label{pl1-gl3}\end{equation}
    and 
    \begin{equation}
        \bar{E}_{1j}:=\{u_j' M_j u_j \le 2 \gamma_e^2 \Gamma [ \frac{2 C_p \log n }{n \bar{\eta}}+ \frac{\bar{r}}{n \bar{\eta}}]\}.\label{pt1-s1a}
        \end{equation} with 
        \begin{equation}
        \bar{E}_1:= \{\max_{1\le j \le p} u_j' M_j u_j \le 2 \gamma_e^2 \Gamma [ \frac{2 C_p \log n }{n \bar{\eta}}+ \frac{\bar{r}}{n \bar{\eta}}]\}.\label{pt1-s2}
        \end{equation}
        By (\ref{pt1-s1})
        \begin{equation}
        E_{1j} \subseteq \bar{E}_{1j}.\label{pt1-s1b}
            \end{equation}
        Clearly, since the upper bounds of the events $ \bar{E}_{1j}, \bar{E}_1$ are the same 
        \begin{equation}
            \cap_{j=1}^p \bar{E}_{1j} = \bar{E}_1.\label{pt1-s1c}
        \end{equation}
        By (\ref{pt1-s1b})(\ref{pt1-s1c})
        \begin{equation}
            \cap_{j=1}^p E_{1j}  \subseteq \bar{E_1}.\label{pt1-s1d}
        \end{equation}
        See  that  by $E_{1j}^c$ as the complement of $E_{1j}$
        \begin{equation}
            P (\cap_{j=1}^p  E_{1j} ) + P (\cup_{j=1}^p  E_{1j}^c) =1.\label{pt1-s2a}
        \end{equation}
        Note that  we have, by (\ref{pt1-s1d}) 
        \begin{equation}
          P (\bar{E}_1) \ge P (\cap_{j=1}^p  E_{1j} )\label{pt1-s1e}
          \end{equation}
so by (\ref{pt1-s2a})(\ref{pt1-s1e}) and 
\begin{eqnarray}
     P (\bar{E}_1) &\ge& 1 - P (\cup_{j=1}^p E_{1j}^c)\nonumber \\
     & \ge & 1 - \sum_{j=1}^p P (E_{1j}^c).\label{pt1-s1f}
     \end{eqnarray}
  Now we calculate the sum of the upper bound probability of $E_{1j}^c$, across $j$.
Note that $u_j$ is a   sub-Gaussian vector, and independent of $Y_{-j}$, and $\hat{B}$ by Assumption 1 and the definition of $\hat{B}$,  with $M_j$ being a symmetric positive semi-definite matrix. 
So $M_j$ is independent of $u_j$.
Then 
Lemma B.1 (with $t=C_p \log n $, $H=M_j, \zeta=u_j$, conditional on $M_j$) allows us to write
\begin{equation}
    P (E_{1j}^c| M_j) = 
    P\left( u_j' M_j u_j > 2 \gamma_e^2 \sigma_{j}^2 [ 2 C_p \| M_j \|_2 \log n + tr (M_j) ] | M_j\right) \le \frac{1}{n^{C_p}}.\label{pt1-s5}
    \end{equation}
    
Next, use the union bound and by Assumption 3, using $\log p \le  (\frac{C_p}{2}) \log n$. By the law of total probability (tower property) applied to (\ref{pt1-s5}), and since the bound in (\ref{pt1-s5}) holds almost surely (for every realization of $M_j$),
\begin{eqnarray}
\sum_{j=1}^p P (E_{1j}^c) & = & \sum_{j=1}^p E\left[ P (E_{1j}^c | M_j)\right]
\le \sum_{j=1}^p \frac{1}{n^{C_p}} \nonumber \\
&\le &
\exp (\log p - C_p \log n) \le \exp (-\frac{C_p}{2} \log n)
= \frac{1}{n^{C_p/2}}
.\label{pt1-s6}
\end{eqnarray}
Use (\ref{pt1-s1f})(\ref{pt1-s6}) in (\ref{pl1-gl2})(\ref{pt1-s2})
\begin{equation}
    P \left( \max_{1 \le j \le p} \| \hat{B} (Y_{-j} \hat{B})^+ u_j \|_2^2 \le 2 
     \gamma_e^2 \Gamma [ \frac{2 C_p \log n }{n \bar{\eta}}+ \frac{\bar{r}}{n \bar{\eta}}]    \right) \ge  1 - \frac{1}{n^{C_p/2}}.\label{pt1-s7}
\end{equation}

		{\bf Step 2.} Now consider the second term on the right side of (A.1). Keeping in mind that,   $A_{-j}' (A_{-j}^{+})' = I_K$ and recalling on (5), while setting 
        $M_{\hat{B}}:=I_{p-1}- P_{\hat{B}}$, we further have		
        \begin{eqnarray*}
		&&\hat{B} (Y_{-j} \hat{B})^+ F  =  \hat{B} (Y_{-j} \hat{B})^+ F A_{-j}' A_{-j}^{+'} = \hat{B} (Y_{-j} \hat{B})^+ (Y_{-j} - U_{-j}) A_{-j}^{+'} \\
		& = & \hat{B} (Y_{-j} \hat{B})^+ Y_{-j} P_{\hat{B}} A_{-j}^{+'} + 	\hat{B} (Y_{-j} \hat{B})^+ Y_{-j} M_{\hat{B}} A_{-j}^{+'} 
		 -  	\hat{B} (Y_{-j} \hat{B})^+  U_{-j}	A_{j}^{+'}
			\end{eqnarray*}
			Use  simple inequality $(a+b+c)^2 \le 3a^2 + 3b^2 + 3c^2$, 
			\begin{eqnarray}
& & \| \hat{B} (Y_{-j} \hat{B})^+ F a_j \|_2^2 \le  3 \| \hat{B} (Y_{-j} \hat{B})^+ Y_{-j} P_{\hat{B}} A_{-j}^{+'} a_j \|_2^2 \nonumber \\
& + & 3 \| \hat{B} (Y_{-j} \hat{B})^+ Y_{-j} (I_{p-1} - P_{\hat{B}}) A_{-j}^{+'} a_j \|_2^2 
+ 3 \| \hat{B} (Y_{-j} \hat{B})^+  U_{-j}	A_{j}^{+'} a_j \|_2^2.\label{pl1-gl7a}
\end{eqnarray}

In Steps 2a, 2b, and 2c below, we separately analyze each of the three terms on the right side of (\ref{pl1-gl7a}).

{\bf Step 2a.} So  
\begin{equation}
\| \hat{B} (Y_{-j} \hat{B})^+ Y_{-j} P_{\hat{B}} A_{-j}^{+'} a_j \|_2^2 		\le \| \hat{B} (Y_{-j} \hat{B})^+ Y_{-j} P_{\hat{B}}\|_2^2 \| A_{-j}^{+'} a_j \|_2^2.\label{pl1-gl8}		
		\end{equation}
		Note that we can write the first term on the right side of (\ref{pl1-gl8}) as 
		\begin{equation}
		\| \hat{B} (Y_{-j} \hat{B})^+ Y_{-j} P_{\hat{B}}\|_2		= \| \hat{B} (\hat{B}' Y_{-j}' Y_{-j} \hat{B})^+ \hat{B}' Y_{-j}' Y_{-j} P_{\hat{B}} \|_2, \label{pl1-gl9}
		\end{equation}
		where we use Moore-Penrose inverse rule (134) in \cite{bsmw22} which is for a rectangular matrix $S$, we have  $(S'S)^+ S' = S^+$, which we input 
		$S= Y_{-j} \hat{B}$ in our case above.
		To simplify the expressions see that by the proof of Lemma 15 of \cite{bbsmw21} $P_{\hat{B}}= V_1 V_1',$ where $\hat{B}= V_1 D V_2'$ is a singular value decomposition.   
		$V_1,V_2$ are orthonormal matrices and with dimensions of $(p-1) \times r_0, q \times r_0$ respectively,  and  $D$ is a  $r_0$ dimensional diagonal matrix, and the  rank of $(\hat{B}) = r_0$.
		Rewrite right side of (\ref{pl1-gl9}) as 
		\begin{equation}
		\| \hat{B} (Y_{-j} \hat{B})^+ Y_{-j} P_{\hat{B}}\|_2 = \| V_1 D V_2' (V_2 D V_1' Y_{-j}' Y_{-j} V_1 D V_2')^+ V_2D V_1' Y_{-j}' Y_{-j} V_1 V_1' \|_2.\label{pl1-gl10}				\end{equation}
		Then in (\ref{pl1-gl10}) right side see that 
		\begin{equation}
		(V_2 D V_1' Y_{-j}' Y_{-j} V_1 D V_2')^+	= V_2 D^{-1} (V_1' Y_{-j}' Y_{-j} V_1)^+ D^{-1} V_2',\label{pl1-gl11}
		\end{equation}
	which is shown in the proof of Lemma 15 of \cite{bbsmw21}.	 Then use (\ref{pl1-gl11}) in (\ref{pl1-gl10}), $V_2'V_2=I_{r_0}$
		\begin{eqnarray*}
		V_1 D V_2' (V_2 D V_1' Y_{-j}' Y_{-j} V_1 D V_2')^+ V_2 D V_1' & = & V_1 D V_2' V_2 D^{-1} [V_1' Y_{-j}' Y_{-j} V_1]^+ D^{-1} V_2' V_2 D V_1' \\
		& = & V_1 [ V_1' Y_{-j}' Y_{-j} V_1]^+ V_1'		
		\end{eqnarray*}
		Use the last expression in (\ref{pl1-gl10}) right side 
		\[ 
		\| V_1 D V_2' (V_2 D V_1' Y_{-j}' Y_{-j} V_1 D V_2')^+ V_2DV_1' Y_{-j}' Y_{-j} V_1 V_1' \|_2	= \| V_1 [V_1' Y_{-j}' Y_{-j} V_1]^+ V_1' Y_{-j} Y_{-j}' V_1 V_1' \|_2.\]	
		Then clearly since $V_1: (p-1) \times r_0$ matrix, and Exercise 10.26 of \cite{abamag2005} shows $[V_1' Y_{-j}' Y_{-j} V_1]^+ V_1' Y_{-j}' Y_{-j} V_1	$ is a symmetric idempotent matrix and all idempotent matrices eigenvalues are either 0 or 1, and since $V_1' V_1 =I_{r_0}$ 
			\[ \| V_1 [V_1' Y_{-j}' Y_{-j} V_1]^+ V_1' Y_{-j}' Y_{-j} V_1 V_1' \|_2 \le \| V_1\|_2 \| [V_1' Y_{-j}' Y_{-j} V_1]^+ V_1' Y_{-j}' Y_{-j} V_1	\|_2 \| V_1' \|_2 \le 1.\]	
			So use this last inequality in (\ref{pl1-gl10})	
			\begin{equation}
		\| \hat{B} (Y_{-j} \hat{B})^+ Y_{-j} P_{\hat{B}}\|_2 \le 1,\label{pl1-gl12}
		\end{equation}	
		Now consider the second term on the right side of (\ref{pl1-gl8}),
		$ \| A_{-j}^{+'} a_j \|_2^2 = a_j' A_{-j}^+ A_{-j}^{+'} a_j = a_j' (A_{-j}' A_{-j})^{-1} a_j,$
		since $A_{-j}^{+}= (A_{-j}' A_{-j})^{-1} A_{-j}^{'}$ with $p-1 > K$. Next 
		 \begin{eqnarray}
		 a_j' A_{-j}^+ (A_{-j}^+)' a_j & = & a_j' \Sigma_f^{1/2} (\Sigma_f^{1/2} A_{-j}' A_{-j} \Sigma_f^{1/2})^{-1} \Sigma_f^{1/2} a_j \nonumber \\
		 & \le & \frac{a_j' \Sigma_f a_j}{\lambda_K (A_{-j} \Sigma_f A_{-j}')},\label{31a}
\end{eqnarray}
where we use first Rayleigh inequality on p.182 of \cite{abamag2005} and then Theorem 2.6.3b of \cite{hj2013}, where $\lambda_K(.)$ is the $K$-th largest eigenvalue for the matrix 
$A_{-j} \Sigma_f A_{-j}': (p-1) \times (p-1)$.  By (6)(7), we try to rewrite (\ref{31a}) right-side. 
\begin{eqnarray}
\lambda_K (A_{-j}' \Sigma_f A_{-j}) & = &\left[ \frac{\lambda_K (A_{-j}' \Sigma_f A_{-j})}{\| \Sigma_{U,-j} \|_2}\right] \| \Sigma_{U,-j} \|_2 \nonumber \\
& = & \xi_j \| \Sigma_{U,-j} \|_2 \ge \bar{\xi} \| \Sigma_{U,-j} \|_2 \nonumber \\
& \ge & \bar{\xi} \beta,\label{45a}
\end{eqnarray}
where we use $\beta:= \min_{1 \le j \le p} \| \Sigma_{U,-j} \|_2  >0$ by Assumption 2(ii).
Then by  Assumption 4(iii)
\begin{equation}
a_j' \Sigma_f a_j \le \| a_j \|_2^2 Eigmax (\Sigma_f) \le C K,\label{46ab}
\end{equation}
by $Eigmax (\Sigma_f) \le C < \infty$ Assumption 3. So the first term on the right side of (\ref{31a}) can be bounded as  as 
\begin{equation}
\max_{1 \le j \le p} \| A_{-j}^{+'} a_j \|_2^2 =\max_{1 \le j \le p}  \left[\frac{ (a_j' \Sigma_f a_j)}{\lambda_K (A_{-j} \Sigma_f A_{-j}')} \right]\le \frac{ C K}{\bar{\xi} \beta}.\label{pl1-gl13}
\end{equation}	
Clearly by (\ref{pl1-gl12})(\ref{pl1-gl13}) in (\ref{pl1-gl8})
	\begin{equation}
\max_{1 \le j \le p} \| \hat{B} (Y_{-j} \hat{B})^+ Y_{-j} P_{\hat{B}} A_{-j}^{+'} a_j \|_2^2 		\le 	\frac{ C K}{\bar{\xi} \beta}.\label{pl1-gl13a}
\end{equation}

{\bf Step 2b.}	Take the second term on the right side of (\ref{pl1-gl7a})

\begin{equation}
\| \hat{B} (Y_{-j} \hat{B})^+ Y_{-j} (I_{p-1} - P_{\hat{B}}) A_{-j}^{+'} a_j \|_2^2 \le \| \hat{B} (Y_{-j} \hat{B})^+ \|_2^2 
\| Y_{-j} (I_{p-1} - P_{\hat{B}}) A_{-j}^{+'} a_j \|_2^2.\label{pl1-gl14}
\end{equation}
See that the first term on the right side in (\ref{pl1-gl14})  by $\| S_b \|_2^2 = \| S_b'S_b \|_2$ for a rectangular matrix $S_b:=\hat{B} (Y_{-j} \hat{B})^+$ and $M_j$ definition in (\ref{pl1-gl2a})
\begin{equation}
\| \hat{B} (Y_{-j} \hat{B})^+ \|_2^2  = \| M_j \|_2 \le \frac{1}{n \hat{\eta}_j},\label{pl1-gl15}
\end{equation}
where the last inequality is by (\ref{pl1-gl6}). Then by definition of $\hat{\Psi}_j$ in (16)
\begin{equation}
\| Y_{-j} (I_{p-1} - P_{\hat{B}}) \|_2^2 = \sigma_1^2 [(Y_{-j} (I_{p-1} - P_{\hat{B}})] = n \hat{\Psi}_j.\label{pl1-gl16}
\end{equation}
Now consider (\ref{pl1-gl13})(\ref{pl1-gl15})(\ref{pl1-gl16}) on the second term-right side of (\ref{pl1-gl7a}), with 
$ \bar{\Psi}:= \max_{1 \le j \le p} \hat{\Psi}_j, \bar{\eta}:=\min_{1 \le j \le p} \hat{\eta}_j,$
 with $\hat{\eta}_j$ defined in (15)
\begin{equation}
\max_{1 \le j \le p}
\| \hat{B} (Y_{-j} \hat{B})^+ Y_{-j} (I_{p-1} - P_{\hat{B}}) A_{-j}^{+'} a_j \|_2^2 \le C \left(\frac{\bar{\Psi}}{\bar{\eta}}\right) \left( \frac{K}{\bar{\xi} \beta}
\right).\label{pl1-gl17}
\end{equation}

{\bf Step 2c.} Consider the third term on the right side of (\ref{pl1-gl7a})

\begin{equation}
\| \hat{B} (Y_{-j} \hat{B})^+  U_{-j}	A_{j}^{+'} a_j \|_2^2 \le \| \hat{B} (Y_{-j} \hat{B})^+ \|_2^2  \| U_{-j}	A_{j}^{+'} a_j \|_2^2,\label{pl1-gl18}	
\end{equation}
where we analyze the following on the right side of (\ref{pl1-gl18}): $
 \| U_{-j}	A_{j}^{+'} a_j \|_2^2.$ Define \[\gamma_{e,j}^2 = \gamma_w^2 a_j' A_{j}^+ \Sigma_{U,-j} A_{j}^{+'} a_j.
 \] 
 Now we simplify $\gamma_{e,j}^2$ term. See that, uniformly over $j$, by Assumption 2 and (\ref{pl1-gl13}) 
  \begin{eqnarray}
\max_{1 \le j \le p}  \gamma_w^2 a_j' A_{j}^+ \Sigma_{U,-j} A_{j}^{+'} a_j & \le & [\max_{1 \le j \le p} Eigmax (\Sigma_{U,-j})] [\max_{1 \le j \le p}
\gamma_w^2 a_j' A_{j}^{+} A_{j}^{+'} a_j] \nonumber \\
& \le & \frac{\gamma_w^2 C \Gamma K }{\bar{\xi} \beta}.\label{pl1-gl20}
\end{eqnarray}

We again show our differences in theory with \cite{bbsmw21}, \cite{bsmw22} papers, by providing a union bound  proof, rather than their pointwise proof.
Then define the events
\begin{equation}
E_{2j}:= \{ \|U_{-j} A_{-j}^{+'} a_j \|_2^2 \le (n + 2 C_p \log n + 2 \sqrt{ C_p n \log n} )\gamma_{e,j}^2 \}\label{pt1-s20}
\end{equation}
and 
\begin{equation}
    \bar{E}_{2j}:= \{ \|U_{-j} A_{-j}^{+'} a_j \|_2^2 \le (n+ 2 C_p \log n + 
    2 \sqrt{ C_p n \log n }) \frac{C \gamma_w^2 \Gamma K }{\bar{\xi} \beta} \}.\label{pt1-s21}
    \end{equation}
    By (\ref{pl1-gl20}) and the explanations above that equation, with (\ref{pt1-s20})(\ref{pt1-s21})
    \begin{equation}
        E_{2j} \subseteq \bar{E}_{2j}.\label{pt2c-1}
    \end{equation}
Next, define 
\begin{equation}
    \bar{E}_2:= \{ \max_{1 \le j \le p} \|U_{-j} A_{-j}^{+'} a_j \|_2^2 \le (n+ 2 C_p \log n + 
    2 \sqrt{ C_p n \log n }) \frac{C \gamma_w^2 \Gamma K }{\bar{\xi} \beta} \}.\label{pt2c-2}
    \end{equation}
    See that by definitions in (\ref{pt1-s21})(\ref{pt2c-2})
    \begin{equation}
        \cap_{j=1}^p \bar{E}_{2j} = \bar{E}_2.\label{pt2c-3}
    \end{equation}
    Then by (\ref{pt2c-1})(\ref{pt2c-3})
    \begin{equation}
        \cap_{j=1}^p E_{2j} \subseteq \bar{E}_2.\label{pt2c-4}
    \end{equation}
Then by (\ref{pt2c-4})
    \begin{equation}
 P (\bar{E}_2)  \ge  
 P ( \cap_{j=1}^p E_{2j}) 
 =  1 - P ( \cup E_{2j}^c) 
 \ge  1 - \sum_{j=1}^p P (E_{2j}^c).\label{pt1-s22}
    \end{equation}

 Since $U_{-j} A_{-j}^{+'} a_j$ is  $\gamma_w \sqrt{a_j' A_{-j}^+ \Sigma_{U,-j} A_{-j}^{+'} a_j}$ sub-Gaussian vector for each $j=1,\cdots,p$ via Assumption 4, apply Lemma B.1
 with $\epsilon_j:= U_{-j} A_{j}^{+'} a_j$ and $H:=I_n$ and $\gamma_{e,j}^2 = \gamma_w^2 a_j' A_{j}^+ \Sigma_{U,-j} A_{j}^{+'} a_j$ with $t=C_p \log n $
 \begin{equation}
 P [ \| U_{-j} A_{-j}^{+'} a_j \|_2^2 > (n + 2 C_p \log n + 2 \sqrt{C_p n \log n} )\gamma_{e,j}^2] \le 1/n^{C_p}.\label{pl1-gl19}
 \end{equation}
 Then take the union bound and with Assumption 3,  $\log p \le  \frac{C_p}{2} \log n$
 \begin{equation}
     \sum_{j=1}^p P (E_{2j}^c) \le \exp (\log p - C_p \log n) 
     \le \exp ( \frac{-C_p}{2} \log n) = \frac{1}{n^{C_p/2}}.\label{pt1-s23}
 \end{equation}
 Next use  (\ref{pt1-s22})(\ref{pt1-s23})(\ref{pt2c-2})
 \begin{equation}
 P ( \max_{1 \le j \le p} \| U_{-j} A_{-j}^{+'} a_j \|_2^2 \le 
(n+ 2 C_p \log n + 
    2 \sqrt{ C_p n \log n }) \frac{C \gamma_w^2 \Gamma K }{\bar{\xi} \beta} ) \ge 1 - \frac{1}{n^{C_p/2}}.\label{pt1-s24}
      \end{equation}

Use (\ref{pl1-gl15}) and (\ref{pt1-s24}) in (\ref{pl1-gl18}) with probability at least $1-1/n^{C_p/2}$ (note that we ignore the smaller order terms than the right side term in (\ref{pl1-gl20}) since they play no role in the subsequent analysis), while incorporating $\gamma_w^2$ into C,
\begin{equation}
\max_{1 \le j \le p} 
 \| \hat{B} (Y_{-j} \hat{B})^+  U_{-j}	A_{j}^{+'} a_j \|_2^2 \le \frac{C K \Gamma }{ \bar{\eta}  \bar{\xi} \beta}.\label{pl1-gl21}
\end{equation}

Combine all Steps 2a-2c results (\ref{pl1-gl13a})(\ref{pl1-gl17})(\ref{pl1-gl21}) in (\ref{pl1-gl7a}) to have, with probability at least $1-1/n^{C_p/2}$, absorbing 3 into the constant $C$, 
 \begin{equation}
 \max_{1 \le j \le p}
 \| \hat{B} (Y_{-j} \hat{B})^+ F a_j \|_2^2  \le \frac{CK }{\bar{\xi} \beta} + \frac{ \bar{\Psi} K }{\bar{\eta} \bar{\xi} \beta} + \frac{C K  \Gamma }{ \bar{\eta} \bar{\xi} \beta}.\label{pl1-gl22}
 \end{equation}
 
 {\bf Step 2d}.  A combination of (\ref{pt1-s7}) (\ref{pl1-gl22}) provides the following upper bound 
 for the first two right side terms on (\ref{pl1-gl1}), with probability at least $1-2/n^{C_p/2}$
 
 \begin{eqnarray}
 \max_{1 \le j \le p}
		\| \hat{B} (Y_{-j} \hat{B})^+ u_j \|_2^2 &+&  \| \hat{B} (Y_{-j} \hat{B})^+ F a_j \|_2^2 \le 2 \gamma_e^2 \Gamma [ \frac{2 \log n }{n \bar{\eta}} + \frac{\bar{r}}{n \bar{\eta}}] \nonumber \\
        & 
		+&\frac{CK }{\bar{\xi} \beta} + \frac{ \bar{\Psi} K }{\bar{\eta} \bar{\xi} \beta} + \frac{ C K  \Gamma }{ \bar{\eta}\bar{\xi} \beta}.\label{pl1-gl23}
 \end{eqnarray}

	{\bf Step 3.} After setting $\bar{a}_j:= \Sigma_f^{1/2} a_j, \bar{A}_{-j}:= A_{-j} \Sigma_f^{1/2}$, $\bar{G}_j:= I_K + \bar{A}_{-j}' \Sigma_{U,-j}^{-1} \bar{A}_{-j}$ definitions, we upper bound the last term on the right side of (\ref{pl1-gl1}) via Lemma 1 in the following way
 \begin{eqnarray*}
\| \alpha_j^*\|_2^2& = & 	\| \Sigma_{U,-j}^{-1} \bar{A}_{-j} \bar{G}_j^{-1} \bar{a}_j \|_2^2 
 =  \bar{a}_j' \bar{G}_j^{-1} \bar{A}_{-j}' \Sigma_{U,-j}^{-1} \Sigma_{U,-j}^{-1} \bar{A}_{-j} \bar{G}_j^{-1} \bar{a}_j \\
	& \le & \frac{\bar{a}_j' \bar{G}_j^{-1} \bar{A}_{-j}' \Sigma_{U,-j}^{-1} \bar{A}_{-j} \bar{G}_j^{-1} \bar{a}_j}{Eigmin (\Sigma_{U,-j})} 
	=  \frac{1}{Eigmin(\Sigma_{U,-j})} \bar{a}_j' \bar{G}_j^{-1} ( \bar{G}_j - I_K) \bar{G}_j^{-1} \bar{a}_j, 
\end{eqnarray*}
 Then 
\begin{eqnarray}
\| \Sigma_{U,-j}^{-1} \bar{A}_{-j} \bar{G}_j^{-1} \bar{a}_j \|_2^2 & \le & 	 \frac{1}{Eigmin(\Sigma_{U,-j})} [ \bar{a}_j' \bar{G}_j^{-1} \bar{a}_j - \bar{a}_j' \bar{G}_j^{-1} \bar{G}_j^{-1} \bar{a}_j]  \nonumber \\
&\le & 	 \frac{1}{Eigmin(\Sigma_{U,-j})} 	  \bar{a}_j' \bar{G}_j^{-1} \bar{a}_j.\label{40a}	 
\end{eqnarray}	
Next, 
$ \bar{a}_j' \bar{G}_j^{-1} \bar{a}_j = \bar{a}_j' [ I_K + \bar{A}_{-j}' \Sigma_{U,-j}^{-1} \bar{A}_{-j}]^{-1} \bar{a}_j.$
Define $H_j:= \bar{A}_{-j}' \Sigma_{U,-j}^{-1} \bar{A}_{-j}$, and $H_j$ is invertible, with rank $K$
\begin{eqnarray}
\bar{a}_j' (I_K + H_j)^{-1} \bar{a}_j& = & \bar{a}_j' H_j^{-1/2}   [ I_K + H_j^{-1} ]^{-1} H_j^{-1/2} \bar{a}_j \nonumber \\
& \le & \frac{\bar{a}_j' H_j^{-1} \bar{a}_j}{Eigmin(I_K + H_j^{-1})} \le \frac{\bar{a}_j' H_j^{-1} \bar{a}_j}{1+ Eigmin(H_j^{-1})} \nonumber \\
&\le& \bar{a}_j' H_j^{-1} \bar{a}_j = \bar{a}_j' [ \bar{A}_{-j}' \Sigma_{U,-j}^{-1} \bar{A}_{-j}]^{-1} \bar{a}_j,\label{41a}
\end{eqnarray}
where we use Rayleigh inequality for the first inequality, and for the last equality we use $H_j$ definition. 

\noindent Next use $\bar{a}_j:= \Sigma_f^{1/2} a_j, \bar{A}_{-j}:= A_{-j} \Sigma_f^{1/2}$ definitions
\[\bar{a}_j' [ \bar{A}_{-j}' \Sigma_{U,-j}^{-1} \bar{A}_{-j}]^{-1} \bar{a}_j = a_j' (A_{-j}' \Sigma_{U,-j}^{-1} A_{-j})^{-1} a_j.\]
Then using the inequality after proof  of (63) in \cite{bsmw22} at the end of Section A.2.3 of \cite{bsmw22}
\begin{equation}
\bar{a}_j' [ \bar{A}_{-j}' \Sigma_{U,-j}^{-1} \bar{A}_{-j}]^{-1} \bar{a}_j \le \| \Sigma_{U,-j} \|_2 a_j' (A_{-j}' A_{-j})^{-1}a_j.\label{42a}
\end{equation}
Combine (\ref{41a})(\ref{42a}) in (\ref{40a}) to have, by (\ref{31a}) 
\begin{equation}
\| \Sigma_{U,-j}^{-1} \bar{A}_{-j} \bar{G}_j^{-1} \bar{a}_j \|_2^2 	\le \left[\frac{\| \Sigma_{U,-j} \|_2}{Eigmin(\Sigma_{U,-j})}\right] \left[ \frac{	a_j' \Sigma_f a_j}{\lambda_K (A_{-j} \Sigma_f A_{-j}')}\right].\label{43}
\end{equation}
Next,
 \begin{eqnarray}
\left[\frac{\| \Sigma_{U,-j} \|_2}{Eigmin(\Sigma_{U,-j})}\right] 
\left[ \frac{	a_j' \Sigma_f a_j}{\lambda_K (A_{-j} \Sigma_f A_{-j}')}\right]
& \le  & \left[\frac{\| \Sigma_{U,-j} \|_2}{Eigmin(\Sigma_{U,-j})}\right] 
\left[ \frac{C K}{\xi_j \| \Sigma_{U,-j} \|_2}
\right] \nonumber \\
& =  & \frac{C K}{Eigmin (\Sigma_{U,-j}) \xi_j} \le \frac{C K}{ \bar{\xi}},\label{46a}
\end{eqnarray}
where we use $\min_{ 1 \le j \le p} Eigmin (\Sigma_{U,-j}) \ge c > 0$ Assumption 2(i) for the last inequality as well as $\bar{\xi}$ definition, and second equality in (\ref{45a}), with (\ref{46ab}). Now we combine Steps 1-3 to have with probability at least $1-2/n^{C_p/2}$ to have the desired result.(ii). Asymptotics follows from (i). %

{\bf Q.E.D.}

{\bf Proof of Corollary 1}. (i). We need to simplify three terms on the upper bound in Theorem 1 are $\bar{r}, \bar{\eta}, \bar{\Psi}$.  Note that 
$\bar{r}:= \max_{1 \le j \le p} \hat{r}_j$. In that respect, 
\begin{equation}
\hat{r}_j = rank( Y_{-j} P_{V_{1, k_j}}) = rank (Y_{-j} V_{1, k_j}) = k_j,\label{pcl1-1}
\end{equation}
where the first equality is definition and the second equality is by (\ref{pl1-gl5}), and the third equality is 
the proof of Corollary 5 in \cite{bbsmw21}. Then the term 
\begin{equation}
\hat{\eta}_j= \frac{1}{n} \sigma_{k_j}^2 (Y_{-j} P_{V_{1,k_j}}) =  \hat{\lambda}_{k_j},\label{pcl1-2}
\end{equation}
where the first equality is definition and the second equality is by the proof of Corollary 5 in \cite{bbsmw21}, and note that 
$\hat{\lambda}_{k_j} = \frac{1}{n} \sigma_{k_j}^2  (Y_{-j})$ by definition. Then 
\begin{equation}
\hat{\Psi}_j = \frac{\sigma_1^2 (Y_{-j} ( I_{p-1} - P_{V_{1,k_j}}) )}{n} =  \hat{\lambda}_{k_j+1},\label{pcl1-3}
\end{equation}
where the first equality is definition, and the second  one is   by the proof of Corollary 5 in \cite{bbsmw21}, and with $\hat{\lambda}_{k_j+1}= \frac{1}{n} \sigma_{k_j+1}^2 (Y_{-j})$. We need to prove (20) to proceed further. First we  define the event 
\begin{equation}
E_{3j}:=\{ \sigma_{K+1}^2 (Y_{-j})/n < C_0 \Delta_{-j}\},\label{defev1}
\end{equation}
and its complement is $E_{3j}^c$. We start by proving
\begin{equation}
 P [ \cap_{j=1}^p E_{3j}] \to 1.\label{pcl1-6}
\end{equation}
In that respect, by using Weyl's inequality 
\begin{equation}
\sigma_{K+1} (Y_{-j}) \le \sigma_{K+1} (F A_{-j}') + \sigma_1 (U_{-j}) = \sigma_1 (U_{-j}),\label{pcl1-7}
\end{equation}
since $\sigma_{K+1} (F A_{-j}') =0$ by $F A_{-j}'$ having rank of $K$. See that 
$\sigma_1^2 (U_{-j})/n= \| U_{-j}' U_{-j} \|_2/n= \| U_{-j} U_{-j}' \|_2/n$ by singular value definition. We provide a definition of $\tilde{U}_{-j}$, which is an  $n \times (p-1)$ matrix, $U_{-j}$ also based on Assumption 4,  with $\tilde{U}_{-jt}': 1 \times (p-1)$ $t$-th row of $\tilde{U}_{-j}$, and we stack these $n$ rows to get $\tilde{U}_{-j}$ and then form $U_{-j}$
	$ U_{-j}:= \tilde{U}_{-j} \Sigma_{U,-j}^{1/2}.$
	Then by Lemma 22 of \cite{bbsmw21} via Assumption 4,  and definition (18)
\begin{equation}
P [ \frac{1}{n} \| \tilde{U}_{-j} \Sigma_{U,-j} \tilde{U}_{-j}'\|_2 \le  C_0 \Delta_{-j} ] \ge 1 - \exp (-n).\label{pcl1-8}
\end{equation} 
Define
$ G_{1j}:= \{\frac{1}{n} \| \tilde{U}_{-j} \Sigma_{U,-j} \tilde{U}_{-j}'\|_2 \le  C_0 \Delta_{-j}\}.$
Also see that by (\ref{pcl1-7})
\begin{equation}
    G_{1j} \subseteq E_{3j},\label{pcl1-8a}
 \end{equation}
for all $j=1,\cdots,p$.
By (\ref{pcl1-8})(\ref{pcl1-8a})
\begin{equation}
P (E_{3j})= P ( \frac{1}{n} \sigma_{K+1}^2 (Y_{-j}) < C_0 \Delta_{-j}) \ge P (G_{1j})  \ge 1 - \exp (-n).\label{pcl1-9}
\end{equation}
So clearly 
$ P( E_{3j}^c) \le \exp(-n).$
To complete the proof of (\ref{pcl1-6})
\[P ( \cap_{j=1}^p  E_{3j}) 
=  1 - P ( (\cap_{j=1}^p  E_{3j})^c) = 1 - P ( \cup_{j=1}^p  E_{3j}^c) 
\ge 1 - \sum_{j=1}^p P (E_{3j}^c).
\]
Next, the union bound argument
\begin{equation}
P ( \cup_{j=1}^p  E_{3j}^c) \le  \sum_{j=1}^p P (E_{3j}^c) \le \exp (\log p -n) \le \exp[ (c'-1) n],\label{pcl1-10}
\end{equation}
given Assumption 3, $\log p \le c' n$ with $c'<1$ a positive constant upper bounded by one. So this clearly shows 
\begin{equation}
 P ( \cap_{j=1}^p  E_{3j}) \ge 1 -  \exp ((c'-1)n).\label{a42a}
 \end{equation}
Then by (17)(19)(\ref{a42a}), we have $P ( \max_{1 \le j \le p} \hat{s}_j < K+1) \ge 1 -\exp((c'-1)n)$  hence 
\begin{equation}
 P (  \max_{1 \le j \le p} \hat{s}_j \le K) \ge 1 - (\exp(c'-1)n),\label{cl1-s1} 
\end{equation}
 with $\hat{s}_j$ being a positive integer. By (\ref{pcl1-1}), $\hat{r}_j = k_j$, and since by $\hat{s}_j$ definition
in (17) we have $k_j = \hat{s}_j$, so 
\begin{equation}
    \hat{r}_j = \hat{s}_j.\label{cl1-s2}
\end{equation}
Then using (\ref{cl1-s1}) with (\ref{cl1-s2})
\begin{equation}
 P ( \max_{1 \le j \le p} \hat{r}_j \le K) 
    \ge 1 - \exp ((c'-1)n).\label{cl1-s3}
\end{equation}
Now we consider $\hat{\Psi}_j$ uniformly in $j$ (in probability).
The uniformity proof has two parts. First, define for each $j=1,\cdots,p$
\begin{equation}
    G_{2j}:= \{ \hat{\lambda}_{\hat{s}_j+1} < C_0 \Delta_{-j}\}.\label{p.0}
\end{equation}
Then first by $\hat{\Psi}_j$ definition in (\ref{pcl1-3}), and since $\hat{s}_j= \hat{k}_j$ by $\hat{s}_j$ definition with (\ref{p.0})
\begin{equation}
    P ( \cap_{j=1} \{ \hat{\Psi}_j < C_0 \Delta_{-j} \} ) = 
    P ( \cap_{j=1}^p \{ \hat{\lambda}_{k_j +1} < C_0 \Delta_{-j} \}) 
    =  P (\cap_{j=1}^p G_{2j}).\label{p.00}
\end{equation}
The first part of the uniformity proof is done. For the second part, from $\hat{s}_j$ definition in (17) we have 
\[ \hat{\lambda}_{\hat{s}_j+1} < C_0 \Delta_{-j}.\]
See that 
\[ P [ \cap_{j=1}^p G_{2j}] \ge 1 - \sum_{j=1}^p P  [G_{2j}^c] =1 \]

That implies
\begin{equation}
    P (\cap_{j=1}^p G_{2j} ) = 1.\label{p.3}
    \end{equation}
    
    So we proved uniformity across $j=1,\cdots, p$ for $\hat{\Psi}_j$ combining (\ref{p.00}) (\ref{p.3}). Now we consider $\hat{\eta}_j$ uniformly in $j$ (in probability).
\begin{equation}
    P (\cap_{j=1}^p \{ \hat{\eta}_j \ge C_0 \Delta_{-j}\})
    = P ( \cap_{j=1}^p \{ \hat{\lambda}_{\hat{s}_j} \ge C_0 \Delta_{-j}\}),\label{p.4}
\end{equation}
since $\hat{s}_j = k_j$ and using $\hat{\eta}_j$ definition in (\ref{pcl1-2}).
Note that since we proved $\max_{1 \le j \le p} \hat{s}_j \le K$ with probability approaching one, and by (17)(19), uniformly over $j$
$ \hat{\lambda}_{\hat{s}_j} \ge \hat{\lambda}_{K}.$
With (\ref{p.4})(19) and the above inequality
\begin{eqnarray}
    P (\cap_{j=1}^p \{ \hat{\eta}_j \ge C_0 \Delta_{-j}\})& \ge & 
    P (\cap_{j=1}^p \{ \hat{\lambda}_{K} \ge C_0 \Delta_{-j}\} 
= P (\cap_{j=1}^p \{ \hat{\lambda}_{K+1} < C_0 \Delta_{-j}\}) \nonumber \\
&\ge& 1- \exp ((c'-1)n),\label{cl1-s5}
\end{eqnarray}
and the last inequality is by (\ref{a42a})(\ref{defev1}), and $\hat{\lambda}_{K+1}:= \sigma_{K+1}^2 (Y_{-j})/n$. Then $C_0 \Delta_{-j}$ in (18) 
     \begin{equation}
        C_0  \Delta_{-j} = c_{\Delta} 
        [ \| \Sigma_{U_{-j}}\|_2+ tr (\Sigma_{U_{-j}})/n] 
         \ge c_{\Delta} \| \Sigma_{U_{-j}\|_2 } \ge \beta,\label{cl1-s6}
          \end{equation}
          since $c_{\Delta} >1$.
          Define 
          $ G_{3j}:=\{ \hat{\eta}_j \ge \beta\}.$
          Then by  (\ref{defev1}) and above event definition, with equality in (\ref{cl1-s5})
          $ \cap_{j=1}^p E_{3j} \subseteq \cap_{j=1}^p G_{3j},$
          for each $j=1,\cdots,p$. So combine with (\ref{cl1-s5})(\ref{cl1-s6})
          we have 
          \begin{equation}
P (\cap_{j=1}^p G_{3j}) =
P ( \cap_{j=1}^p \{ \hat{\eta}_j \ge \beta \} ) \ge P (\cap_{j=1}^p E_{3j}) 
\ge 1 - \exp ((c'-1)n).\label{cl1-s7}
          \end{equation}
          Furthermore, using (\ref{p.00})(\ref{cl1-s5}), 
          uniformly in $j=1,\cdots,p,$ with probability approaching one
          \begin{equation}
              \frac{\hat{\Psi}_j}{\hat{\eta}_j} \le 1.\label{ratio1}
          \end{equation}
 Our proofs here are uniform and detailed extensions of pointwise results of  proofs of Corollaries 5-6 in Sections B.2.1-B.2.2 of \cite{bbsmw21}. (ii). Note that asymptotic case is obtained from (i). {\bf  Q.E.D.}

	{\bf Proof of Corollary 2}. 
	(i). We consider first $p - 1 < n$ scenario (in which RRE effectively becomes OLS) and again proceed by simplifying three terms, $\bar{r}$, $\bar{\eta}$, $\bar{\Psi}$, that appear in the upper bound in Theorem 1.    
    In that respect $\hat{B}= I_{p-1}$ by definition of RRE.  Then clearly $\hat{r}_j= rank (Y_{-j} )= p-1$ since $p-1<n$ in this Corollary. Hence 
	\begin{equation}
	\bar{r}= \max_{1 \le j \le p} \hat{r}_j < p.\label{pcl2-1}
		\end{equation}
		Also since $I_{p-1} - P_{I_{p-1}} = 0$ we have 
		\begin{equation}
\bar{\Psi} = 0. \label{pcl2-2}
\end{equation}
This last equality shows that RRE has one less bias term in the upper bound compared to adaptive PCR estimator or compared to the general upper bound in Theorem 1. Use Corollary 11.6.7 in \cite{bern2018} (with $i=p-1, j=1$ there) or (11.6.18) of \cite{bern2018}
for the first inequality below, with $\Sigma_{-j,-j}$ as defined in   Appendix B,
\begin{eqnarray}
\hat{\eta}_j & = & \frac{1}{n} \sigma_{p-1}^2 (Y_{-j}) = \frac{1}{n} \sigma_{p-1}^2 (Y_{-j} \Sigma_{-j,-j}^{-1/2} \Sigma_{-j,-j}^{1/2}) \nonumber \\
& \ge  &\frac{1}{n} \sigma_{p-1}^2 (Y_{-j} \Sigma_{-j,-j}^{-1/2}) \sigma_{p-1}^2 (\Sigma_{-j,-j}^{1/2}) \nonumber \\
 & = & [ \frac{1}{n} \sigma_{p-1}^2 (Y_{-j} \Sigma_{-j,-j}^{-1/2})
] [ Eigmin (\Sigma_{-j, -j})] 
 \ge  c [ \frac{1}{n} \sigma_{p-1}^2 (Y_{-j} \Sigma_{-j,-j}^{-1/2})
],\label{pcl2-3}
\end{eqnarray}
where  the last equality is by the definition of singular value. 
To get the last inequality in (\ref{pcl2-3}),  clearly by $\Sigma_{-j,-j}, \Sigma_{U,-j}$ definitions, by using (5) and by Assumptions 2-3
$ \min_{1 \le j \le p} Eigmin (\Sigma_{-j,-j}) \ge \min_{1 \le j \le p} Eigmin (\Sigma_{U,-j}) \ge c.$
So we obtain 
\begin{equation}
    \min_{1 \le j \le p} \hat{\eta}_j \ge c \min_{1 \le j \le p} 
    [ \frac{1}{n}\sigma_{p-1}^2 (Y_{-j} \Sigma_{-j,-j}^{-1/2})].\label{a68a}
\end{equation}
To analyze (\ref{pcl2-3})(\ref{a68a}), define $D_j:=\{ \frac{1}{n} \sigma_{p-1}^2 (Y_{-j} \Sigma_{-j,-j}^{-1/2}) \ge C \}$. Define a positive constant $c^{''}>1$, 
\begin{eqnarray}
P (\min_{1 \le j \le p}\frac{1}{n}\sigma_{p-1}^2 (Y_{-j} \Sigma_{-j,-j}^{-1/2}) \ge C )
&= & P ( \cap_{j=1}^p D_j) 
 =  1 - P [ (\cap_{j=1}^p D_j)^c]  = 1 - P ( \cup_{j=1}^p D_j^c) \nonumber \\
& \ge & 1 - \sum_{j=1}^p P (D_j^c) 
 \ge  1 -  \exp (\log p - C_p \log n) 
 \nonumber \\
& \ge & 1 -  \exp (\frac{-C_p}{2} \log n)\nonumber \\
& = & 1 - \frac{1}{n^{C_p/2}},\label{cl2-s1}
\end{eqnarray}
where to get the second  inequality, with $Y_{j-1}: n \times (p-1)$, and $n>p-1$ here, we use    Theorem 5.39 of \cite{v2012} with Corollary 10 proof in \cite{bbsmw21} via Assumption 4, and applying a union bound (with $t_1 =\sqrt{\log n}$ in Theorem 5.39 of \cite{v2012}). To get the second inequality, we also   benefit from  Assumption 3(i)
$\log p \le (C_p/2) \log n  $ with $0< C_p<2$. For sufficiently large $n$, Assumption 3(i), 
$p \le n^{C_p/2}$
with $0< C_p < 2$ implies 
the condition $p \log n \le c_0 n $, with $c_0 \in (0,1)$ a positive constant as  in Corollary 10 of \cite{bbsmw21}.

Next, we use (\ref{pcl2-3})(\ref{a68a}) and (\ref{cl2-s1}) with $D_j$ definition
\begin{equation}
 P ( \min_{1 \le j \le p} \hat{\eta}_j \ge C) \ge 1 - \frac{1}{n^{C_p/2}}.\label{cl2-s2}
    \end{equation}
  Now substitute (\ref{pcl2-1})(\ref{pcl2-2})(\ref{cl2-s2}) in the upper bound in Theorem 1 to have the result. (ii). Asymptotics is derived by using (i).

{\bf Q.E.D}

{\bf Proof of Corollary 3}. 
(i). There are again three terms that we consider, but we have $p-1>n$ in this RRE case. So 
$
\bar{r}= \max_{1 \le j \le p} rank (Y_{-j}) =n.$
Then we have $\bar{\Psi}=0$ as in Corollary 2 due to $\hat{B}= I_{p-1}$. The last term is 
$\hat{\eta}_j = \frac{1}{n} \sigma_n^2 (Y_{-j})$, which is the $n$-th largest singular value of $Y_{-j}: n \times (p-1)$ matrix. %
Also we assume $ tr (\Sigma_{U,-j}) / \| \Sigma_{U,-j}\|_2 > C n$, and $\tilde{U}_{-j}$ has independent entries. 
Use Lemma B.2 
\begin{equation}
  P ( \hat{\eta}_j \ge tr (\Sigma_{U_{-j}}/n)) 
\ge 1 -  4  \exp(-C_p n).\label{a72a}
\end{equation}
Also  we have by definition $\delta_n:= \min_{1 \le j \le p} tr (\Sigma_{U,-j})/n$. Define 
$G_{4j}:= \{ \hat{\eta}_j \ge tr (\Sigma_{U,-j}/n)\}$, $  G_{5j}:= \{ \hat{\eta}_j \ge  \delta_n\}.$
Note that for each $j=1,\cdots,p$
\begin{equation}
    G_{4j} \subseteq G_{5j},
\mbox{hence} 
\quad 
    \cap_{j=1}^p G_{4j} \subseteq \cap_{j=1}^p G_{5j}.\label{pcl3-s1}
\end{equation}
Note that by $G_{5j}$ definition, 
\begin{equation}
    \cap_{j=1}^p G_{5j} = \{ \min_{1 \le j \le p} \hat{\eta}_j \ge  \delta_n\},\label{pcl3-s2}
\end{equation}
since the lower bounds are the same $ \delta_n$.
Next we have the following proof
\begin{eqnarray}
P (\min_{1 \le j \le p} \hat{\eta}_j \ge  \delta_n ) & = & P (\cap_{j=1}^p G_{5j}) 
 \ge  P ( \cap_{j=1}^p G_{4j} ) = 1 - P ( \cup_{j=1}^p \{ \hat{\eta}_j  < tr (\Sigma_{U,-j})/n\})\nonumber \\
& \ge & 1 - \sum_{j=1}^p P ( \{ \hat{\eta}_j  < tr (\Sigma_{U,-j})/n\})
\ge  1 - \exp (\log 4 p - C_p  n) \nonumber \\
& \ge & 1 - \exp (-C_p n/2),\label{pcl3-1}
\end{eqnarray}
where the first equality is by (\ref{pcl3-s2}), and the second  inequality is by (\ref{pcl3-s1}), and the third inequality is by (\ref{a72a}), and the last inequality is by Assumption $3^{*}$(i). The desired result follows by incorporating these three terms in the upper bound of Theorem 1. (ii). Asymptotics follows from (i).

{\bf Q.E.D.}

Before the following proof, we proceed with three inequalities that are used for rectangular matrices. Take a generic rectangular matrix $M$ and a vector $x$.

\begin{equation}
\| M x \|_2 \le \| M \|_2 \| x\|_2,\label{i1}
\end{equation}
 by (11.4.4) of \cite{bern2018}. Then by (11.3.15) of \cite{bern2018} where for two generic rectangular matrices $M,L$ 
 \begin{equation}
 \| M L \|_2 \le \| M\|_2 \| L\|_2,\label{i2}
 \end{equation}
We form the following matrix norm inequality for a generic rectangular matrix $M: m_1 \times m_2$
\begin{eqnarray}
\| M \|_2 \le \| M \|_F &= &[\sum_{i=1}^{m_1} \sum_{j=1}^{m_2} | M_{ij} |^2 ]^{1/2} \nonumber \\
& \le & \sqrt{m_1 m_2} \max_{1 \le i \le m_1} \max_{1 \le j \le m_2} | M_{ij} |= \sqrt{m_1 m_2} \| M \|_{\infty}.\label{61}
\end{eqnarray}
where $\|.\|_F $ is the Frobenius norm and we get the first inequality by (4.4) of \cite{v2019}.

{\bf Proof of Theorem 2}.

 See that 
\begin{equation}
y_{jt} = a_j' f_t  + u_{jt},\label{eq1}
\end{equation}
and 
\begin{equation}
Y_{-jt} = A_{-j}  f_t  + U_{-jt},\label{eq2}
\end{equation}

(i). Note that we need to estimate the $j$-th main diagonal term for the precision matrix of asset returns. In that sense the key element is the reciprocal of that term which is defined in (11), and the population version in (B.11) in Online Appendix. We use triangle inequality
to consider the estimation error by (11)(B.11)
\begin{equation}
| \tilde{\tau}_j^2 - \tau_j^2| \le | \frac{1}{n} \sum_{t=1}^n [y_{jt}^2 - E (y_{jt} )^2] | +
\left| [n^{-1} \sum_{t=1}^n y_{jt} Y_{-jt}'] \tilde{\alpha}_j - E (y_{jt} Y_{-jt} )'\alpha_{j}^*
\right|.\label{47}
\end{equation}
Next add and subtract to the second right side term $E [(y_{jt}  Y_{-jt})' ]\tilde{\alpha}_j$, by iid nature of data,
\begin{eqnarray}
| \tilde{\tau}_j^2 - \tau_j^2| &\le &| \frac{1}{n} \sum_{t=1}^n [y_{jt}^2 - E (y_{jt})^2] |  
+ \left| n^{-1} \sum_{t=1}^n [y_{jt} Y_{-jt}' - E (y_{jt}  Y_{-jt})' \tilde{\alpha}_j 
\right| \nonumber \\ &+& \left|E [(y_{jt}   Y_{-jt})'] (\tilde{\alpha}_j - \alpha_{j}^*) 
\right|.\label{48}
\end{eqnarray}
Take uniform bounds over $j=1,\cdots,p$ and use Cauchy-Schwartz inequality for the second and third right side terms on (\ref{48})
\begin{eqnarray}
\max_{1 \le j \le p} | \tilde{\tau}_j^2 - \tau_j^2| &\le &
\max_{1 \le j \le p} | \frac{1}{n} \sum_{t=1}^n [y_{jt}^2 - E (y_{jt} ) ^2] |  \nonumber \\
& + & \max_{1 \le j \le p} \| n^{-1} \sum_{t=1}^n [ y_{jt} Y_{-jt}' - E (y_{jt}  Y_{-jt})']\|_2  \max_{1 \le j \le p} \|\tilde{\alpha}_j\|_2 \nonumber \\
& + & \max_{1 \le j \le p} \| E [(y_{jt} Y_{-jt} )'] \|_2 
\max_{1 \le j \le p} \| \tilde{\alpha}_j - \alpha_{j}^*\|_2.\label{49}
\end{eqnarray}

Each right side term in (\ref{49}) will be analyzed as a step in the proof.

{\bf Step 1.} Start with the first right side term on (\ref{49}), and by (2)
\[ y_{jt}^2 = a_j' f_t f_t' a_j + 2 a_j' f_t u_{jt} + u_{jt}^2,\]
and by Assumption 1(iv), and using  (\ref{eq1})
\[ E (y_{jt} )^2 = a_j' \Sigma_f a_j + E	 (u_{jt}^2).\]
Use triangle inequality by the last two equalities, and noting $\Sigma_f:= E (f_t f_t )'$ due to zero-mean $f_t$
\begin{eqnarray}
\max_{1 \le j \le p} | \frac{1}{n} \sum_{t=1}^n [y_{jt}^2 - E (y_{jt} )^2]|
& \le & \max_{1 \le j \le p} | a_j' [\frac{1}{n} \sum_{t=1}^n (f_t f_t'-  \Sigma_f)]a_j | \nonumber \\
& +&  \max_{1 \le j \le p} | \frac{1}{n} \sum_{t=1}^n [u_{jt}^2 - E (u_{jt}^2)]| \nonumber \\
&+& \max_{1 \le j \le p} 2 | a_j' [n^{-1}\sum_{t=1}^n f_t u_{jt} ]|.\label{50}
\end{eqnarray}
Take the first right side term in (\ref{50}) above, and use  Cauchy-Schwarz inequality first and then (\ref{i1}) 
\begin{equation}
\max_{1 \le j \le p} | a_j' [\frac{1}{n} \sum_{t=1}^n (f_t f_t'-  \Sigma_f)]a_j | \le 
\left[ \max_{1 \le j \le p} \| a_j \|_2^2 \right] \| \frac{1}{n} \sum_{t=1}^n [f_t f_t' - \Sigma_f] \|_{2}.\label{51}
\end{equation}
We consider the second term on the right side of (\ref{51}), with $F: n \times K$ matrix, 
and $ \| \Sigma_f \|_2 \le C < \infty$ by Assumption 3(ii)
\begin{equation}
\| \frac{F'F}{n} - \Sigma_f \|_2 \le C [ \sqrt{\frac{K}{n}} + \sqrt{\frac{\log n}{n}}] \| \Sigma_f \|_2,\label{pla2-1}
\end{equation}
by (5.26) of \cite{v2012}  since rows of $F$ are independent subgaussian vectors, with probability at least $1-2/n^c$ via Assumption 1(iii), Assumption 4(ii), and note that $K < n$. By Assumption 4(iii)
\begin{equation}
\max_{1 \le j \le p} \| a_j \|_2^2 \le C K.\label{53} 
\end{equation}
Use (\ref{pla2-1})-(\ref{53}) in (\ref{51}) and Assumption 3 to have
\begin{equation}
\max_{1 \le j \le p} | a_j' (\frac{1}{n} \sum_{t=1}^n [f_t f_t' - \Sigma_f ] a_j | = O_p \left(\max(\frac{K^{3/2}}{n^{1/2}}, K \sqrt{\frac{\log n}{n}})\right).\label{54}
\end{equation}
Then in (\ref{50}) consider the second right side term
\begin{equation}
\max_{1 \le j \le p} | \frac{1}{n} \sum_{t=1}^n [u_{jt}^2 - E (u_{jt}^2)]|= O_p (\sqrt{\frac{\log p}{n}}).\label{55}
\end{equation}
We now show how we obtain (\ref{55}).
Note that by Corollary 2.8.3 of \cite{v2019}, for $t_1>0$, and the union bound 
\[ P \left[ \max_{1 \le j_1 \le p } \max_{1 \le j_2 \le p} | \frac{1}{n} \sum_{t=1}^n [u_{j_1,t} u_{j_2,t} - E (u_{j_1,t} u_{j_2,t})]|>t_1
\right]
\le p^2 exp \left(-c min(\frac{t_1^2}{C_{\psi}^2}, \frac{t_1}{C_{\psi}})n\right),\]
 where we use $u_{j_1,t}$ and $u_{j_2,t}$ are both subgaussian random variables, so $u_{j_1,t} u_{j_2,t}$ are subexponential random variable (independent across $t=1,\cdots,n$) by Lemma 2.7.7 of \cite{v2019} and exercise 2.7.10 of \cite{v2019} shows centered subexponential are also subexponential random variables, and  we have 
 \[ \max_{1 \le j_1 \le p} \max_{1 \le j_2 \le p}  \| u_{j_1,t} u_{j_2,t} \|_{\psi} \le C_{\psi} < \infty\]
  where 
$C_{\psi}$ are the upper bound constant  for the subexponential norm in Definition 2.7.5 of \cite{v2019}. Next
$
2 p^2 exp \left(-c min(\frac{t_1}{C_{\psi}^2}, \frac{t_1}{C_{\psi}}) n \right)  =  exp \left(\log (2 p^2) - c \frac{t_1^2 n}{C_{\psi}^2}\right)$,
with $t_1:= C \frac{\sqrt{\log p}}{\sqrt{n}}$ and with sufficiently large $n$ (i.e.$ \sqrt{n} \ge \frac{C}{C_{\psi}} (\log p)^{1/2}$). Set 
$ c_{\psi}:= c \frac{C^2}{C_{\psi}^2} - 2 > 0$
 to have 
$ exp [ \log 2 + 2 \log p - c \frac{C^2}{C_{\psi}^2} \log p ] = \frac{2}{p^{c_{\psi}}}.$
Then
\[ P \left(\max_{1 \le j_1 \le p } \max_{ 1 \le j_2 \le p} | \frac{1}{n} \sum_{t=1}^n u_{j_1,t} u_{j_2,t} - E u_{j_1,t} u_{j_2,t}| > C \frac{\sqrt{\log p}}{\sqrt{n}}
\right) \le \frac{2}{p^{c_{\psi}}}.\]
The asymptotics in (\ref{55}) are obtained after taking $n \to \infty, p \to \infty$. Consider the third right side term in (\ref{50}). Lemma 2.7.7  and Corollary 2.8.3 of \cite{v2019} together with Assumptions 1, 4 and   Holder's inequality give
\begin{eqnarray}
\max_{1 \le j \le p} | a_j' (\frac{1}{n}\sum_{t=1}^n f_t u_{jt} )| & \le & \max_{1 \le j \le p} \| a_j \|_1 \max_{1 \le j \le p} \| \frac{1}{n} \sum_{t=1}^n f_t u_{jt} \|_{\infty} \nonumber \\
& = & O_p ( K \sqrt{\frac{\log (Kp)}{n}}) = O_p ( \frac{K \sqrt{\log p}}{\sqrt{n}}),\label{55a} 
\end{eqnarray}
where the last equality uses $p-1>K$. By (\ref{54})-(\ref{55a}) we have the first right side term in (\ref{49})
\begin{equation}
\max_{1 \le j \le p} | \frac{1}{n} \sum_{t=1}^n [y_{jt}^2 - E (y_{jt} )^2] | = O_p \left( \max( \frac{K^{3/2}}{n^{1/2}}, K \sqrt{\frac{\max(\log p, \log n)}{n}}) \right).\label{55ab}
\end{equation}

{\bf Step 2.}  We consider the second right side term in (\ref{49}). By (2)(3) and Assumption 1(iv)
\[ y_{jt} Y_{-jt}' = a_j' f_t f_t' A_{-j}' + a_j' f_t U_{-jt}' + u_{jt} f_t' A_{-j}' + u_{jt} U_{-jt}',\]
and by (\ref{eq1}) (\ref{eq2})
\[  E (y_{jt} Y_{-jt} )'= a_j' \Sigma_f  A_{-j}'.\]
Use these last two equations  to see that the following term 

\[ \max_{1 \le j \le p} \| \frac{1}{n} \sum_{t=1}^n [y_{jt} Y_{-jt}' - E (y_{jt} Y_{-jt} )'] \|_2,\]
is upper bounded by 
 \begin{eqnarray} 
&&\max_{1 \le j \le p} \| a_j'\left( \frac{1}{n} \sum_{t=1}^n [ f_t f_t' - \Sigma_f ] \right)A_{-j}' \|_2 
 + \max_{1 \le j \le p} \|( \frac{1}{n} \sum_{t=1}^n u_{jt} f_t') A_{-j}' \|_2 \nonumber \\
&+& \max_{1 \le j \le p} \| a_j' ( \frac{1}{n} \sum_{t=1}^n f_t U_{-jt}' ) \|_2 
 +  \max_{ 1 \le j \le p} \| \frac{1}{n} \sum_{t=1}^n u_{jt} U_{-jt}' \|_2.\label{58} 
\end{eqnarray}
Consider the first term on the right side of (\ref{58}) by (\ref{i1})(\ref{i2})
 \begin{eqnarray}
 & & \max_{1 \le j \le p} \| a_j' ( \frac{1}{n} \sum_{t=1}^n [f_t f_t' - \Sigma_f ] A_{-j}' \|_2 
\le \max_{1 \le j \le p} \| a_j \|_2   
 \max_{1 \le j \le p} 
\left\| \left(\frac{1}{n} \sum_{t=1}^n [ f_t f_t' - \Sigma_f ]\right)
A_{-j}' \right\|_2 \nonumber \\
& \le &  ( \max_{1 \le j \le p} \| a_j \|_2 )\| \frac{1}{n} \sum_{t=1}^n [ f_t f_t' - \Sigma_f]\|_2 
 \times  (\max_{1 \le j \le p} \| A_{-j} ' \|_2) \nonumber \\
& = & O (K^{1/2}) O_p ( \max (\frac{K^{1/2}}{n^{1/2}}, \frac{\sqrt{\log n}}{n^{1/2}})) O (\sqrt{p}) 
 =  O_p ( max (\frac{K \sqrt{p}}{n^{1/2}}, \frac{ \sqrt{p} \sqrt{K \log n}}{n^{1/2}})),\label{58a}
\end{eqnarray}
and the rates are by (\ref{pla2-1})(\ref{53}) and Assumption 3. Then consider the second term on the right side of (\ref{58})

\begin{eqnarray}
\max_{1 \le j \le p} \|( \frac{1}{n} \sum_{t=1}^n u_{jt} f_t') A_{-j}' \|_2 & \le & \max_{1 \le j \le p} \| ( \frac{1}{n} \sum_{t=1}^n u_{jt} f_t')\|_2 \max_{1 \le j \le p} \| A_{-j}' \|_2 \nonumber \\
& \le & \sqrt{K} \max_{1 \le j \le p} \| ( \frac{1}{n} \sum_{t=1}^n u_{jt} f_t')\|_{\infty}  \max_{1 \le j \le p} \| A_{-j}' \|_2 \nonumber \\
& = & O_p (\sqrt{K} \frac{\sqrt{\log p}}{\sqrt{n}}) O (\sqrt{p}),\label{60} 
\end{eqnarray}
where we use (\ref{i1}) for the first inequality since $u_{jt} f_t': 1 \times K$ vector and $A_{-j}': K \times p-1$ matrix, and the second inequality is by simple vector norm inequality tying $l_2, l_{\infty}$ vector norms, and the rates are by Assumption 3(iii) $\max_{1 \le j \le p} \| A_{-j} ' \|_2 = O (\sqrt{p})$ and Lemma 2.7.7 and Corollary 2.8.3. of \cite{v2019} which provides $\| n^{-1} \sum_{t=1}^n u_{jt} f_t' \|_{\infty} = O_p (\sqrt{\log p/n})$.
 Now third term on the right side of (\ref{58})

\begin{eqnarray}
\max_{1 \le j \le p} \|a_j' ( \frac{1}{n} \sum_{t=1}^n f_t U_{-jt}') \|_2 & \le & \max_{1 \le j \le p} \| a_j \|_2 \max_{1 \le j \le p} \| \frac{1}{n} \sum_{t=1}^n f_{t} U_{-jt}' \|_2 \nonumber \\
& \le & \max_{1 \le j \le p} \| a_j \|_2 \sqrt{K p} \max_{1 \le j \le p} \| \frac{1}{n} \sum_{t=1}^n f_t U_{-jt}'\|_{\infty} \nonumber \\
& = & O (\sqrt{K})  O_p ( \frac{\sqrt{K p (\log p)}}{\sqrt{n}}) = O_p ( K \sqrt{p} \frac{\sqrt{\log p}}{\sqrt{n}}),\label{62}
\end{eqnarray}
where we use for the first inequality (\ref{i1})
and then for the second inequality we use (\ref{61}) and for the rates we use Assumption 5(iii), and Lemma 2.7.7 and Corollary 2.8.3 of \cite{v2019} with Assumption 4 since $f_t U_{-jt}'$ is $K \times p-1$ matrix.  Then for the fourth term on the right side of (\ref{58}) via Corollary 2.8.3 of \cite{v2019} and Assumption 4
and simple vector norm inequality

\begin{equation}
\max_{1 \le j \le p} \| \frac{1}{n} \sum_{t=1}^n u_{jt} U_{-jt}' \|_2 \le \sqrt{p} \max_{1 \le j \le p} \| \frac{1}{n} \sum_{t=1}^n u_{jt} U_{-jt}' \|_{\infty} = O_p (\sqrt{p} \frac{\sqrt{\log p}}{\sqrt{n}}).\label{63}
\end{equation}

Relying on  (\ref{58a})(\ref{60})(\ref{62})(\ref{63}), and keeping in mind that  (\ref{62}) has the largest rate  by Assumptions 3 and 5(i), and  we have the following for  the second right side term in (\ref{49})
\begin{equation}
\max_{1 \le j \le p} \|  \frac{1}{n} \sum_{t=1}^n [y_{jt} Y_{-jt}' - E (y_{jt}  Y_{-jt} )']\|_2 = O_p ( K \sqrt{p} \frac{\sqrt{\log p}}{\sqrt{n}}).\label{64}
\end{equation}

\noindent We find it useful to recall the rate definition
\begin{equation}
 r_{w1}:= \max \left( \sqrt{\frac{\log n + \bar{r}}{n \bar{\eta}}} , \sqrt{\frac{K }{\bar{\xi} }}, \sqrt{ \frac{K \bar{\Psi}}{\bar{\xi} \bar{\eta} }}, \sqrt{\frac{K}{\bar{\xi} \bar{\eta}}}\right),\label{64a}
 \end{equation}

\noindent  The proof of Theorem 1-Step 2d gives
\begin{equation}
\max_{1 \le j \le p} \| \tilde{\alpha}_j \|_2 = O_p (r_{w1}).\label{65}
\end{equation}
Combine (\ref{64})(\ref{65}) for the second right side term in (\ref{49})
\begin{eqnarray}
& & \max_{1 \le j \le p} \|  \frac{1}{n} \sum_{t=1}^n [y_{jt} Y_{-jt}' - E (y_{jt}  Y_{-jt} )']\|_2   \max_{1 \le j \le p} \| \tilde{\alpha}_j \|_2 \nonumber \\
&=&
 O_p ( (K \sqrt{p \log p/n} )r_{w1}) \label{66}
\end{eqnarray}

{\bf Step 3.} Now consider the third term on the right side of (\ref{49}). The key term is 
$E [(y_{jt} Y_{-jt} ) '] = a_j \Sigma_f A_{-j}',$
 where we use Assumption 1 and (2)(3). Since we have for rectangular matrices (\ref{i1})(\ref{i2})

\begin{equation}
\max_{1 \le j \le p} \| E (y_{jt}  Y_{-jt}) ' \|_2  =  \max_{1 \le j \le p} \| a_j' \Sigma_f A_{-j} ' \|_2 = d_{2n},
\label{67}
\end{equation}
 By (\ref{67}), the third term on the right side of (\ref{49}) is 
via Theorem 1
\begin{equation}
\max_{1 \le j \le p} \| E (y_{jt}  Y_{-jt} )' \|_2 \max_{1 \le j \le p} \| \hat{\alpha}_j - \alpha_{j}^*\|_2 
= O_p (  d_{2n} r_{w1}).\label{68}
\end{equation}

{\bf Step 4.} We combine (\ref{55ab})(\ref{66})(\ref{68}) with Assumption 5(i) on the right side of (\ref{49}) to have the desired result.

{\bf Q.E.D.}

(ii).
 By Assumption  5(ii)-(iii) we obtain the consistency. {\bf Q.E.D.}

(iii). By (B.11) in Appendix B, and matching (13) in the main text,
$ \tau_j^2:= E [(y_{jt} )^2] - E [(y_{jt}  Y_{-jt} )'] \alpha_{j}^*.$
Then by Cauchy-Schwarz inequality
\begin{eqnarray}
\max_{1 \le j \le p} E [(y_{jt}  Y_{-jt} )'] \alpha_{j}^* 
& \le & \left[ \max_{1 \le j \le p} \| E [(y_{jt}  Y_{-jt})']\|_2 
\right] \nonumber \\
& \times& \left[ \max_{1 \le j \le p} \| \alpha_{j}^* \|_2
\right] \nonumber \\
& \le&  \frac{C K^{1/2}  d_{2n}}{\sqrt{\bar{\xi}}} \to 0,\label{pla4iii-1}
\end{eqnarray}
where we use Step 3 of Theorem 2 together with Step 3 of Theorem 1 proof, and (\ref{67}), and Assumption 5(iii).
Next by using (2), Assumptions 1-3
\begin{eqnarray*}
\min_{1 \le j \le p} E (y_{jt} )^2 &\ge& min_{1 \le j \le p} a_j' \Sigma_f a_j + \min_{1 \le j \le p} E (u_{jt}^2) \\
& \ge & Eigmin (\Sigma_f ) \min_{1 \le j \le p} \| a_j \|_2^2 + \min_{1 \le j \le p} \sigma_{j}^2 \ge c >0.
\end{eqnarray*}
The last two results provide the desired result.
{\bf Q.E.D.}

	{\bf Proof of Theorem 3}. First of all 
	\[ \max_{1 \le j \le p} \| \tilde{\Theta}_j' - \Theta_{j}' \|_2 \le 
	\max_{1 \le j \le p} \left| \frac{1}{\tilde{\tau}_j^2} - \frac{1}{\tau_j^2} 
	\right| + \max_{1 \le j \le p} \| \frac{\tilde{\alpha}_j}{\tilde{\tau}_j^2} - \frac{\alpha_{j}^*}{\tau_j^2}\|_2.\]
	Then add and subtract $\alpha_{j}^*/\tilde{\tau}_j^2$ to the right hand side of the immediately above inequality, and via triangle inequality and arranging
	
	\begin{eqnarray}
\max_{1 \le j \le p} \| \tilde{\Theta}_j' - \Theta_{j}' \|_2 &\le&  \max_{1 \le j \le p} \left| \frac{1}{\tilde{\tau}_j^2} - \frac{1}{\tau_j^2} 
	\right| 
	+  \max_{1 \le j \le p} \| \tilde{\alpha}_j - \alpha_{j}^* \|_2  \max_{1 \le j \le p } [\frac{1}{\tilde{\tau}_j^2}] \nonumber \\
&	+ & \max_{1 \le j \le p} \| \alpha_{j}^* \|_2 \max_{1 \le j \le p} \left| \frac{1}{\tilde{\tau}_j^2} - \frac{1}{\tau_j^2} 
	\right|.\label{pthm1-1}
		\end{eqnarray}
		Note that  since $\min_{1 \le j \le p} \tilde{\tau}_j^2 \ge \min_{1 \le j \le p} \tau_j^2 - \max_{1 \le j \le p} | \tilde{\tau}_j^2 - \tau_j^2|$
		\[
		\max_{1 \le j \le p} \left(\frac{1}{\tilde{\tau}_j^2} \right) \le  \frac{1}{\min_{1 \le j \le p} \tau_j^2 - \max_{1 \le j \le p} | \tilde{\tau}_j^2 - \tau_j^2|} 
		 \le  \frac{1}{c - o_p (1)},
		\]
  by Theorem 2(ii)(iii) under Assumptions 1, 5. Then by (A.53)  and Assumption 5(iii) 
  \begin{equation}
   \max_{1 \le j \le p} \| \alpha_{j}^* \|_2 = O (\frac{\sqrt{K}}{\sqrt{\bar{\xi}}}) = o(1).\label{a87a}
   \end{equation}
   So the third term on right side of (\ref{pthm1-1}) converges to zero faster in probability  than the first right side term, and the second term converges to zero faster in probability than the first term, given Theorem 2. So the main rate in (\ref{pthm1-1}) is governed by the first (main-diagonal) term, consistent with Theorem 2(i) and with the discussion following Theorem 2 in the main text.
   {\bf Q.E.D}
	
	 We now provide three corollaries for specific subcases of our general estimator. Note that we provide the specific case of adaptive PCR estimator of $\tau_j^2$, which is defined as $\hat{\tau}_{j,  PCR}^2$. In that respect the rate $r_{w1}$ simplifies considerably. In adaptive PCR case these rates are with wpa1
	\begin{equation}
	 r_{w1,PCR}= \max \left( \sqrt{\frac{\log n +K }{n }}, \sqrt{\frac{K}{\bar{\xi} }}
	\right),\label{pcr-1}
	\end{equation}

	See that rate (\ref{pcr-1}) can be obtained by seeing $\bar{r} \le K, \bar{\Psi}/\bar{\eta} \le 1, \bar{\eta} \ge \beta$ wpa1 as shown in Corollary 1. Corollary \ref{coa1} below follows through replacing $r_{w1}$ in Theorem 2 with $r_{w1,PCR}$.
	
	\begin{corollary}\label{coa1}
	 (i).   Under Assumptions 1-5(i) and using $r_{w1,PCR}$ instead of $r_{w1}$,  
	
	 \begin{eqnarray*}
	 \max_{1 \le j \le p} | \hat{\tau}_{PCR,j}^2 - \tau_j^2| &=&
	   O_p \left( K (\sqrt{p \log p/n }) r_{w1,PCR}
	 \right)\\
&	 +& O_p \left(  d_{2n} r_{w1,PCR}
	 \right)+ O_p \left( max (\frac{K^{3/2}}{n^{1/2}}, K  \frac{\max(\sqrt{\log p}, \sqrt{\log n})}{n^{1/2}})
   \right).
	\end{eqnarray*}
	
	(ii). Under Assumptions 1-5,  
	 \[ \max_{1 \le j \le p} | \hat{\tau}_{PCR,j}^2 - \tau_j^2|	= o_p (1).\]

	(iii). Under Assumptions 1-5,   
		\[ \min_{1 \le j \le p} \tau_j^2 \ge c - o_p (1).\]
	\end{corollary}

	Here we consider the second subcase  $p-1<n$ scenario of  RRE estimator of $\tau_j^2$. We define it as   $\hat{\tau}_{RRE,j}^2$. 
		 We define the following rate which will be used instead of $r_{w1}$,  wpa1,  which can be deduced from Corollary 2 proof,
		 \begin{equation}
		  r_{RRE-l}:= \max( \sqrt{\frac{\log n + p}{n}
	}, \sqrt{\frac{K}{\bar{\xi}}}).\label{RREl}
	\end{equation}

	\begin{corollary}\label{coa2}
	 (i).Under Assumptions 1-2, and Assumption 3(i) with $0< C_p < 2$, Assumptions 
     4-5(i) and using $r_{RRE-l}$ instead of both $r_{w1}$, 
	$p/n = a_{2n} \to 0$,

	\begin{eqnarray*}
	 \max_{1 \le j \le p} | \hat{\tau}_{RRE,j}^2 - \tau_j^2| &=& 
	   O_p \left( K (\sqrt{ a_{2n} \log p }) r_{RRE-l}
	 \right)\\
&	 + &O_p \left( d_{2n} r_{RRE-l}
	 \right) + 
	  O_p \left( max (\frac{K^{3/2}}{n^{1/2}}, K  \frac{\max(\sqrt{\log p}, \sqrt{\log n})}{n^{1/2}})
   \right)
	\end{eqnarray*}
	
	(ii). Under Assumptions 1-5 with the conditions in (i)
	 \[ \max_{1 \le j \le p} | \hat{\tau}_{RRE,j}^2 - \tau_j^2|	= o_p (1).\]

	(iii). Under Assumptions 1-5 with the conditions in (i)
		\[ \min_{1 \le j \le p} \tau_j^2 \ge c - o_p (1).\]
	\end{corollary}

	We now consider the third subcase  of RRE estimator when $p-1>n$. 
	Note that this subcase will be different than the others, and we use an estimator of $\tau_j^2$ which is different due to interpolation property of RRE in this case. First let us show the interpolation and how it affects the estimator $\tilde{\tau}_j^2$ proposed, and the need to modify. In case of $p-1>n$, RRE estimator becomes the $l_2$ norm interpolator. 
	 Define the minimum norm interpolator as in (1) of \cite{bsmw22}, for each $j=1,\cdots, p$
	 \[ \hat{\alpha}_j:= argmin \{ \| \alpha_{j} \|_2: \| Y_{-j} \alpha_{j} - y_j \|_2= \min_{\gamma_{j}}  \| Y_{-j} \gamma_{j} - y_j \|_2.\},\]
	 Lemma 28 of \cite{bsmw22} shows 
	 $ \hat{\alpha}_j = Y_{-j}^+ y_j = \hat{\alpha}_{RRE,j}.$
	  Then when $p-1>n$,
	 \[Y_{-j} \hat{\alpha}_j - y_j = 
	 Y_{-j} Y_{-j}^+ y_j - y_j = y_j - y_j =
	 0,\]
	 with $Y_{-j} Y_{-j}^+= I_n $ by (136) of \cite{bsmw22}. This is interpolation, and we cannot use the formula for $\tilde{\tau}_j^2$ since  with interpolation it can be shown, $\tilde{\tau}_j^2=0$ in RRE in (11), with $p-1>n$ case. So we propose the following estimator, which can be considered a subcase of general $\tilde{\tau}_j^2$. Define
	 \begin{equation}
	  \hat{\tau}_{INT,j}^2:= \frac{y_j'y_j}{n}.\label{ca3-1}
	\end{equation}

	\begin{corollary}\label{coa3}	
	 Under Assumptions 1-2, Assumptions $3^{*}(i)$, 3(ii), 4 
	
	(i). \[
	 \max_{1 \le j \le p} | \hat{\tau}_{INT,j}^2 - \tau_j^2| = O_p \left( \frac{\sqrt{K} d_{2n}}{\sqrt{\bar{\xi}}}
	 \right) +  O_p \left( max (\frac{K^{3/2}}{n^{1/2}}, K  \frac{\sqrt{\log p}}{n^{1/2}})
   \right).
	\]
	
	(ii). With added to (i),  Assumptions  5(ii) with $\sqrt{K} d_{2n}/\sqrt{\bar{\xi}} \to 0$
	 \[ \max_{1 \le j \le p} | \hat{\tau}_{INT,j}^2 - \tau_j^2|	= o_p (1).\]

	(iii). Using Assumptions in (ii)
    \[ \min_{1 \le j \le p} \tau_j^2 \ge c - o_p (1).\]
	\end{corollary}

	{\bf Proof of Corollary \ref{coa3}}. (i). By (\ref{13})(\ref{ca3-1}) and triangle inequality
	\begin{eqnarray*}
	 \max_{1 \le j \le p} | \hat{\tau}_{INT,j}^2 - \tau_j^2| &\le& \max_{1 \le j \le p} | \frac{1}{n} \sum_{t=1}^n  [y_{jt}^2 - E (y_{jt} )^2 ]| \\
&	+& \max_{1 \le j \le p} \| E (y_{jt} Y_{-jt} )' \|_2 \max_{1 \le j \le p} \| \alpha_{j}^*\|_2.
	\end{eqnarray*}
	Next, Step 1 in the proof of Theorem 2 with (\ref{55ab}), and (\ref{pla4iii-1}) provides the result. (ii). Given Assumptions in part (i)  with $\sqrt{K} d_{2n}/\sqrt{\bar{\xi}} \to 0$ provides the consistency. (iii). This is the same as in the proof of Theorem 2 (iii).
	{\bf Q.E.D.}

	\setcounter{equation}{0}
	\setcounter{lemma}{0}
	\setcounter{table}{0}
	\renewcommand{\theequation}{B.\arabic{equation}}
	\renewcommand{\thelemma}{B.\arabic{lemma}}
	\renewcommand{\thecorollary}{B.\arabic{corollary}}
	\renewcommand{\thetable}{\thesection.\arabic{table}}
	\renewcommand{\thethm}{B.\arabic{thm}}
	\begin{center}
		\uppercase{{\bf Appendix B}}
	\end{center}

	In this part, we provide a precision matrix formula and then provide two lemmata on tail inequalities.  %

	{\bf Precision Matrix Formula Derivation}
	
	Precision matrix  relation to a hidden factor model context is established here for the first time in the literature and in detail. %
	Define all assets in the portfolio  as  $y_t=(y_{jt}, Y_{-jt}')': p \times 1$. In case of $j=3$, this represents $y_{3t}$ in the third cell in the vector $y_t$ and $Y_{-3t}$ represents elements $1,2, 4,\cdots,p$ in $y_t$.   
	
	{\bf Proof of Lemma 1}.
	 We will provide formula for $j$ th row and then stack all rows to get an expression, since $E y_{jt}=0, E Y_{-jt}=0$. Before that define $\Sigma_{j,j}:= E [(y_{jt})^2]$ and $\Sigma_{j,-j}:=E [(y_{jt} Y_{-jt})']: 1 \times p-1$ and represents $j$ th row of $\Sigma$ with $j$ th element missing, and 
	by symmetricity of $\Sigma$, $\Sigma_{j,-j}=(\Sigma_{-j,j})'$. Specifically,  define $\Sigma_{-j,j}:=E [ (Y_{-jt}   y_{jt}) ]: p-1 \times 1$, as the $j$ th column of $\Sigma$ where $j$ th element is missing. 
	In the same way we define $\Sigma_{-j,-j}:= E [ (Y_{-jt} Y_{-jt})']: p-1 \times p-1$ where this matrix is $j$ the row and column deleted from $\Sigma$. By Section 2.1 of \cite{yuan2010},  the main diagonal elements are given by  the scalar
	\begin{equation}
	\Theta_{j,j}:= ( \Sigma_{j,j}- \Sigma_{j,-j} \Sigma_{-j, -j}^{-1} \Sigma_{-j,j})^{-1}.\label{5}
	\end{equation}
	and the $j$ th row without the $j$ th element is:
	\begin{equation}
	\Theta_{j,-j}:=- \Theta_{j,j} \Sigma_{j,-j} \Sigma_{-j,-j}^{-1}.\label{6}
	\end{equation}
		Now we relate these quantities to a linear regression context. Define the following  minimizer
	$ \alpha_{j}^*:= argmin_{\alpha_{j} \in R^{p-1}} E [ (y_{jt}   - Y_{-jt} ' \alpha_{j}]^2,$
	with given zero mean $(y_{jt}), (Y_{-jt})$ and finite second moments for $y_{jt}, Y_{-jt}$, Lemma 27 in \cite{bsmw22} shows that 
	\begin{equation}
	\alpha_{j}^* = \Sigma_{-j,-j}^+ \Sigma_{-j,j},\label{7}
	\end{equation}
	where $\Sigma_{-j,-j}^+$ is the Moore-Penrose inverse of $\Sigma_{-j,-j}$. 
	First, 
	\begin{equation}
     \Sigma_{-j,-j}= E [ (Y_{-jt}  Y_{-jt})']= A_{-j} \Sigma_f A_{-j}' + \Sigma_{U,-j},\label{b3a}
     \end{equation}
	by (3) and Assumption 1(iv). Then 
	\begin{eqnarray*}
	\min_{ 1 \le j \le p} Eigmin( \Sigma_{-j,-j})& =& \min_{1 \le j \le p} Eigmin (A_{-j} \Sigma_f A_{-j}' + \Sigma_{U,-j}) \\
	&\ge& min_{1 \le j \le p} Eigmin (\Sigma_{U,-j}) \ge c >0,
	\end{eqnarray*}
	by Assumption 2-3. 
	 (\ref{7}) simplifies to
	\begin{equation}
	\alpha_{j}^*= \Sigma_{-j,-j}^{-1} \Sigma_{-j,j}.\label{8}
	\end{equation}
	Define $\eta_{jt}:= y_{jt}  - Y_{-jt} ' \alpha_{j}^*$ for $j=1,\cdots, p$. Now consider
	\begin{eqnarray}
	E [ (Y_{-jt}   \eta_{jt}] & = & E [ (Y_{-jt}  y_{jt} ) ] - E [ (Y_{-jt}  Y_{-jt} )'] \alpha_{j}^* \nonumber \\
	& = & \Sigma_{-j,j} - \Sigma_{-j,-j} \alpha_{j}^* 
	=  \Sigma_{-j,j} - \Sigma_{-j,-j} \Sigma_{-j,-j}^{-1} \Sigma_{-j,j} =0,\label{9}
		\end{eqnarray}
		where the first equality uses $\eta_{jt}$ definition, and the second equality uses definitions of moment matrices, and the third one uses (\ref{8}). Given (\ref{9}) we can write the following infeasible linear regression model,
		\begin{equation}
		y_{jt}= Y_{-jt}  ' \alpha_{j}^* + \eta_{jt}.\label{10}
		\end{equation}
		Now we relate (\ref{6}) using (\ref{8}). Then 
		\begin{equation}
		\Theta_{j,-j}= - \Theta_{j,j} (\alpha_{j}^*)',\label{11}
		\end{equation}
	and then define $\tau_j^2:= E \eta_{jt}^2$. See that by (\ref{9})(\ref{10})
	\[ E [y_{jt}] ^2 = (\alpha_{j}^*)'  E [(Y_{-jt} )  (Y_{-jt} )'] \alpha_{j}^* + E [\eta_{jt}^2],\]
	which implies by definitions $\Sigma_{j,j}:= E [ y_{jt}]^2, \Sigma_{-j,-j}:= E [ (Y_{-jt} Y_{-jt} )']$ 
	\begin{eqnarray}
	\tau_j^2 & = & \Sigma_{j,j} - (\alpha_{j}^*)' \Sigma_{-j,-j} \alpha_{j}^* 
	=  \Sigma_{j,j}- \Sigma_{j,-j} \Sigma_{-j,-j}^{-1} \Sigma_{-j,j} \label{12a} \\
	& = & 1/\Theta_{j,j},\label{12}
	\end{eqnarray}
	where we use (\ref{8}) for the second equality and (\ref{5}) for the third one. Clearly, also by (\ref{8})(\ref{12a})
	\begin{equation}
	\tau_j^2 = \Sigma_{j,j} - \Sigma_{j,-j} \alpha_{j}^*.\label{13}
	\end{equation}
	 We can form the $j$ th row of precision matrix by using $j$ the main diagonal term, via (\ref{12})(\ref{13})
	\begin{equation}
	\Theta_{j,j} = \frac{1}{\Sigma_{j,j} - \Sigma_{j,-j} \alpha_{j}^*},\label{14}
	\end{equation}
	and the rest of that row, via (\ref{11})(\ref{12})
	\begin{equation}
	\Theta_{j,-j} = \frac{-(\alpha_{j}^*)'}{\tau_j^2}.\label{15}
	\end{equation}
	 We start simplifying $\alpha_{j}^*$ formula in case of hidden factors as in (2)(3). The proof for (i) here  is also a detailed version of the proof of Lemma 7 and (57) of \cite{bsmw22}. This is repeated to show that our scenario fits into their framework. Note that $\Sigma_f:= E (f_t f_t )'$.	 In that respect in (\ref{8}), by (2)-(3)
	 \begin{eqnarray}
	 \Sigma_{-j,j} = E [ (Y_{-jt} y_{jt} ) ] 
	 &=& E \{[ A_{-j} f_t  + U_{-jt}][(f_t ' a_j + u_{jt}]\} \nonumber \\
	 & = & A_{-j} \Sigma_f  a_j + A_{-j} E[ f_t   u_{jt}] \nonumber \\
	 &+& E [U_{-jt} f_t'] a_j + E [U_{-jt} u_{jt}] \nonumber \\
	 & =&  A_{-j} \Sigma_f a_j = \bar{A}_{-j} \bar{a}_j,\label{20}
	 \end{eqnarray}
	 by Assumption 1(iv). Then the same analysis in (\ref{20}) provides with $\bar{A}_{-j}:= A_{-j} \Sigma_f^{1/2}$,
	 	 \begin{equation}
	 \Sigma_{-j,-j}= E [ (Y_{-jt} Y_{-jt} )'] = A_{-j} \Sigma_f A_{-j}' + \Sigma_{U,-j} = \bar{A}_{-j} \bar{A}_{-j}^{'} + \Sigma_{U,-j}.\label{app-21}
	 \end{equation}
	 Using Sherman-Morrison-Woodbury formula in p.19 of \cite{hj2013}
	 \begin{eqnarray}
	 \Sigma_{-j,-j}^{-1} & = & (A_{-j} \Sigma_f  A_{-j}' + \Sigma_{U,-j})^{-1} \nonumber \\
	 & = &  \Sigma_{U,-j}^{-1} - \Sigma_{U,-j}^{-1} A_{-j} (\Sigma_f^{-1} + A_{-j}' \Sigma_{U,-j}^{-1} A_{-j})^{-1} A_{-j}' \Sigma_{U,-j}^{-1} \nonumber \\
	 & = &
	  \Sigma_{U,-j}^{-1} - \Sigma_{U,-j}^{-1} \bar{A}_{-j} ( I_K + \bar{A}_{-j}' \Sigma_{U,-j}^{-1} \bar{A}_{-j})^{-1} \bar{A}_{-j}' \Sigma_{U,-j}^{-1},\label{22}
	 \end{eqnarray}
	  where we use Assumption 3, $\Sigma_f$ is positive definite. Note also that 
	 \begin{equation}
	 \bar{A}_{-j}^+ \bar{A}_{-j} = \bar{A}_{-j}' (\bar{A}_{-j}^+)' = I_K,\label{23}
	 \end{equation}
	 since $A_{-j}, \Sigma_f$ has full rank K by Assumption 3,  so (135) of \cite{bsmw22} provides the last result (\ref{23}). Define $\bar{a}_j:= \Sigma_f^{1/2} a_j$ and (\ref{8})(\ref{20})
	 \begin{eqnarray}
	 \alpha_{j}^* &=& \Sigma_{-j,-j}^{-1} \bar{A}_{-j} \bar{a}_j 
	  =  \Sigma_{-j,-j}^{-1} [ \bar{A}_{-j} \{ \bar{A}_{-j}' (\bar{A}_{-j}^+)'\} \bar{a}_j] \nonumber \\
	 & = & \Sigma_{-j,-j}^{-1} [ \Sigma_{-j,-j} - \Sigma_{U,-j}] (\bar{A}_{-j}^+)' \bar{a}_j 
	  =  [ I_{p-1} - \Sigma_{-j,-j}^{-1} \Sigma_{U,-j}] (\bar{A}_{-j}^+)' \bar{a}_j,\label{24}	
	 \end{eqnarray}
	 where we use (\ref{23}) for the second equality and for the third equality (\ref{app-21}) and $\bar{A}_{-j}$ definition. By using $\bar{G}_j$ definition before Lemma 1
	 	 and using (\ref{22}) on right side of  (\ref{24}) for $\Sigma_{-j,-j}^{-1}$
	 \begin{eqnarray}
	 I_{p-1} - \Sigma_{-j,-j}^{-1} \Sigma_{U,-j} & = & I_{p-1} - [ \Sigma_{U,-j}^{-1} - \Sigma_{U,-j}^{-1} \bar{A}_{-j} \bar{G}_j^{-1} \bar{A}_{-j}' \Sigma_{U,-j}^{-1}] \Sigma_{U,-j} \nonumber \\
	 & = & \Sigma_{U,-j}^{-1} \bar{A}_{-j} \bar{G}_j^{-1} \bar{A}_{-j}'\label{25}
	 \end{eqnarray}
	 Use (\ref{25}) in (\ref{24}) 
	 \begin{equation}
	 \alpha_{j}^* = \Sigma_{U,-j}^{-1} \bar{A}_{-j} \bar{G}_j^{-1} \bar{A}_{-j}' (\bar{A}_{-j}^+)' \bar{a}_j 
	  =  \Sigma_{U,-j}^{-1} \bar{A}_{-j} \bar{G}_j^{-1} \bar{a}_j,\label{26}
	 \end{equation}
	 by (\ref{23}) for the second equality. One of the other elements in the precision matrix is the reciprocal of the main diagonal term, $\tau_j^2$ in (\ref{13}). We provide a formula for that in terms of hidden factor regression model, which will be new. In that respect, by (2)(3), Assumption 1
	\begin{equation}
	\Sigma_{j,j}:=E (y_{jt} )^2= a_j' \Sigma_f a_j + \sigma_j^2= \bar{a}_j' \bar{a}_j + \sigma_j^2.\label{27}
		\end{equation}
		Similarly, using (2)(3), Assumption 1
		\begin{equation}
		\Sigma_{j,-j}:= E (y_{jt})(Y_{-jt} )' = \bar{a}_j' \bar{A}_{-j}'.\label{28}
		\end{equation}
	Then combine (\ref{26})(\ref{27})(\ref{28}) in (\ref{13})
    \begin{equation}
        \tau_j^2 = (\bar{a}_j' \bar{a}_j + \sigma_j^2) - \bar{a}_j' \bar{A}_{-j}' \Sigma_{U,-j}^{-1} \bar{A}_{-j}
        \bar{G}_{j}^{-1} \bar{a}_j.\label{28a}
    \end{equation}
    (i)-(ii). By (\ref{26}) and (\ref{28a}), and using them in (\ref{12})(\ref{15})  we have the desired result.
	{\bf Q.E.D.}

	{\bf Tail Inequalities}
	
	 Here we first provide a tail inequality for quadratic form of a mean-zero subgaussian random vector. This is Lemma 21 in \cite{bbsmw21}. 
\begin{lemma}\label{lb1}
    Let $\zeta$ be a zero mean  $d$ dimensional $\gamma_e$ sub-Gaussian random vector. For all symmetric positive-semi definite matrices $H$, and all $t \ge 0$,
    \[ P \left( \zeta' H \zeta >  \gamma_e^2 (\sqrt{tr (H)} + \sqrt{2 \| H \|_2 t }
)^2    \right) \le exp(-t),\]
 \end{lemma}
where $tr(H)$ represents trace of $H$, and $\gamma_e$ subgaussian random vector is defined in Assumption 4. 

Now, we provide a lemma for a lower bound inequality for singular value of a specific random matrix for RRE when $p-1>n$. The following is  Proposition 14 of \cite{bsmw22}.
\begin{lemma}\label{lb2}
 Under Assumptions $3^{*}(i)$, Assumption 3(ii), Assumption 4, with some $C>0$ large enough, 
 $re (\Sigma_{U,-j}) > C n $, for each $j=1,\cdots,p$, and $\tilde{U}_{-j}$ has independent entries, in the case of $p-1>n$
 \[ P \left( \sigma_n^2 (Y_{-j}) \ge C tr (\Sigma_{U,-j})\right) \ge 1 - 4 exp (-C_p n) .\]
 \end{lemma}

{\bf Discussion of a Remark in Theorem 3}

In this part, we show a lower bound for the rate that is used in Theorem 3, $d_{2n}/\sqrt{{\bar{\xi}}}$ which will simplify.  First $d_{2n}:= \max_{1 \le j \le p} \| a_j' \Sigma_f A_{-j}' \|_2$, and
\[ {\bar{\xi}}:= \min_{1 \le j \le p} \left[ \frac{\lambda_K (A_{-j} \Sigma_f A_{-j}')}{\| \Sigma_{U,-j} \|_2} \right],\]
 which are defined just before Assumption 5, and in (7) in main text. Start with 
 \[ \| a_j' \Sigma_f A_{-j}' \|_2^2 = \| \bar{a}_j' \bar{A}_{-j}' \|_2^2, \]
 with $\bar{a}_j':= a_j' \Sigma_f^{1/2} , \bar{A}_{-j}':= \Sigma_f^{1/2} A_{-j}'$. For each $j=1, \cdots, p$
 \[  \| \bar{a}_j' \bar{A}_{-j}' \|_2^2 \ge Eigmin (\bar{A}_{-j}' \bar{A}_{-j}) \| a_j \|_2^2 \ge Eigmin (A_{-j}' A_{-j}) Eigmin (\Sigma_f) \| a_j \|_2^2.
 \]
 Now, since $A_{-j}$ is full rank, we have $Eigmin ( A_{-j}' A_{-j}) \ge c > 0$, with $A_{-j}: p-1 \times K$ matrix, and $K < p-1$. Since $Eigmin (A_{-j}' A_{-j}) \ge c > 0$,  and $Eigmin (\Sigma_f) \ge c > 0$ as well,
 \[ d_{2n} \ge \sqrt{\min_{1 \le j \le p}  Eigmin (A_{-j}' A_{-j}) } Eigmin (\Sigma_f) \max_{1 \le j \le p} \| a_j \|_2.\]
 
 We obtain the following upper bound for ${\bar{\xi}}$
\[ \sqrt{{\bar{\xi}} }\le [Eigmax (\Sigma_f)]^{1/2}[ \max_{1 \le j \le p } Eigmin (A_{-j}' A_{-j})]^{1/2}/\beta^{1/2}.\]

Then  since the minimum and maximum eigenvalues of $\Sigma_f$ is positive constant
\[ \frac{d_{2n}}{\sqrt{{\bar{\xi}}}} \ge  c_g \frac{\sqrt{\min_{1 \le j \le p}  Eigmin (A_{-j}' A_{-j}) }}{\sqrt{\max_{1 \le j \le p}  Eigmin (A_{-j}' A_{-j}) }} \max_{1 \le j \le p} \|a_j \|_2^2,\] with $c_g >0$ is a generic positive constant
and if the minimum eigenvalues of the same order, this rate assumption in Theorem 3, $d_{2n}/\sqrt{\bar{\xi}} \to 0$  implies factor loadings converge to zero. So there is a  possibility of weak factor loading.

	\setcounter{equation}{0}
	\setcounter{lemma}{0}
	\setcounter{table}{0}
	\renewcommand{\theequation}{C.\arabic{equation}}
	\renewcommand{\thelemma}{C.\arabic{lemma}}
	\renewcommand{\thecorollary}{C.\arabic{corollary}}
	\renewcommand{\thetable}{C.\arabic{table}}
	\renewcommand{\thethm}{C.\arabic{thm}}
	\begin{center}
		\uppercase{ {\bf Appendix C}}
	\end{center}

\begin{table}[!t]
	\centering
	\caption{Simulation Results for the Precision Matrix Estimation with $t$-distributed errors}
    \begin{adjustbox}{max width=\textwidth}	
	\begin{threeparttable}	
\begin{tabular}{c|cccccccc}
			\hline
			\hline
			\multicolumn{1}{c}{Method} & \multicolumn{2}{c}{$n = 100$} &       & \multicolumn{2}{c}{$n = 200$} &       & \multicolumn{2}{c}{$n = 400$} \bigstrut\\
			\cline{2-3}\cline{5-6}\cline{8-9}          
			\multicolumn{1}{c}{} & $p = n/2$ & $p = 3n/2$ &       & $p = n/2$ & $p = 3n/2$ &       & $p = n/2$ & $p = 3n/2$ \bigstrut\\
			\hline
			\multicolumn{9}{c}{$K = 3$} \bigstrut\\
			\hline
			PCR-3F & 0.1203 & 0.1051 &       & 0.1035 & 0.0896 &       & 0.0897 & 0.0784 \bigstrut[t]\\
    		PCR-5F & 0.1203 & 0.1030 &       & 0.1050 & 0.0911 &       & 0.0939 & 0.0825 \\
    		PCR-20F & 0.1363 & 0.1194 &       & 0.1085 & 0.0921 &       & 0.0956 & 0.0829 \\
    		PCR-adaptive & 0.1199 & 0.1033 &       & 0.1058 & 0.0917 &       & 0.0950 & 0.0831 \\
    		RRE   & 1.0728 & 0.1819 &       & 1.0095 & 0.1610 &       & 0.9186 & 0.1447 \bigstrut[b]\\ \hline
    		POET  & 0.1451 & 0.1214 &       & 0.1186 & 0.0984 &       & 0.1053 & 0.0878 \bigstrut[t]\\
            NL-LW & 0.1498 & 0.1250 &       & 0.1304 & 0.1099 &       & 0.1153 & 0.0989 \\
    		NDW   & 0.1417 & 0.1300 &       & 0.1232 & 0.1133 &       & 0.1081 & 0.0992
            \bigstrut[b]\\
			\hline
			\multicolumn{9}{c}{$K = 5$} \bigstrut\\
			\hline
			PCR-3F & 0.1381 & 0.1198 &       & 0.1257 & 0.1054 &       & 0.1120 & 0.0964 \bigstrut[t]\\
    		PCR-5F & 0.1203 & 0.1033 &       & 0.1078 & 0.0897 &       & 0.0909 & 0.0782 \\
    		PCR-20F & 0.1329 & 0.1162 &       & 0.1089 & 0.0920 &       & 0.0949 & 0.0826 \\
    		PCR-adaptive & 0.1192 & 0.1027 &       & 0.1071 & 0.0917 &       & 0.0945 & 0.0828 \\
    		RRE   & 1.0543 & 0.1800 &       & 0.9760 & 0.1601 &       & 0.9094 & 0.1443 \bigstrut[b]\\ \hline
    		POET  & 0.1372 & 0.1124 &       & 0.1181 & 0.0954 &       & 0.0996 & 0.0841 \bigstrut[t]\\
    		NL-LW & 0.1472 & 0.1249 &       & 0.1314 & 0.1102 &       & 0.1151 & 0.0987 \\
            NDW   & 0.1465 & 0.1354 &       & 0.1277 & 0.1179 &       & 0.1107 & 0.1020
            \bigstrut[b]\\
			\hline
			\hline
		\end{tabular}
		\vspace*{-0.3cm}
		\begin{tablenotes}
			\singlespacing
			\item \leavevmode\kern-\scriptspace\kern-\labelsep 
			Note: The quantities in the table correspond to estimation error for the precision matrix in terms of the spectral norm.
		\end{tablenotes}
	\end{threeparttable}
    \end{adjustbox}%
	\label{tab:sim_results_t_dist}%
\end{table}%

In this part C, we provide additional result tables for the simulation and empirical application and a discussion of literature and hidden factors. 

Specifically, in Table~\ref{tab:sim_results_t_dist}, we provide the Monte Carlo results with $t$-distributed errors, which are qualitatively very similar to the ones obtained with Gaussian errors as reported in Table 2 in the main text.

Moreover, the out-of-sample portfolio results of the empirical application without transaction costs are reported in Table~\ref{tab_appl_ntc_p2}. 
The results here are similar to the ones with transaction costs, but as expected Sharpe Ratios are higher without transaction costs.
In addition, Table~\ref{tab_appl_tc_p1} presents the results including transaction costs for the shorter out-of-sample period spanning January 2005 to December 2019. Compared to a full span of time, Table 3 in main text, including 2025 there, this sub-span provides mixed results. With small-medium size portfolios, top Sharpe Ratios are lower in this period between January   2005-December 2019.

\begin{table}[!t]
	\centering
	\caption{Out-of-sample forecasting results without transaction costs: January 2005--December 2025}
    \begin{adjustbox}{max width=\textwidth}
	\begin{threeparttable}
			\begin{tabular}{c|cccccccc}
				\hline
				\hline
				Model & EW    & PCR-3 & PCR-5 & PCR-adaptive & RRE   & POET  & NL-LW & RN-FM \bigstrut\\
				\hline
				\multicolumn{9}{c}{$p = 10$} \bigstrut\\
				\hline
				AV    & 0.089 & 0.104 & 0.111 & 0.097 & 0.103 & 0.089 & 0.045 & 0.097 \bigstrut[t]\\
				SD    & 0.166 & 0.183 & 0.183 & 0.176 & 0.182 & 0.166 & 0.140 & 0.176 \\
				SR    & 0.533 & 0.567 & \textbf{0.605} & 0.551 & 0.568 & 0.533 & 0.321 & 0.552 \\
				p-value SR & 0.299 & 0.334 & -     & 0.341 & 0.081 & 0.298 & 0.053 & 0.218 \\
				PT    & 0.096 & 0.271 & 0.269 & 0.257 & 0.271 & 0.096 & 0.253 & 0.261 \bigstrut[b]\\
				\hline
				\multicolumn{9}{c}{$p = 50$} \bigstrut\\
				\hline
				AV    & 0.078 & 0.238 & 0.221 & 0.225 & 0.165 & 0.079 & 0.089 & 0.093 \bigstrut[t]\\
				SD    & 0.142 & 0.202 & 0.201 & 0.198 & 0.175 & 0.141 & 0.122 & 0.139 \\
				SR    & 0.553 & \textbf{1.182} & 1.101 & 1.134 & 0.944 & 0.559 & 0.731 & 0.672 \\
				p-value SR & 0.009 & -     & 0.258 & 0.478 & 0.142 & 0.012 & 0.046 & 0.028 \\
				PT    & 0.069 & 0.644 & 0.700 & 0.647 & 1.147 & 0.079 & 0.395 & 0.244 \bigstrut[b]\\
				\hline
				\multicolumn{9}{c}{$p = 100$} \bigstrut\\
				\hline
				AV    & 0.080 & 0.222 & 0.184 & 0.166 & 0.047 & 0.081 & 0.056 & 0.089 \bigstrut[t]\\
				SD    & 0.146 & 0.313 & 0.295 & 0.269 & 0.272 & 0.143 & 0.129 & 0.142 \\
				SR    & 0.546 & \textbf{0.709} & 0.622 & 0.617 & 0.174 & 0.569 & 0.434 & 0.627 \\
				p-value SR & 0.579 & -     & 0.330 & 0.345 & 0.022 & 0.641 & 0.238 & 0.787 \\
				PT    & 0.068 & 1.535 & 1.460 & 1.385 & 6.215 & 0.149 & 1.803 & 0.222 \bigstrut[b]\\
				\hline
				\multicolumn{9}{c}{$p = 200$} \bigstrut\\
				\hline
				AV    & 0.082 & 0.164 & 0.104 & 0.083 & 0.112 & 0.073 & 0.068 & 0.088 \bigstrut[t]\\
				SD    & 0.150 & 0.436 & 0.309 & 0.275 & 0.146 & 0.131 & 0.111 & 0.145 \\
				SR    & 0.549 & 0.376 & 0.335 & 0.301 & \textbf{0.764} & 0.561 & 0.616 & 0.606 \\
				p-value SR & 0.210 & 0.175 & 0.128 & 0.075 & -     & 0.230 & 0.402 & 0.337 \\
				PT    & 0.064 & 2.577 & 1.772 & 1.986 & 1.619 & 0.240 & 0.515 & 0.190 \bigstrut[b]\\
				\hline
				\hline
			\end{tabular}
		\vspace*{-0.3cm}
		\begin{tablenotes}
			\singlespacing
			\item \leavevmode\kern-\scriptspace\kern-\labelsep 
			Note: Our principal component regression with a fixed number of latent factors equal to 3 or 5, as well as the adaptive procedure for selecting the number of factors (PCR-3, PCR-5, PCR-adaptive) and the ridgeless regression estimator (RRE) are compared to the equally weighted portfolio (EW), the POET estimator of \cite{fan2013} (POET), the non-linear shrinkage estimator of \cite{lw2017} (NL-LW) and the residual based nodewise regression factor model of \cite{caner2022} (RN-FM). The table shows the portfolio average return (AV), standard deviation (SD), Sharpe ratio (SR), the p-value of the \cite{ledoit2008robust} test for equal Sharpe ratios, where for each portfolio size $p$ the method with the largest SR is used as benchmark and tested against all remaining methods (p-value SR), and the portfolio turnover (PT).
		\end{tablenotes}
	\end{threeparttable}
    \end{adjustbox}%
	\label{tab_appl_ntc_p2}%
\end{table}%

\begin{table}[!t]
	\centering
	\caption{Out-of-sample forecasting with transaction costs: January 2005--December 2019}
    \begin{adjustbox}{max width=\textwidth}
	\begin{threeparttable}
			\begin{tabular}{c|cccccccc}
				\hline
				\hline
				Model & EW    & PCR-3 & PCR-5 & PCR-adaptive & RRE   & POET  & NL-LW & RN-FM \bigstrut\\
				\hline
				\multicolumn{9}{c}{$p = 10$} \bigstrut\\
				\hline
				AV         & 0.034 & 0.047 & 0.048 & 0.045 & 0.042 & 0.034 & 0.014 & 0.035 \bigstrut[t]\\
				SD         & 0.132 & 0.133 & 0.134 & 0.128 & 0.135 & 0.132 & 0.120 & 0.130 \\
				SR         & 0.261 & \textbf{0.357} & 0.355 & 0.352 & 0.310 & 0.261 & 0.120 & 0.271 \\
				p-value SR & 0.381 & - & 0.971 & 0.929 & 0.201 & 0.388 & 0.092 & 0.154 \bigstrut[b]\\
				\hline
				\multicolumn{9}{c}{$p = 50$} \bigstrut\\
				\hline
				AV         & 0.066 & 0.183 & 0.166 & 0.163 & 0.121 & 0.067 & 0.071 & 0.075 \bigstrut[t]\\
				SD         & 0.132 & 0.176 & 0.167 & 0.165 & 0.162 & 0.132 & 0.108 & 0.124 \\
				SR         & 0.501 & \textbf{1.045} & 0.991 & 0.990 & 0.748 & 0.510 & 0.658 & 0.600 \\
				p-value SR & 0.081 & - & 0.631 & 0.618 & 0.131 & 0.080 & 0.141 & 0.122 \bigstrut[b]\\
				\hline
				\multicolumn{9}{c}{$p = 100$} \bigstrut\\
				\hline
				AV         & 0.069 & 0.123 & 0.074 & 0.065 & 0.004 & 0.074 & 0.029 & 0.075 \bigstrut[t]\\
				SD         & 0.138 & 0.257 & 0.198 & 0.179 & 0.260 & 0.138 & 0.111 & 0.130 \\
				SR         & 0.499 & 0.479 & 0.372 & 0.360 & 0.016 & 0.536 & 0.256 & \textbf{0.575} \\
				p-value SR & 0.072 & 0.799 & 0.501 & 0.472 & 0.061 & 0.383 & 0.167 & - \bigstrut[b]\\
				\hline
				\multicolumn{9}{c}{$p = 200$} \bigstrut\\
				\hline
				AV         & 0.077 & 0.131 & 0.057 & 0.040 & 0.096 & 0.072 & 0.060 & 0.078 \bigstrut[t]\\
				SD         & 0.142 & 0.382 & 0.255 & 0.215 & 0.134 & 0.126 & 0.101 & 0.132 \\
				SR         & 0.544 & 0.343 & 0.224 & 0.187 & \textbf{0.714} & 0.573 & 0.588 & 0.587 \\
				p-value SR & 0.461 & 0.338 & 0.160 & 0.096 & - & 0.526 & 0.567 & 0.574 \bigstrut[b]\\
				\hline
				\hline
			\end{tabular}
		\vspace*{-0.3cm}
		\begin{tablenotes}
			\singlespacing
			\item \leavevmode\kern-\scriptspace\kern-\labelsep 
			Note: Our principal component regression with a fixed number of latent factors equal to 3 or 5, as well as the adaptive procedure for selecting the number of factors (PCR-3, PCR-5, PCR-adaptive) and the ridgeless regression estimator (RRE) are compared to the equally weighted portfolio (EW), the POET estimator of \cite{fan2013} (POET), the non-linear shrinkage estimator of \cite{lw2017} (NL-LW) and the residual based nodewise regression factor model of \cite{caner2022} (RN-FM). The table shows the portfolio average return (AV), standard deviation (SD), Sharpe ratio (SR) and the p-value of the \cite{ledoit2008robust} test for equal Sharpe ratios, where for each portfolio size $p$ the method with the largest SR is used as benchmark and tested against all remaining methods (p-value SR).
		\end{tablenotes}
	\end{threeparttable}
    \end{adjustbox}%
	\label{tab_appl_tc_p1}%
\end{table}%

\clearpage
\setcounter{figure}{0}
\renewcommand{\thefigure}{C.\arabic{figure}}

To further investigate the role of the idiosyncratic errors in double descent in RRE, we consider two alternative specifications for $\Sigma_U$. In the first alternative specification, the diagonal entries of $\Sigma_U$ are drawn independently from $\text{Unif}(0.1,0.3)$. In the second specification, the idiosyncratic error variance is scaled with $p$ such that $tr(\Sigma_U) / n = 2$ is  constant as the dimensionality increases. These two experiments allow us to distinguish between the role of the overall scale of the idiosyncratic error covariance and the role of its interaction with the quantity $\delta_n$, which depends on the dimensionality.

\subsection*{C.1. Smaller, constant idiosyncratic variances}

Under the first alternative error specification, the double descent pattern remains present, although the scale of the idiosyncratic error covariance is substantially smaller than under the baseline error specification. The results are shown in Figure \ref{fig:rre_dd_mc2} and Table \ref{tab:rre_dd_mc2}, with the latter including an in-depth subgraph showing regime of $p>n$. In this specification, $Eigmax(\Sigma_U)$ remains approximately constant at $0.3$ for all values of $p$, compared with values ranging from $8.706$ to $20.709$ under the baseline specification. Hence, the variance component of the theoretical bound receives a substantially smaller weight throughout the entire range of $p$. Nevertheless, the reduction in the variance component associated with the increase in $\delta_n$ remains operative. In particular, in the post-interpolation region, $\delta_n$ increases from $0.198$ at $p=200$ to $0.998$ at $p=1000$. Since the relevant variance contribution contains the factor $\sqrt{\frac{\log n+n}{n\delta_n}},$ this increase in $\delta_n$ continues to reduce the variance component as $p$ increases beyond $n$. The fact that $Eigmax(\Sigma_U)$ is small and essentially constant does not eliminate this mechanism. It merely reduces its overall scale. Thus, the variance contribution can still decline sufficiently strongly after the interpolation region to generate the observed double descent pattern. However, here the bias is reduced a lot due to a large increase in signal-to-noise ratio $\bar{\xi}$ with the increase in $p$. So the double descent is affected also by bias decrease.

\begin{figure}[!t]
    \centering
    \caption{Maximum Absolute Linear-Combination Error of the RRE Precision Matrix Estimator under the First Alternative Specification of $\Sigma_U$ for Different Values of $p$ ($n=200$, $K=3$)}
    \includegraphics[width=0.75\linewidth]{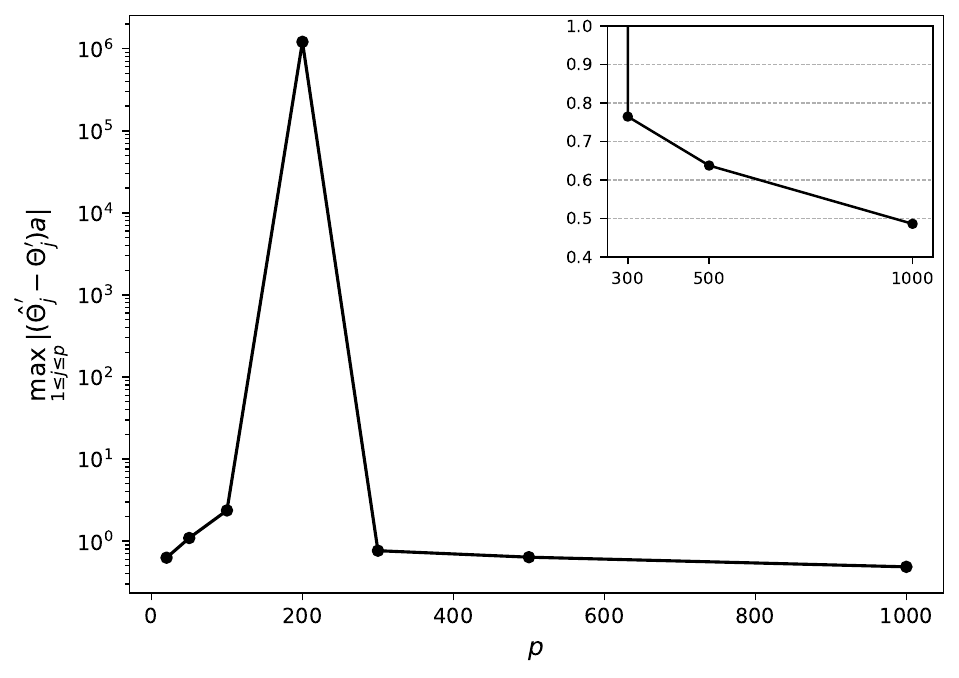}
    \label{fig:rre_dd_mc2}
\end{figure}
\begin{table}[!t]
	\centering
	\caption{Quantities Entering the Convergence Rate of the RRE Precision Matrix Estimator under the First Alternative Specification of $\Sigma_U$ for Different Values of $p$ ($n=200$, $K=3$)}
    \begin{adjustbox}{max width=\textwidth}
	\begin{threeparttable}
    \begin{tabular}{cccccc}
			\hline
			\hline
			$p$ & $\max_{1 \le j \le p} \| a_j\|_2^2$  & $Eigmax(\Sigma_U)$ & $d_{2n}$ & $\bar{\xi}$ & $\delta_n$ \bigstrut\\
            20 & 7.267  & 0.291 & 26.383  & 50.067   & 0.018 \\
            50 & 8.539  & 0.296 & 48.070  & 182.618  & 0.049 \\
            100 & 9.840 & 0.298 & 76.132  & 407.986  & 0.099 \\
            200 & 11.004 & 0.299 & 115.710 & 885.943  & 0.198 \\
            300 & 11.508 & 0.299 & 146.141 & 1366.015 & 0.298 \\
            500 & 12.325 & 0.300 & 200.507 & 2316.174 & 0.499 \\
            1000 & 13.169 & 0.300 & 293.768 & 4691.124 & 0.998 \bigstrut[b]\\
			\hline
			\hline
		\end{tabular}
		\vspace*{-0.3cm}
		\begin{tablenotes}
			\scriptsize
			\singlespacing
			\item \leavevmode\kern-\scriptspace\kern-\labelsep 
			Note: Key quantities entering the convergence rate of the RRE precision matrix estimator under the first alternative data generating process for the idiosyncratic errors, with diagonal elements of $\Sigma_U$ drawn independently from $\mathrm{Unif}(0.1,0.3)$. The table reports the maximum row norm of the factor loadings, the largest eigenvalue of the idiosyncratic covariance matrix $Eigmax(\Sigma_U)$, and the quantities $d_{2n}$, $\bar{\xi}$, and $\delta_n$ appearing in the theoretical convergence rate.
		\end{tablenotes}
	\end{threeparttable}
	\end{adjustbox}%
	\label{tab:rre_dd_mc2}%
\end{table}%
\subsection*{C.2. Constant total noise relative to sample size}

The results for the second alternative error specification are depicted in Figure \ref{fig:rre_dd_mc3} and Table \ref{tab:rre_dd_mc3}. Here, the diagonal elements of $\Sigma_U$ are scaled such that $tr(\Sigma_U)/n=2$. As a consequence, $Eigmax(\Sigma_U)$ decreases sharply with $p$, from $29.028$ at $p=20$ to only $0.599$ at $p=1000$, while $\delta_n$ remains essentially constant, increasing only from $1.855$ to $1.997$. Hence, two changes occur simultaneously. First, the overall weight of the idiosyncratic variance component falls strongly as $p$ increases. Second, and more importantly for the double descent mechanism described in Corollary 6, the increase in $\delta_n$ that generated the post-interpolation reduction in the variance component under the baseline and first alternative error specification is largely absent. In the first alternative error specification in Figure \ref{fig:rre_dd_mc2}, $Eigmax(\Sigma_U)$ is approximately constant, so the increase in $\delta_n$ remains the main source of a reduction in the variance contribution for $p>n$. Consequently, the mechanism responsible for the double descent pattern remains effective despite the substantially lower noise level. In the second specification, in Figure \ref{fig:rre_dd_mc3}, by contrast, the idiosyncratic variance scale already decreases rapidly with $p$ well before the interpolation region, while $\delta_n$ is almost unchanged once $p$ increases. The variance component is therefore progressively weighted down as dimensionality rises, but this change is not driven by an increase in $\delta_n$. As a result, there is no sufficiently strong change in the variance contribution specifically for $p>n$ to generate the post-interpolation decline characteristic of double descent. Consequently, the double descent pattern disappears under the second alternative specification, where the estimation error no longer declines after the interpolation region, as observed under the baseline and first alternative specifications.

The remaining quantities are broadly consistent with this interpretation. The effective dimension $d_{2n}$ increases from $26.824$ to $296.379$ in the second specification, while $\bar{\xi}$ increases sharply from $0.523$ to $2339.823$. Thus, as in the other specifications, the increase in effective dimension is partly offset by the improvement in the signal-to-noise measure. The key difference across the experiments therefore lies not in the behavior of these bias-related quantities, but in the interaction between the idiosyncratic variance scale and $\delta_n$.

\begin{figure}
    \centering
    \caption{Maximum Absolute Linear-Combination Error of the RRE Precision Matrix Estimator under the Second Alternative Specification of $\Sigma_U$ for Different Values of $p$ ($n=200$, $K=3$)}
    \includegraphics[width=0.75\linewidth]{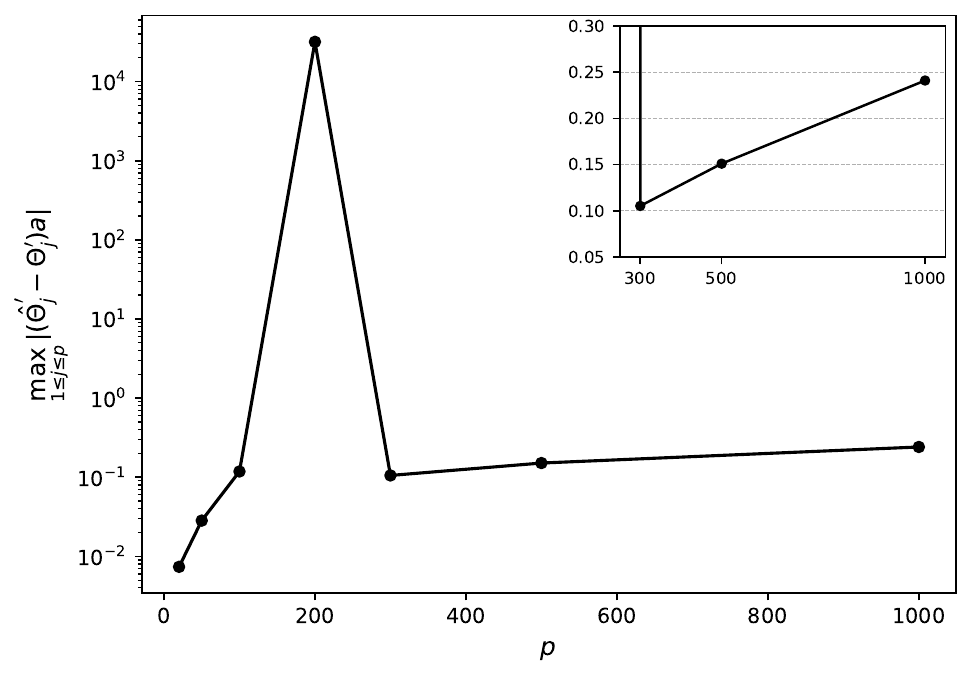}
    \label{fig:rre_dd_mc3}
\end{figure}

\begin{table}[!t]
	\centering
	\caption{Quantities Entering the Convergence Rate of the RRE Precision Matrix Estimator under the Second Alternative Specification of $\Sigma_U$ for Different Values of $p$ ($n=200$, $K=3$)}
    \begin{adjustbox}{max width=\textwidth}
	\begin{threeparttable}
    \begin{tabular}{cccccc}
			\hline
			\hline
			$p$ & $\max_{1 \le j \le p} \| a_j\|_2^2$  & $Eigmax(\Sigma_U)$ & $d_{2n}$ & $\bar{\xi}$ & $\delta_n$ \bigstrut\\
            20 & 7.198  & 29.028 & 26.824  & 0.523    & 1.855 \\
            50 & 8.577  & 11.851 & 48.741  & 4.606    & 1.941 \\
            100 & 9.800 & 5.971  & 76.142  & 20.541   & 1.970 \\
            200 & 10.733 & 2.991 & 115.700 & 88.599   & 1.985 \\
            300 & 11.561 & 1.999 & 147.121 & 202.094  & 1.990 \\
            500 & 12.271 & 1.200 & 198.048 & 579.463  & 1.994 \\
            1000 & 13.151 & 0.599 & 296.379 & 2339.823 & 1.997 \bigstrut[b]\\
			\hline
			\hline
		\end{tabular}
		\vspace*{-0.3cm}
		\begin{tablenotes}
			\scriptsize
			\singlespacing
			\item \leavevmode\kern-\scriptspace\kern-\labelsep 
			Note: Key quantities entering the convergence rate of the RRE precision matrix estimator under the second alternative data generating process for the idiosyncratic errors, where the error variance is scaled with $p$ such that $tr(\Sigma_U) / n = 2$ is  constant as the dimensionality increases. The table reports the maximum row norm of the factor loadings, the largest eigenvalue of the idiosyncratic covariance matrix $Eigmax(\Sigma_U)$, and the quantities $d_{2n}$, $\bar{\xi}$, and $\delta_n$ appearing in the theoretical convergence rate.
		\end{tablenotes}
	\end{threeparttable}
	\end{adjustbox}%
	\label{tab:rre_dd_mc3}%
\end{table}%

\newpage

\clearpage

\clearpage
\setcounter{table}{0}
\renewcommand{\thetable}{D.\arabic{table}}
\section*{Appendix D: Extended Literature Comparison}
This appendix collects the detailed comparison moved from the original introduction. The brief positioning statement is retained in the main paper. The distinctions below should be read within each paper's assumptions and loss function.

\begin{table}[!t]
\centering
\caption{Relation to selected regression and precision-matrix methods}
\label{tab:relatedwork}
\begin{threeparttable}
\footnotesize
\renewcommand{\arraystretch}{1.25}
\begin{adjustbox}{max width=\textwidth}
\begin{tabular}{>{\raggedright\arraybackslash}p{2.7cm}>{\raggedright\arraybackslash}p{2.1cm}>{\raggedright\arraybackslash}p{2.0cm}>{\raggedright\arraybackslash}p{2.1cm}>{\raggedright\arraybackslash}p{4.5cm}}
\hline\hline
Paper & Target & Factor treatment & Structural restriction & Relation to this paper \\
\hline
\cite{bbsmw21} & Regression prediction & Latent representation & Factor regression & Source of the projection class and error decomposition; here the target is all precision rows. \\
\cite{bsmw22} & Interpolating prediction & Latent representation & Factor regression & Related minimum-norm analysis; precision estimation additionally requires diagonal normalization. \\
\cite{caner2019} & Precision matrix and portfolios & No factor plug-in & Sparse precision rows & Penalized row regressions; here the estimator is dense and unpenalized. \\
\cite{canerkock2018} & Parameter inference & Observed regressors & Sparse nuisance structure & Debiasing and confidence regions; here the target is the precision matrix itself. \\
\cite{cd2025} & Return precision matrix & Nonlinear fit using observed factors & Nonlinear factor model & Estimates a factor-related function; here latent factors are used in the analysis without a plug-in fit. \\
\hline\hline
\end{tabular}
\end{adjustbox}
\end{threeparttable}
\end{table}

Several studies in the statistics literature have examined the estimation of precision matrices.
An excellent survey is provided by \cite{crz2016}, which presents neighborhood- and penalized-based 
approaches. In Section 3 of their paper, they provide two major approaches to precision matrix 
estimation. The first class is neighborhood-based and is closely related to nodewise regression, 
where a lasso regression is performed for each variable against all remaining variables as regressors. Alternatives to lasso, such as the Dantzig selector, can also be used within this framework (see, e.g., \cite{cl2011} and \cite{caner2019}). The second major approach is based on penalized likelihood estimation. As shown by \cite{blp2020}, lasso-type penalties in this context can induce shrinkage bias and lead to 
overly sparse precision matrix estimates, with many zero entries in each row. In a penalized 
likelihood-based setting, they reformulate the problem as a mixed-integer optimization problem, 
which is solvable using convex optimization techniques. This approach produces better results than 
lasso-based approaches. Nevertheless, sparsity remains a central structural assumption underlying these two major classes of precision matrix estimators.

A distinct approach, commonly used in finance and statistics, derives the precision matrix by 
inverting the covariance matrix using the Sherman--Morrison--Woodbury (SMW) formula. In this 
framework, sparsity is assumed in the covariance matrix of the idiosyncratic errors, while asset 
returns are modeled through a factor structure (see, e.g., \cite{fan2013} and \cite{caner2022}). Our approach 
is different from the approaches discussed in the literature. First of all, we are not assuming 
exact sparsity in the precision matrix of the outcomes. 
Second, the SMW formula is used only in deriving the population formula (Appendix B, (B.16)), never as a plug-in step in the estimator itself.

This allows us to avoid imposing a particular structural restriction on the outcomes underlying the covariance inversion. Instead we use a new precision matrix formula without
sparsity and rely on hidden factors to generate each outcome. This effectively ensures that we do
not benefit from the SMW matrix inversion formula.
 Thus, our framework combines factor-model-estimation-free dense estimators with high-dimensional consistency through hidden factor-based dimension reduction, without requiring factor estimation or factor identification.

Another approach is nodewise based. The nodewise regression based estimator of
\cite{caner2019} uses sparsity of the precision matrix of outcomes. This restriction effectively assumes that a potentially large number of entries of the inverse covariance matrix are exactly zero. In case of normally distributed data, sparsity of the precision matrix implies that a large number of conditional correlations between outcomes are zero, which seems very restrictive in fields like finance and economics.  Its theory is relatively simple, and uses lasso based techniques. Our method uses regressions of each outcome on the remaining outcomes, but does not use lasso penalization or impose row sparsity. Nodewise regression as shown in \cite{mein2006} uses lasso based row estimation of the precision matrix. As stressed above, equations (2.1) and (2.2) of \cite{caner2019} employ lasso based penalized estimation, compared with our unpenalized dense estimator in (10) of Definition 1, which is designed to accommodate a dense precision matrix.
The  rate of convergence of \cite{caner2019} is $\bar{s}\sqrt{\log p /n}$, where $\bar{s}$ is the maximum number of nonzero elements per row of $\Theta$; if $\bar{s}=p$, that estimator is not consistent, whereas ours remains consistent (see the discussion after Corollary 4).   There is also no finite sample upper bounds  in \cite{caner2019} linking the estimator to the signal-to noise and complexity.

The residual nodewise regression estimator of \cite{caner2022} crucially depends on observed factors, and imposes a sparse precision matrix of the errors. The technique used in  \cite{caner2022} cannot be extended to hidden factors, since their observed-factor structure is fit by OLS, which
is not applicable to hidden factor structures that must instead be handled via the projection-based estimators in Definition 1. Also \cite{caner2022} does not link its rate of convergence to a signal-to-noise ratio, whereas our rate $r_{w1}$ in (9) is explicitly governed by the signal-to-noise ratio.

We discuss \cite{cd2025} separately and in detail, since it is the paper similar to ours in motivation: where they invoke a signal-to-noise-ratio narrative for a weak factor regime, and evaluate the model performance on comparable equity portfolio applications. However, the factors are observed in \cite{cd2025}, specifically the outcomes are explained  via an unknown non-linear function of the observed factors.
This somewhat similar setup is a feature of the sub-literature on factor-based precision matrix estimation in finance rather than a borrowing from one paper to the other.   The two papers differ, however, in how the precision matrix is actually built from that similar decomposition. \cite{cd2025} estimates the unknown non-linear function of observed factors directly with a deep neural network, forms a dense residual covariance plus a low-rank factor component, and inverts the result with a novel matrix inversion formula in equations (5.2) and (5.3) of their paper due to zero eigenvalue of the covariance matrix of non-linear factors. Consequently, their estimator is consistent only when the number of outcomes does not exceed the sample size ($p \le n$), and its performance depends heavily on the quality of the non-linear factor fit. Our estimator never estimates the unobserved factors: the latent factor representation in (1) is used exclusively inside the proof of Theorem 1 and is not used as a plug-in component of the estimator itself. We likewise do not invoke the matrix inversion formula in equations (5.2) and (5.3) of \cite{cd2025}. The RRE special case of our class remains consistent for $p$ growing much faster than $n$ (Corollary 3), a regime \cite{cd2025} does not cover. The two approaches are therefore best read as complementary: \cite{cd2025} trades a nonparametric factor-estimation step for potentially sharper rates when $p \le n$, while our class trades that estimation step away entirely at the cost of the signal-to-noise conditions in Assumption 5.

\cite{canerkock2018} desparsifies a conservative lasso to obtain an honest confidence region for a single low-dimensional parameter in a high-dimensional model; the sparsity and the debiasing correction are both specific to that scalar-parameter target and do not apply when the object of interest is the full $p\times p$ dense precision matrix, which is what we estimate here.

Unlike nodewise regression of \cite{mein2006}, our method does not impose any sparsity. Nodewise
regression uses an $l_1$-norm based penalization to form the rows of the precision matrix. Doing
so, the underlying assumption is the imposition of sparsity in the precision matrix, which may be
interpreted as a strong assumption. An excellent paper by \cite{fan2013} provides a statistical
high-dimensional analysis of the covariance matrix of the outcomes via hidden factor-based
estimation. Similar to our approach, it relies on hidden factors, but uses a thresholding based
principal components estimator for the covariance matrix, and then inverts this through the SMW
formula. The key difference is that our estimator never requires a plug-in estimate of the factors, which are used only for the proof of consistency.

Our paper is related to \cite{bbsmw21} and \cite{bsmw22}.  There are three major differences between our paper and their papers. 
First, their papers analyze 
a general linear class of estimators in hidden factor models in a linear regression model. They consider only one outcome variable and 
how it can be predicted with this new general class of estimators. Our analysis, on the other hand, 
considers the precision matrix estimation of outcome variables in linear regression via hidden 
factors. We use a linear linkage between outcome variables to obtain 
a precision matrix estimator.  Specifically, equations (11), (12), and (13) are forming the new precision matrix estimator and clearly show how different this is from only a part of our off-diagonal term which is (10) and is used in \cite{bbsmw21}.

The second difference is a theoretical difference, and is related to the estimation of the 
main diagonal terms in the precision matrix.  The proof technique and ideas in \cite{bbsmw21} and 
\cite{bsmw22} do not simply extend to this case.  This off-diagonal term estimation proof is new and forms Theorem 2 of our paper. The precision matrix estimation is different from 
the risk analysis of a single outcome asset. We also provide an explicit connection in our 
estimation error upper bounds to the signal-to-noise ratio. 

Third, our theoretical results provide uniform guarantees over all $j=1,\cdots, p$ outcomes, when $p$ may be growing. Our uniformity proofs are a new addition to theory, and can be seen in the proof of Theorem 1, between (A.7) and (A.18) in Step 1, and in Step 2c, they can be seen in (A.36) and (A.47). Also, for sub-cases of our estimators, the uniformity arguments are used in the proofs of Corollaries 1-3.

There is a recent paper that analyzes the precision matrix in the context of minimizing the portfolio variance in finance \cite{cdsz26}. Note that our precision matrix estimator uses a ridgeless regression estimator for the off-diagonal term estimation based on the precision matrix formula. This is different from ridgeless and ridgelet regression estimators in \cite{cdsz26}. In their cases, ridgeless uses the pseudo-inverse of the sample second moments of returns as a precision matrix estimator, whereas the ridgelet precision matrix estimator solves the zero-variance portfolio problem. In a creative and novel way, they show that the ridgelet has desirable properties.  Their approach is tailored to solving a portfolio variance problem, whereas we start with the population precision matrix formula and, for the numerator, use a ridgeless estimator directly for the off-diagonal terms.  Their ridgelet estimator is similar to a linear shrinkage estimator of \cite{lw2004}. To keep the estimator names brief, we refer to our ridgeless estimator as ridgeless from now on and denote it as RRE; this should not be confused with the estimators in \cite{cdsz26}, as discussed. Their estimators are specialized to solve a key financial problem. Our estimators are general and depend on the population precision matrix formula, which is tied to the hidden factor structure.

Another important class of estimators is based on shrinkage. For example, \cite{lw2004} proposes a linear 
shrinkage estimator for covariances in an iid data setting, motivated in part by the invertibility problems that arise for the sample covariance matrix in high dimensions. Their estimator for covariance is dense and model-free, hence making a major contribution. However, unlike our estimators, the estimator converges to a 
modified version of the covariance consistently, not to the covariance matrix itself. \cite{lw2017} 
propose a non-linear shrinkage estimator. This estimator specifically analyzes out-of-sample 
variance loss in an iid data setting, and develops a modified covariance matrix estimator that 
increases the small eigenvalues and decreases the large ones. It is a model-free dense estimator, but the out-of-sample variance loss limit is different from the population one. Unlike our 
estimator, the precision matrix estimator in their paper is not consistent.

\section*{Appendix E: Discussion of Assumptions}
The formal assumptions are stated in the main paper. This appendix preserves the extended discussion and examples from its original model section.%

\subsection*{E.1. Discussion of Assumption 1}
Assumption 1 is a standard convention used in hidden factor  literature (see, e.g., \cite{bbsmw21} and \cite{bsmw22}.\footnote{Observe that (i)-(iii) implies that rows of $F, Y$, and $U$ are iid. Note also that independence of errors implies $Eu_{jt}U_{-jt} = 0$, which is a sparsity restriction
on the covariance matrix of errors.} This is another assumption often present in the literature  (see, e.g.,  \cite{fan2008}) and amounts to a diagonal error covariance matrix, i.e. an exact rather than approximate factor structure. This is a genuine scope restriction: it rules out the approximate factor model of \cite{cr1983}, in which $\Sigma_U$ is only required to have bounded eigenvalues and the idiosyncratic components may be weakly cross-sectionally correlated, which is the standard assumption in the approximate-factor asset pricing literature.  We impose the exact-factor version here because our proof of consistency, through Assumption 4 below, relies on sub-Gaussian concentration inequalities for quadratic forms in $\tilde{U}_{-j,t}$, and the cleanest version of these inequalities is available when $\Sigma_{U,-j}$ is diagonal so that $\Sigma_{U,-j}^{1/2}$ and its inverse are directly available in closed form; extending our estimator and its non-asymptotic bounds to the \cite{cr1983} approximate factor case, e.g. by allowing $\Sigma_U$ to be a bounded-eigenvalue matrix with sparse or banded off-diagonal entries, would require new concentration arguments for the diagonal-term estimator in Theorem 2 and is left for future work. Relaxing this
part of the assumption is certainly an interesting topic well worth exploring further.

 Relaxing the iid structure of data for financial applications is a good idea; however, at this point, our proofs, via Assumption 4 below, also rely on sub-Gaussian-iid-based concentration inequalities for quadratic forms.     We are not imposing sparsity on the links between outcome variables, as is done in the nodewise regression context.

\subsection*{E.2. Discussion of Assumption 2}
Assumption 2(i) is also standard in high-dimensional statistics literature. It
allows for non-singular noise matrices, and large noise magnitudes. Since $\Sigma_{U,-j}$ is diagonal and diagonal entries are error variances, we use a common upper bound for a single variance in (i) and maximum variances over errors in (ii). We note for clarity that, because $\Sigma_{U,-j}$ is diagonal under Assumption 1, the condition $\min_j Eigmin(\Sigma_{U,-j}) \ge c$ in (i) is implied by the individual-variance bound $\min_j \sigma_j^2 \ge c$. (ii) shows how there may be differences in asset error variances, and this may affect bounds in finite samples. Note that $\beta$ is the second highest variance in $\Sigma_U$ due to diagonal nature of covariance matrix of errors.  Asymptotically, our results show that $\Gamma, \beta$ distinction plays no role.

\subsection*{E.3. Discussion of Assumption 3}
With $C_p>2$, we can obtain $p>n$, and we cover high-dimensional cases. In case of 
    $0<C_p \le 2$, we cover low dimensional cases, $p \le n$. This assumption also allows $p = \gamma n$, with $\gamma \in [0, \infty)$.
    When $p,n$ grow together, this means $p$ can grow at most at rate 
    $n^{C_p/2}$. This is for adaptive principal components estimator as well as OLS estimator with $0 < C_p < 2$.
    	
	We allow the spectral norm of $A_{-j}: (p-1) \times K$ matrix to be diverging in $n$, reflective of the high-dimensional nature of $A_{-j}$, where both dimensions are allowed to grow. Specifically in (ii), we allow for pervasive factors with $\max_{j} \| A_{-j} \|_2 = C \sqrt{p}$, $C $ being a generic positive constant as well as weak factor structures if $\max_j \| A_{-j} \|_2 = O (d_{1n})$, with $d_{1n}/\sqrt{p}\to0$ with $p \to \infty, n \to \infty$. Hence, our assumption can also cover  weak factor loadings. We state the bound in terms of $\max_{1 \le j \le p} \| A_{-j}\|_2$, the maximum over the $p$ row-deleted loading matrices, rather than $\|A\|_2$ for the full loading matrix, because $A_{-j}$, not $A$, is exactly what appears in the $j$-th row's precision-matrix formula in Lemma 1 and in the uniform-in-$j$ bounds of Theorem 1; the two are of the same order, since $\|A_{-j}\|_2 \le \|A\|_2$ for every $j$.

\subsection*{E.4. Discussion of Assumption 4}
We have  a discussion about the role of the signal-to-noise ratio in a specific scenario. Let us assume a weaker factor structure in Assumption 3(ii), which is $\max_j \| A_{-j}\|_2 = O (d_{1n})$.
    Since $A_{-j}: (p-1) \times K$  for all matrices $j=1,\cdots,p$, via Assumption 3, if the largest eigenvalue of $A_{-j}' A_{-j}$   grows at the same rate $K$-th largest eigenvalue,   we can assume $\lambda_K (A_{-j}' A_{-j}) \asymp d_{1n}^2$ and since we impose $ \max_{1 \le j \le p } \| \Sigma_{U,-j} \|_2 < \Gamma$ by Assumption 3
	then
	$\lambda_K (A_{-j} \Sigma_f  A_{-j}') \ge Eigmin (\Sigma_f) \lambda_K (A_{-j} A_{-j}') \ge c \lambda_K (A_{-j}' A_{-j}) \asymp d_{1n}^2 .$
	With these additional lower-eigenvalue conditions, the illustrative lower bound is (8) in the main paper.
	
	Depending on how fast/slow $d_{1n}$ grows the signal-to-noise ratio lower bound in (8) can fluctuate between fairly small and fairly large quantities.

\subsection*{E.5. Discussion of Assumption 5}
While the meaning of  Assumption 5 will become clearer when we get to our final results, a few introductory remarks regarding the origin of the assumption are in place. First,  Assumption 5(i)
is a technical assumption that facilitates the proof of Theorem 2  for the finite
sample results. In the case of $p<n$, we need more than one factor.
It is not needed when $p\ge n$. We do not require $K$ to grow without bound: Assumption 5(i) is stated only for the $n>p$ case and is compatible with a fixed, finite $K$, since $K > \log n /\log p$ can hold for a constant $K$ once $p$ is large enough relative to $n$, for example $p=n/2$ case, where $K>1$ with large $n,p$. When $K$ is allowed to grow with $n$ and $p$, as in Assumption 5(ii)-(iii), this is not a pathology specific to our setting but a standard device in the weak- and growing-factor asymptotics literature (see, e.g., \cite{Onatski2012} and \cite{Uematsu2022}), which allows the framework to nest both classical low-dimensional factor models with fixed $K$ and settings in which the effective number of pervasive and weak factors expands with the cross-sectional dimension.

Assumption 5(ii) relates to the tradeoff between the number of factors and the time
span. In particular, it points out that the number of factors should be much smaller than the time span. Assumption 5(iii) involves
stochastic terms and relates to consistency and rate of convergence of the off-diagonal
terms estimates. As we will see later on, the role of Assumption 5(iii) will be discussed separately for adaptive PCR and
RRE estimators. But to provide some basic intuition in a simple subcase, take $p>n$, let $K=1$, and 
$\| \Sigma_{U,-j}\|_2 = I_{p-1}$, with $var(f_t)=1$, and set $r_{w1} := 1 / \sqrt{\xi}$
\[ d_{2n} r_{w1} = \frac{\max_{j} \sqrt{\sum_{i \neq j} (A_{i,-j} a_j)^2}}
{\min_j \sqrt{\sum_{i \neq j} A_{i,-j}^2
}} \to 0
,\]
 where $A_{i,-j}$ is $i$-th element of vector $A_{-j}$ (see that this is a vector since $K=1$). So this is basically factor loadings correlations among themselves declining to zero.

\end{document}